\documentclass[%
reprint,
 amsmath,amssymb,
 aps,
]{revtex4-2}

\usepackage{amsmath,amssymb}
\usepackage{graphicx}
\usepackage{dcolumn}
\usepackage{bm}
\usepackage{physics}
\usepackage{comment}
\usepackage{ulem}
\usepackage{xcolor}

\newcommand{\YN}[1]{{\color{red} #1}}
\begin{document}

\preprint{APS/123-QED}

\title{Frequency-Dependent Squeezing for Gravitational-Wave Detection through Quantum Teleportation}

\author{Yohei Nishino$^{1,2}$}
 \email{yohei.nishino@grad.nao.ac.jp}
\author{Stefan Danilishin$^{3,4}$}
\author{Yutaro Enomoto$^{5}$}
\author{Teng Zhang$^{6}$}

\affiliation{$^1$Department of Astronomy, University of Tokyo, Bunkyo, Tokyo 113-0033, Japan,}
\affiliation{$^2$Gravitational Wave Science Project, National Astronomical Observatory of Japan (NAOJ), Mitaka City, Tokyo
181-8588, Japan,}
\affiliation{$^3$Department of Gravitational Waves and Fundamental Physics, Maastricht University, 6200 MD, Maastricht, The Netherlands,}
\affiliation{$^4$Nikhef – National Institute for Subatomic Physics, Science Park 105,
1098 XG Amsterdam, The Netherlands,}
\affiliation{$^5$Department of Applied Physics, The University of Tokyo, Bunkyo,
Tokyo, 113-0033, Japan,}
\affiliation{$^6$The University of Birmingham, School of Physics and Astronomy, and
Institute of Gravitational Wave Astronomy, University of Birmingham,
Edgbaston, Birmingham, B15 2TT, United Kingdom.}

\date{\today}

\begin{abstract}
Ground-based interferometric gravitational wave detectors are highly precise sensors for weak forces, limited in sensitivity across their detection band by quantum fluctuations of light. Current and future instruments address this limitation by injecting frequency-dependent squeezed vacuum into the detection port, utilizing narrow-band, low-loss optical filter cavities for optimal rotation of the squeezing ellipse at each signal frequency. This study introduces a novel scheme of such vacuum injection employing the principles of quantum teleportation, which works the same as arbitrary number of filter cavities without additional kilometer-scale infrastructure. We applied this scheme to detuned signal recycled-Fabry-P\'{e}rot--Michelson interferometers, which is the baseline design of the low-frequency detector within the Einstein Telescope xylophone detector. It is shown that  our scheme achieves broadband suppression of quantum noise without requiring additional filter cavities or modifications to the core optics of the main interferometer.
\end{abstract}

\maketitle


\section{\label{sec:level1}Introduction}

\begin{figure*}
    \centering
    \includegraphics[scale=0.45]{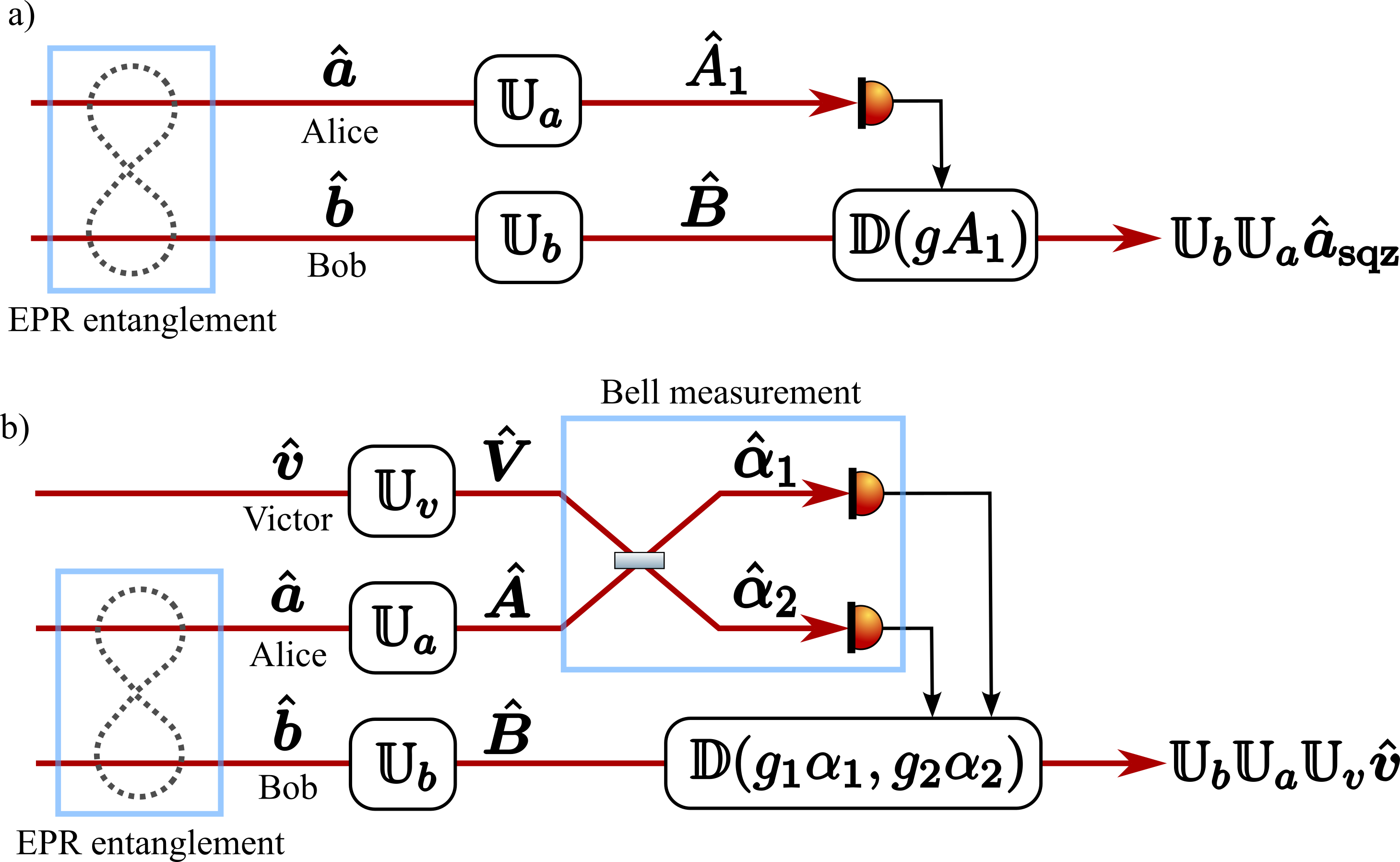}
    \caption{
    Schematics of a) EPR steering and b) quantum teleportation. We use two-photon formalism, where each beam is described by quadrature amplitude operators, where subscripts 1 and 2 denote amplitude and phase quadrature, respectively. All transformations of the beams are represented by transfer matrices, and $\pmb{\hat{a}}_\mathrm{sqz}$ is a squeezed vacuum. The Bell observables here are defined as $\hat{\alpha}_1=(\hat{V}_1-\hat{A}_1)/\sqrt{2}$ and $\hat{\alpha}_2 = (\hat{V}_2+\hat{A}_2)/\sqrt{2}$. When Victor, Alice and Bob's state undergo physical transformations, \YN{$\mathbb{U}_v$, $\mathbb{U}_a$, which are assumed to be phase rotations, and $\mathbb{U}_b$, which corresponds to the ponderomotive squeezing, the teleported state is transformed accordingly to $\mathbb{U}_{b}\mathbb{U}_{a}\mathbb{U}_{v}\pmb{\hat{v}}$ accordingly.}
    }
    \label{Fig_CV_teleportation}
\end{figure*}

In 2015, the Laser Interferometric Gravitational-wave Observatory (LIGO)~\cite{Aasi_2015} achieved a milestone by detecting the first gravitational wave from a binary black hole (BBH) merger~\cite{PhysRevLett.116.061102}. This heralded the beginning of the gravitational-wave astronomy, following which the LIGO-Virgo-KAGRA~\cite{Acernese_2015,KAGRA_2019} collaboration has identified over 90 gravitational wave events~\cite{PhysRevX.9.031040, PhysRevX.11.021053, theligoscientificcollaboration2021gwtc3, PhysRevD.101.083030}. 
Furthermore, the third generation detectors, \textit{i.e.} the Cosmic Explorer~\cite{Abbott_2017} and the Einstein Telescope~\cite{Hild_2011}, striving for tenfold greater sensitivity, will empower the exploration of gravitational wave signals from the events spanning the entire history of the universe~\cite{Pieroni_2022, Maggiore_2020}. This endeavor will illuminate unresolved inquiries in fundamental physics and cosmology~\cite{Sathyaprakash_2012,Shankaranarayanan_2022,sathyaprakash2019cosmology}.

Gravitational wave detectors serving as highly precise displacement measurement instruments, are limited by quantum noise across most of their frequency band. At low frequencies, the optimal sensitivity of a conventional detector faces the constraints of the standard quantum limit (SQL) \cite{1968JETP.26.831}, as a natural consequence of Heisenberg's uncertainty principle~\cite{PhysRevLett.45.75}. To overcome the SQL, scientists have proposed a variety of technologies based on the quantum non-demolition measurement principle~\cite{PhysRevLett.40.667, RevModPhys.68.1}. 
The mainstream approach for quantum noise reduction in the existing and future detectors is frequency-dependent squeezing (FDS) injection~\cite{PhysRevD.65.022002, PhysRevD.23.1693}. The injected squeezed vacuum is produced by means of degenerate parametric down-conversion in nonlinear optical crystals. The squeezing angle at each frequency is optimized by passive optical cavities to cancel out the rotation of the squeezing ellipse due to ponderomotive effect within the interferometer. 
The current gravitational-wave detectors employ a single filter cavity for their tuned-broadband configurations~\cite{MIZUNO1993273}, as it is sufficient for broadband compensation of the ponderomotive squeezing angle in the narrow-band approximation~\cite{PhysRevD.66.122004, PhysRevD.81.122002}. However, when this approximation breaks down, a larger number of filter cavities become necessary (see prior studies~\cite{PhysRevD.68.042001, PhysRevD.98.044044} for example). In either case, as the scale of the detectors increases, these filter cavities, which are currently at the hundred-meter scale~\cite{PhysRevX.13.041021, PhysRevLett.131.041403}, will need to reach the kilometer scale.

The installation of kilometer-scale filter cavities entails substantial costs and technological difficulties. To tackle this challenge, several approaches have been proposed to achieve the required sensitivity without the need for filter cavities. These include the use of electromagnetically induced transparency \cite{PhysRevA.73.053810}, entangled light and negative-mass atomic spin ensembles \cite{2018_negative_mass_spin_GWD,2019_negative_mass_spin_GWD}, and small-scale optomechanical filters \cite{PhysRevLett.113.151102}. The EPR (for Einstein-Podolski-Rosen) or \textit{conditional} squeezing has been proposed by Ma, \textit{et al.}~\cite{Ma_2017}, and experimentally demonstrated in~\cite{PhysRevD.96.062003, 2020NaPho..14..223Y, S_dbeck_2020, PhysRevResearch.3.043079}. In this scheme, by applying the concept of EPR steering~\cite{PhysRevLett.98.140402, PhysRevA.80.032112} to gravitational wave detectors, the main Fabry--P\'{e}rot-Michelson interferometer can be repurposed as a single filter cavity (see Fig.~\ref{Fig_CV_teleportation}a).

In this paper, we propose a generalized scheme of the EPR squeezing, employing the concept of quantum teleportation. The quantum teleportation of an optical state has been a well-established technique in the field of continuous-variable quantum information processing since it was first demonstrated by Furusawa \textit{et al}. based on the Braustein--Kimble scheme ~\cite{doi:10.1126/science.282.5389.706,PhysRevLett.80.869}. Through our scheme, one can realize arbitrary number of operations of quadrature-angle rotation without the use of filter cavities.

The teleportation procedure in the Braustein--Kimble scheme involves Alice, the sending station; Bob, the receiving platform which shares EPR-entangled photons~\cite{PhysRev.47.777} and Victor, that brings an unknown quantum state and initiates the process (see Fig.~\ref{Fig_CV_teleportation}b). Alice conducts the Bell measurement involving Victor's and her own photons, transmitting the outcome to Bob through the classical communication channel. Bob then displaces his photon based on Alice's information, leading to the successful teleportation of Victor's quantum state. By adding physical operations to Alice, Bob, and Victor's paths, described as $\mathbb{U}_{a,b,v}$ in Fig~\ref{Fig_CV_teleportation}b, one can also manipulate the final teleported state as \YN{$\mathbb{U}_{b}\mathbb{U}_{a}\mathbb{U}_{v}\pmb{\hat{v}}$.} 
Furthermore, Bell measurement extends the limit of participating modes from 2 to an arbitrary number. \YN{By plugging the steered state or the teleported state to the initial state of Victor in a new teleportation protocol, $\pmb{\hat{v}}$, via Bell measurement, one can increase the number of equivalent filter cavities to an arbitrary number (see detailed discussion in Appendix B). }

In the following sections, we illustrate how QT squeezing can be implemented in future detectors, which necessitates multiple filter cavities. Specifically, we focus on the low-frequency interferometer of the Einstein Telescope's xylophone detector (ETLF). In the ETLF, the round-trip phase in the signal extraction cavity is slightly detuned to connect the suspended mirrors by the so-called optical spring~\cite{1968JETP.26.831,PhysRevD.65.042001}, which enables overcoming the Standard Quantum Limit (SQL). Additionally, for a narrow-band interferometer such as the ETLF, detuning offers a flexibility to boost the detector’s sensitivity~\cite{Hild_2011}.

\begin{figure}
    \centering
    \includegraphics[scale=0.5]{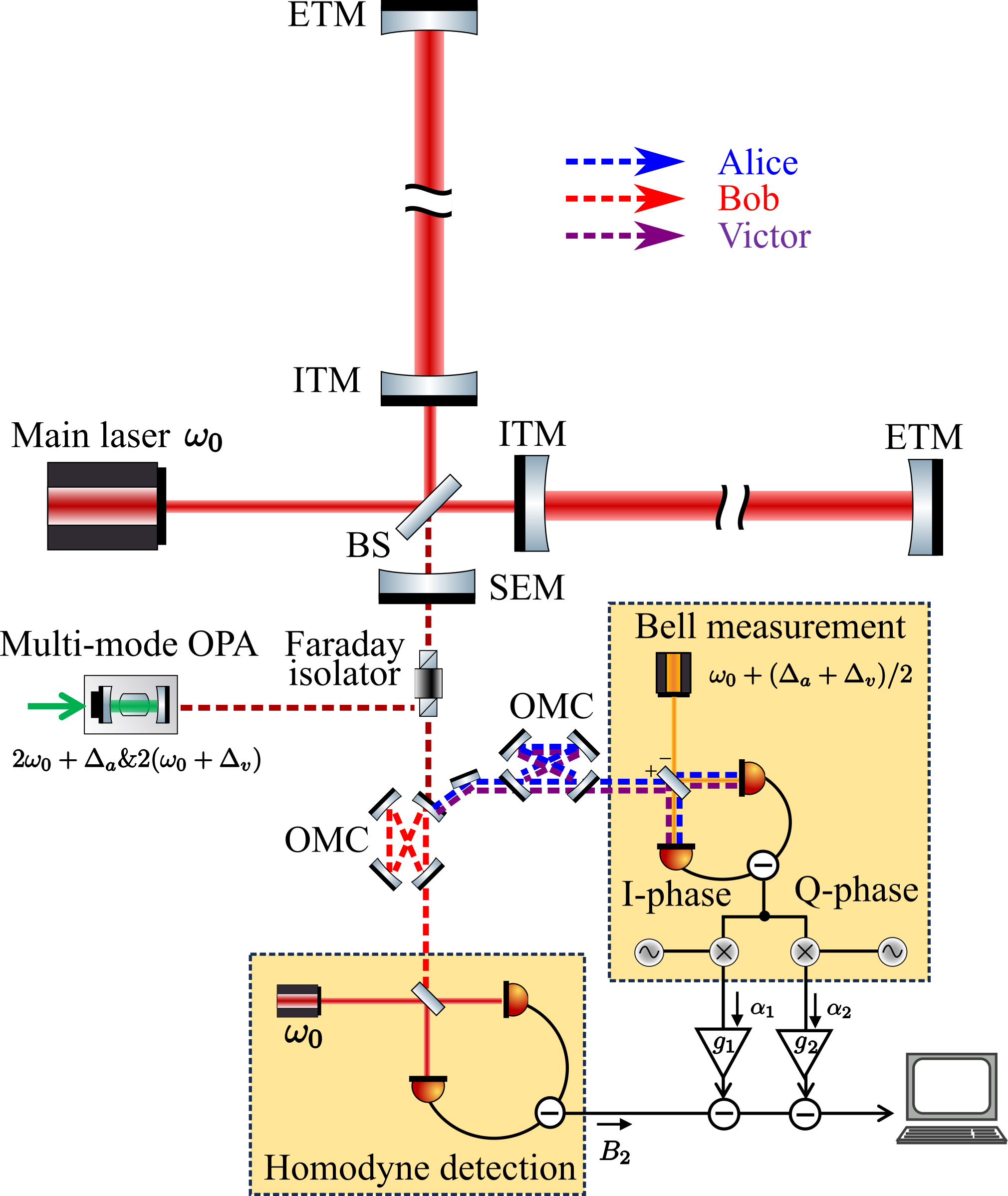}
    \caption{Configuration of QT squeezing. The OPA in the anti-symmetric port is pumped at two frequencies, $2\omega_0+\Delta_a$ and $2(\omega_0+\Delta_v)$, generating entanglement at the sideband frequencies symmetrically (see also Fig.~\ref{fig:continous_fields}b). The former pumping results in a two-mode EPR entanglement, Alice and Bob, while the latter forms a squeezed state, Victor. Three beams are injected through a Faraday isolator. The central part consists of a signal-recycled Fabry--P\'{e}rot Michelson Interferometer, which includes the beam splitter (BS), input-test mass (ITM), end-test mass (ETM) and signal-extraction mirror (SEM). The interferometer is pumped at the frequency $\omega_0$, matching Bob's frequency. The output is spectrally separated by an output mode cleaner (OMC). Bob's beam is collapsed at the homodyne detection, while Victor and Alice is detected through Bell measurement:  two beams are combined with the local oscillator (LO) fields with the LO angle $\xi_\mathrm{LO}$, and subsequently detected by two photo detectors. The outputs are subtracted from each other and demodulated by two demodulation angles. The two sets of measurement data are combined using the optimal filter gain, $(g_1\ g_2)$, and finally we achieve quantum-noise suppression.}
    \label{fig:configuration_Bell_measurement}
\end{figure}

To realize broadband noise suppression for such a detuned configuration, two filter cavities are required: one is for compensating the optical detuning resonance in the interferometer, the other for compensating the so called optical spring resonance from the ponderomotive rigidity effect. We associate the main-interferometer response including the ponderomotive squeezing~\cite{PhysRevD.65.022002} with $\mathbb{U}_b$ in Fig~\ref{Fig_CV_teleportation}b, and quadrature rotations by passive cavities with $\mathbb{U}_{v,a}$.
Fig.~\ref{fig:configuration_Bell_measurement} shows the implementation.
Three beams, Victor, Alice and Bob, are injected into the interferometer from the anti-symmetric port. The interferometer, acting as an empty cavity for Victor and Alice, provides an optimized frequency-dependent quadrature rotation by tuning their center frequencies and macroscopic lengths of the arm and signal extraction cavity, which functions as filter cavities. Bob's frequency is matched to the main laser from the symmetric port, then he sees the interferometer as an active cavity including ponderomotive squeezing~\cite{PhysRevD.65.022002}. The output is spectrally separated into two detection ports by the output mode cleaners; Bob's beam get collapsed via homodyne detection, while Victor and Alice are detected through Bell measurement. By applying Wiener filters to the two outputs of Bell measurement, quantum noise suppression is achieved as shown in Fig~\ref{fig:ss_QT_and_QN_enhancement}.
In following sections, we will present the details of each step mentioned above.

\section{Quantum state preparation}\label{sec2.1}
\begin{figure}
    \centering
    \includegraphics[scale=0.6]{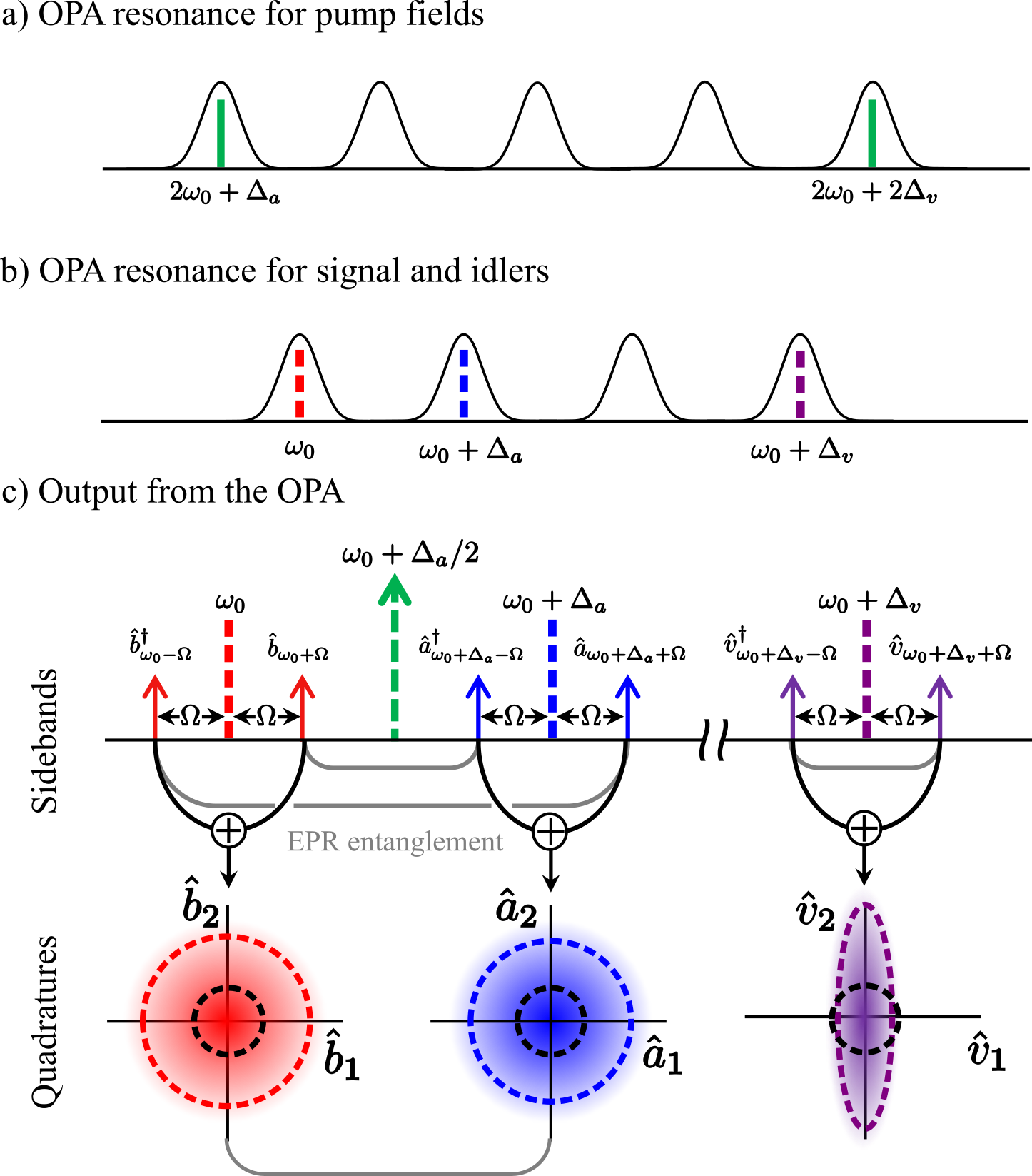}
    \caption{\YN{The OPA resonance for a) two pump fields and b) the signal and idlers. In this specific case two pump fields are separated by 5 FSR of the OPA cavity, and frequency difference between Victor and Bob is three time larger than that of Alice and Bob. c) Schematics of three beams. Top and bottom panels show fields in the sideband and quadrature pictures, respectively. The OPA is pumped at two frequencies, $2\omega_0+\Delta_a$ and $2\omega_0+2\Delta_v$, generating entanglement at the sideband frequencies symmetrically. The former pumping results in a two-mode EPR entanglement, Alice and Bob, while the latter forms a squeezed state, Victor.}}
    \label{fig:continous_fields}
\end{figure}

The initial quantum states of three beams, Alice, Bob, Victor are prepared with a multi-mode squeezer\,\cite{PhysRevD.101.124052}. The entangled beams, Alice and Bob, are centred at frequencies $\omega_0+\Delta_a$ and $\omega_0$, created by driving the OPA with a pumping beam at frequency $2\omega_0+\Delta_a$; the squeezed beam, Victor at frequency $\omega_0+\Delta_v$, is created by the second pumping beam at frequency $2(\omega_0+\Delta_v)$ (see Fig.~\ref{fig:continous_fields}a). \YN{Here $\Delta_a$ is supposed to be a free spectral range (FSR) of the OPA cavity away from Bob's frequency to keep resonant conditions of the entangled beam, and $\Delta_a-2\Delta_v$ equals to the multiples of the FSR of the OPA cavity (see Fig.~\ref{fig:continous_fields}).}

In two-photon formalism~\cite{PhysRevA.31.3093, PhysRevA.31.3068}, the zero-mean random fluctuations of light field is described by a 2D-vector consisting the operators of amplitude and phase quadrature, and for Alice, Bob and Victor, there is $\pmb{\hat{a}}=\{\hat{a}_{1,\Omega},\,\hat{a}_{2,\Omega}\}^{\rm T}, \pmb{\hat{b}}=\{\hat{b}_{1,\Omega},\,\hat{b}_{2,\Omega}\}^{\rm T}, \pmb{\hat{v}}=\{\hat{v}_{1,\Omega},\,\hat{v}_{2,\Omega}\}^{\rm T}$, 
where $\Omega$ is sideband frequency, the superscript $\rm T$ stands for transpose. Below we omit the superscript $\Omega$ for simplicity. 

As shown by Duan \textit{et al.} \cite{2000_Duan}, the strength of the EPR-entanglement of Alice's and Bob's state can be expressed in terms of the spectral densities of the four EPR-operators, $(\hat{a}_{i}\pm\hat{b}_{i})/\sqrt{2}$, as
\begin{align}
    S_{(\hat{a}_{1}\pm\hat{b}_{1})/\sqrt{2}} = e^{\pm 2r},\ S_{(\hat{a}_{2}\pm\hat{b}_{2})/\sqrt{2}} = e^{\mp 2r}
\end{align}
where $r$ signifies the squeezing factor. When $r\rightarrow\infty$, the noise spectra of $S_{(\hat{a}_{1}-\hat{b}_{1})/\sqrt{2}}$ and $S_{(\hat{a}_{2}+\hat{b}_{2})/\sqrt{2}}$ approach zero, corresponding to the original EPR entanglement~\cite{PhysRev.47.777}. 
In a more general situation, with the quadrature $\hat{a}_{-\theta}=\hat{a}_1 \cos \theta-\hat{a}_2 \sin \theta$ measured, the quadrature $\hat{b}_{\theta}=\hat{b}_1\cos \theta+\hat{b}_2\sin \theta$ is conditionally squeezed and vice versa. The spectral density of the conditional squeezed field reads
\begin{equation}\label{eq:Stheta}
S_{\hat{b}_{\theta}\hat{b}_{\theta}}^{\hat{a}_{-\theta}}=1/{\rm cosh}(2r)\,, \ S_{\hat{b}_{\pi/2+\theta}\hat{b}_{\pi/2+\theta}}^{\hat{a}_{-\theta}}=\rm {cosh}(2r)\,.
\end{equation}
The amplitude (phase) quadrature of Victor experiences (anti-)squeezing as
\begin{align}
    S_{\hat{v}_1\hat{v}_1}=e^{-2r},\ S_{\hat{v}_2\hat{v}_2}=e^{2r}.
\end{align}
Throughout this work, we assume uniform squeeze factors for each pumping frequency for the sake of simplicity.

\section{Noise suppression through quantum teleportation}
After passing through the main interferometer, the read out phase quadrature of Bob represented by the observable $B_2$, can be written as,
\begin{equation}
\begin{split}
    \hat{B}_2  = \Gamma e^{i\beta_b}(\hat{b}_1\cos\theta_b-\hat{b}_2\sin\theta_b), \label{eq:B2}
    \end{split}
\end{equation}
where $\Gamma$, $\theta_b$ and $\beta_b$ represent the frequency-dependent gain, quadrature rotation, and phase shift from the ponderomotive squeezing as defined in \cite{PhysRevD.104.062006} (see also Appendix A). In order to squeeze $\hat{B}_2$, we need to displace its photon at the input stage by applying quadrature rotation $-\theta_b$. This can be prepared by applying a phase rotation to its entangled pair Alice, $\theta_a$ and displacing Victor's state with quadrature rotation, $\theta_v$, where there needs $\theta_a+\theta_v=-\theta_b$. 
By configuring detunings of Alice and Victor with respect to the main interferometer, the interferometer can function as a empty optical cavity for Victor and Alice and provide desired quadrature rotations at its output. 

\subsection{Bell measurement}
The teleportation of Victor's state to Bob requires a Bell measurement between Alice and Victor. In the Bell measurement, a local oscillator at frequency $\omega_0+(\Delta_a+\Delta_v)/2$ is combined with Alice and Victor by a half beam splitter. Subsequently, the two output beams are detected by two photo detectors and their photon currents are demodulated at frequency $(\Delta_v-\Delta_a)/2$. By setting the local oscillator angle to $-\pi/2$ and properly choosing the demodulation phase, Bell observables can be carried out (details are available in Appendix C):
\begin{align}
    \pmb{\hat{\alpha}} = \begin{pmatrix}
       \hat{\alpha}_1\\
       \hat{\alpha}_2
    \end{pmatrix}
    = \frac{1}{\sqrt{2}} \begin{pmatrix}
        \hat{V}_1-\hat{A}_1 \\
        \hat{V}_2+\hat{A}_2
    \end{pmatrix},
\end{align}
where $\hat{A}_1=e^{i\beta_a}(\hat{a}_1 \cos \theta_a - \hat{a}_2 \sin\theta_a)$, $\hat{A}_2= e^{i\beta_a} (\hat{a}_1 \sin \theta_a + \hat{a}_2 \cos \theta_a )$ and $\hat{V}_1= e^{i\beta_v} (\hat{v}_1 \cos \theta_v - \hat{v}_2 \sin\theta_v)$, $\hat{V}_2=e^{i\beta_v} (\hat{v}_1 \sin \theta_v   + \hat{v}_2 \cos \theta_v) $ are the quadratures of the Alice and Victor's beams after passing through the interferometer. $\beta_v$ and $\beta_a$ are the averaged phases of both quadratures.

\subsection{Post processing}
The classical communication channel can be built with displacement operation or equivalently achieved through post-processing by combining $\hat{B}_2$ with $g_1 \hat{\alpha}_1$ and $g_2 \hat{\alpha}_2$, namely,
\begin{align}
    \hat{B}_2^{\rm tel}=\hat{B}_2-g_1\hat{\alpha}_1-g_2\hat{\alpha}_2,
\end{align}
where $g_1, g_2$ are filter gains, whose optimal values can be derived by
\begin{align}
    g_1 &=  \frac{S_{\hat{B}_2\hat{\alpha}_1}S_{\hat{\alpha}_2\hat{\alpha}_2}-S_{\hat{\alpha}_2\hat{\alpha}_1}S_{\hat{B}_2\hat{\alpha}_2}}{S_{\hat{\alpha}_1\hat{\alpha}_1}S_{\hat{\alpha}_2\hat{\alpha}_2}-|S_{\hat{\alpha}_1\hat{\alpha}_2}|^2},\\ 
    g_2 &= \frac{S_{\hat{B}_2\hat{\alpha}_2}S_{\hat{\alpha}_1\hat{\alpha}_1}-S_{\hat{\alpha}_1\hat{\alpha}_2}S_{\hat{B}_2\hat{\alpha}_1}}{S_{\hat{\alpha}_1\hat{\alpha}_1}S_{\hat{\alpha}_2\hat{\alpha}_2}-|S_{\hat{\alpha}_1\hat{\alpha}_2}|^2},
\end{align}
(see more details in Appendix A).
In an idealized case absence of imperfections, we can obtain the eventual noise spectrum density of $B_2^{\rm tel}$,
\begin{equation}
    \begin{split}
    S^\mathrm{tel}_{\hat{B}_2\hat{B}_2} = |\Gamma|^2 \frac{1+e^{-2r}\cosh 2r}{e^{-2r}+\cosh 2r} \xrightarrow{r\gg1} |\Gamma|^2 \frac{3}{e^{2r}} \label{eq:spectrum}
    \end{split}
\end{equation}
(see detailed derivations in Appendix A). Taking into account that the gravitational-wave signal sidebands are not affected throughout the post-processing, Eq.(\ref{eq:spectrum}) indicates that the strain sensitivity is improved by a factor of $3/e^{2r}$ in power across a wide range of frequencies with sufficient squeezing, which corresponds to 4.8 dB penalty discussed in Sec. V.
 
\section{Sensitivity comparison}\label{sec3}
\begin{figure}
    \centering
    \includegraphics[scale=0.22]{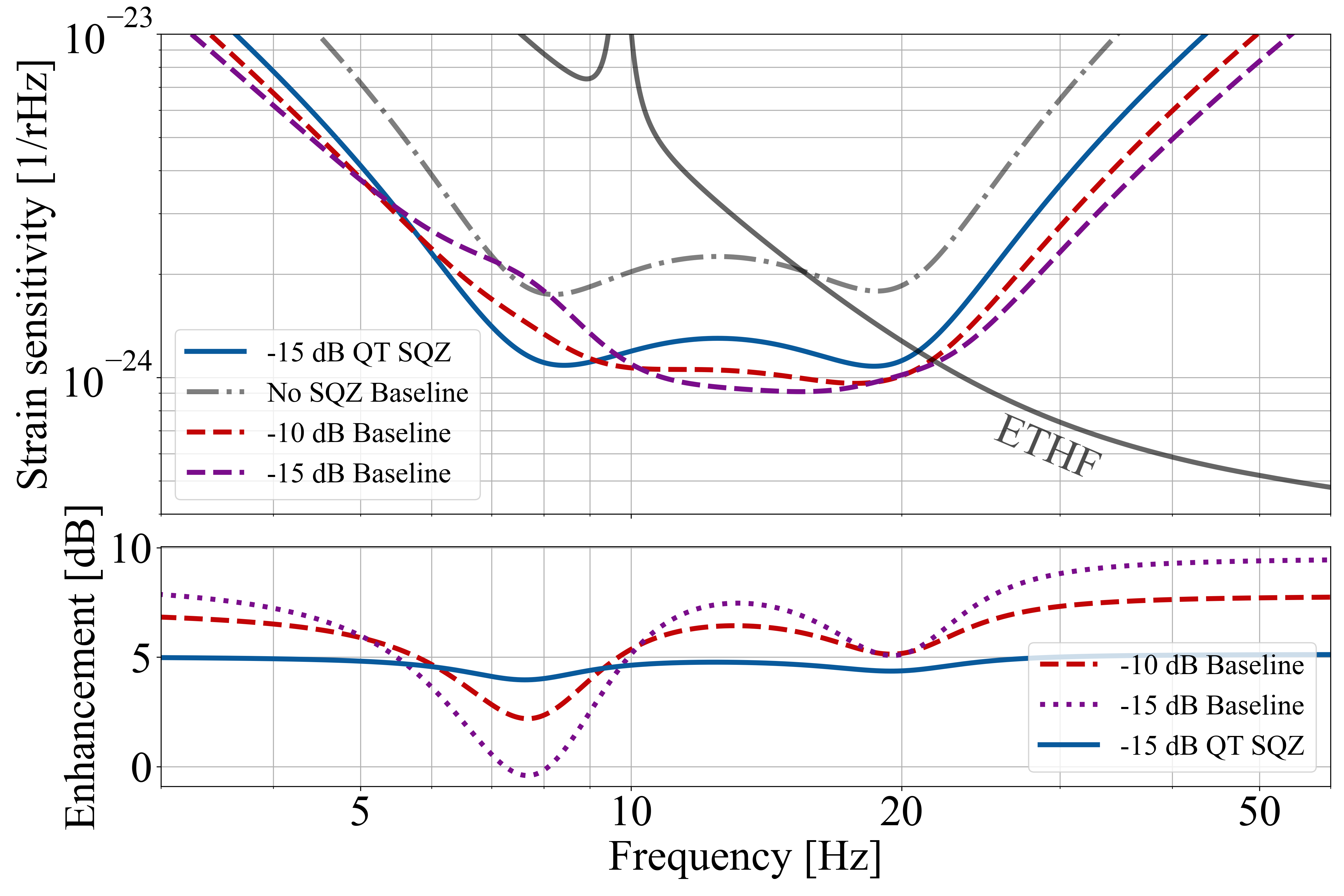}
    \caption{Top: Quantum-noise limited strain sensitivity of ETLF.  The dashed and solid curves represent the baseline FDS with -10 dB and QT squeezing, respectively, with different levels of squeezing. The dotted curve is the baseline FDS with -15 dB squeezing. We also show the sensitivity of the ETHF, which covers the frequency region above 20 Hz. Bottom : Quantum noise-enhancement in power-spectral density compared to the non-squeezed case of the baseline FDS (shown in the dashed-dotted curve in the top panel).
    }
    \label{fig:ss_QT_and_QN_enhancement}
\end{figure}
In the top panel of Fig~\ref{fig:ss_QT_and_QN_enhancement}, we show the quantum noise-limited sensitivity of ETLF~\cite{Hild_2011} with the conventional frequency-dependent squeezing (hereafter called the baseline FDS) and QT squeezing. As a comparison we also plotted the enhancement factor compared to non-squeezed case of the baseline FDS (denoted as No SQZ baseline). As shown, the detuned configuration creates a dip of optical spring at very-low frequency ($<10$ Hz), enabling to broaden the total detector's bandwidth to lower frequency than a tuned configuration. The parameters considered, including imperfections such as losses and phase noises, are consistent with the current design of ETLF employing filter cavities, except for the squeezing level as shown in Table~\ref{tab1}. Detuning frequencies $\Delta_a$ and $\Delta_v$ are also shown in the table, while the method of parameter searching and details of macroscopic length tuning are discussed in Appendix D.

\begin{table}[h]
\caption{Parameters for squeezing in ETLF}\label{tab1}%
\begin{tabular}{p{3.5cm}|p{2.cm}<{\centering}|p{2.cm}<{\centering}}
    Parameters & Baseline FDS & QT SQZ  \\ \hline\hline
    Detuning of the SEC & \multicolumn{2}{c}{0.75 rad} \\ \hline
    Filter cavity length & 1 km & - \\ \hline
    Arm round trip loss & \multicolumn{2}{c}{45 ppm} \\ \hline
    SEC loss & \multicolumn{2}{c}{1000 ppm} \\ \hline
    Injection loss & \multicolumn{2}{c}{4 \%} \\ \hline
    Readout loss & \multicolumn{2}{c}{3 \%} \\ \hline
    FC round-trip loss & 20 ppm &  - \\ \hline
    Squeezer noise RMS\footnotemark[1] & \multicolumn{2}{c}{10 mrad} \\ \hline
    Local oscillator RMS & \multicolumn{2}{c}{10 mrad} \\ \hline
    SEC length RMS & \multicolumn{2}{c}{1 pm} \\ \hline
    Filter cavity length RMS & 1 pm & - \\ \hline
    Detuning $\Delta_a$  & - & $\sim 319$~MHz \\ \hline
    Detuning $\Delta_v$ & - & $\sim 962$~MHz \\ \hline
    Squeezing level & -10 dB & -15 dB \\ \hline
\end{tabular}
\footnotetext[1]{Root mean square}
\end{table}

In general, QT squeezing exhibits sensitivity levels that are inferior to those of the baseline FDS as shown in most frequency range in the lower panel of Fig~\ref{fig:ss_QT_and_QN_enhancement}. This disparity arises from the 4.8\,dB penalty inherent in QT squeezing as indicated by Eq.~(\ref{eq:spectrum}) and there are also threefold optical losses summing from three optical paths. 

However, the performance of QT squeezing can potentially surpass that of baseline FDS around the optical-spring resonance. The uneven sensitivity enhancement in the baseline scenarios arises from the dephasing of the squeezed vacuum. As studied in \cite{PhysRevD.90.062006, PhysRevD.104.062006}, the dephasing of squeezed beam will lead coupling of the noise fluctuation from anti-squeezed quadrature into the squeezed quadrature. Around the resonance, where the effective mechanical susceptibility of test mass strengthens due to the optical spring effect, the interferometer creates more significant ponderomotive squeezing onto the quantum fields interacting with the mirrors and the dephasing from the filter cavity becomes large. Therefore, the sensitivity is more susceptible to dephasing effect in particular at this frequency band, and the optimal input squeezing level is limited to $\sim\ 10$ dB (see the dashed and dotted curves in Fig~\ref{fig:ss_QT_and_QN_enhancement}). Such dephasing can stem from length fluctuation of optical paths and optical losses in detuned cavities. In the baseline FDS, the filter cavity length is constrained to 1\,km due to infrastructure limitations, as also employed in \cite{2023_ET_CoBA_paper}. In contrast, QT squeezing leverages the stability and length of the 10-km-long arm cavities as filter cavities. The length fluctuation of the arm cavities is well-suppressed by multi-stage suspension and control systems. In the particular case of ETLF with parameters in Table~\ref{tab1}, QT squeezing exhibits better sensitivity than the baseline FDS at 8\,Hz, as shown in Fig.~\ref{fig:ss_QT_and_QN_enhancement}.

As a filter cavity, the 10\,km interferometer has lower effective loss compared with 1\,km filter cavity, evaluated through the term loss per unit length~\cite{PhysRevD.90.062006}. We take the 1000 ppm loss from SEC into account, and it turns out the SEC loss is mitigated by the low transmissivity of the input-test masses (see Appendix E for more details). Those result in less effective loss and dephasing, thus allows higher input squeezing level. By accommodating -15 dB squeezing to compensate the 4.8 dB penalty, we can achieve $\sim$ 5\,dB sensitivity improvement over the whole frequency band. In this particular case, higher squeezing level up to -17 dB allows to improve the sensitivity across the overall frequency range (see details in Fig.~\ref{fig:QN_enhancement_various_SQZ_levels} in Appendix E).

\begin{figure}
    \centering
    \includegraphics[scale=0.28]{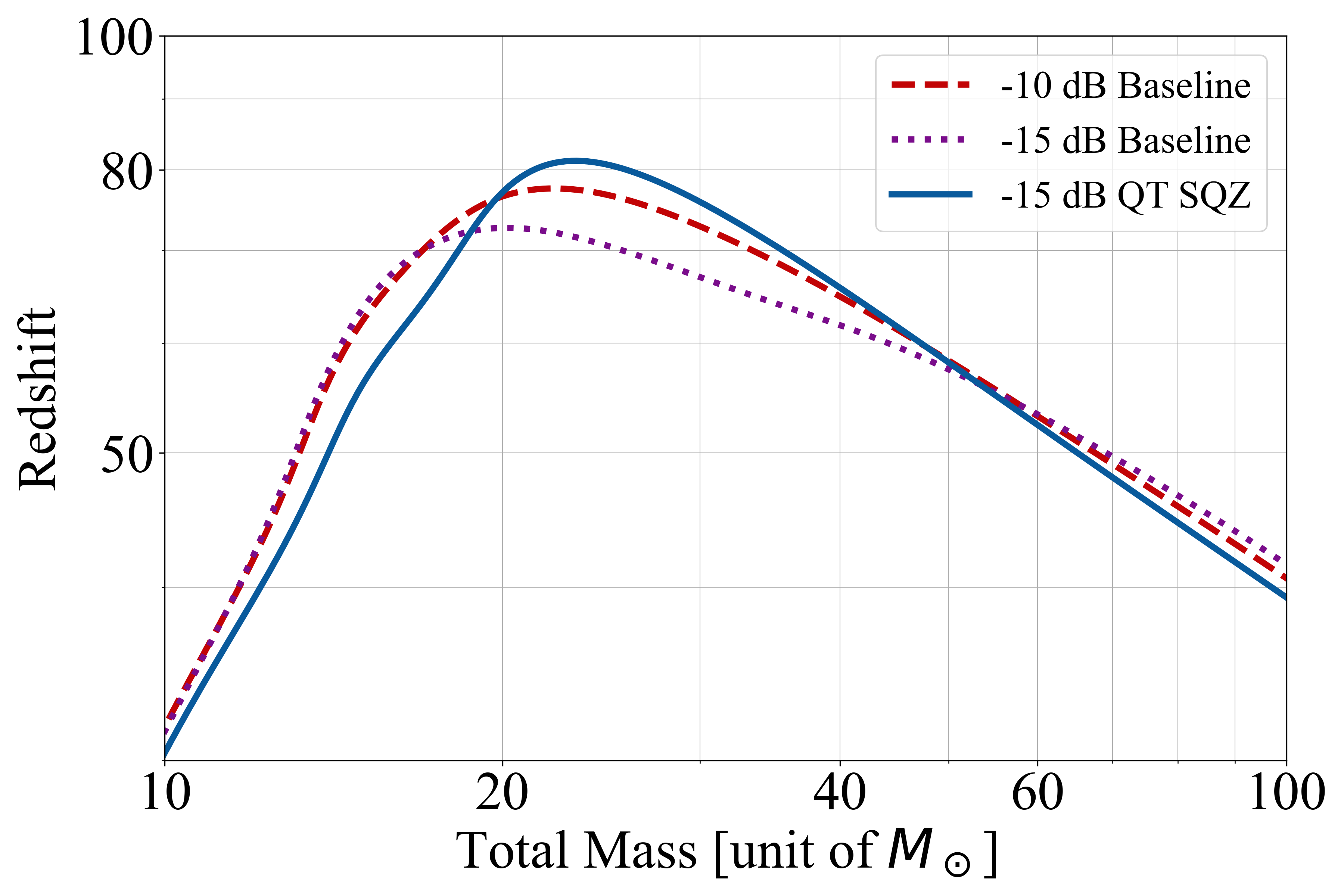}
    \caption{Detection horizon of ET with three squeezing schemes in ETLF, i.e., the baseline FDS with -10 and -15 dB and QT squeezing with -15 dB squeezing.}
    \label{fig:horizon}
\end{figure}

As a figure of merit, we plot the detection horizon of equal-mass non-spinning compact binary coalescence in Fig~\ref{fig:horizon}. The detection criterion is set at a signal-to-noise ratio of 8. The overall power spectral densities integrate classical and quantum noise of ETLF, as well as the total noise of ETHF, which covers the sensitivity above 20 Hz as shown in Fig.~\ref{fig:horizon}. The horizon plot indicates that the QT squeezing achieves performance almost equivalent to the baseline FDS; more precisely, QT squeezing shows slightly better maximum horizon at the cost of the mass-range below $20\ M_\odot$.

\section{Discussions}\label{sec4}
Our scheme attains frequency-dependent squeezing by replacing the arbitrary number of external filter cavities, which would otherwise require kilometer-scale additional vacuum tunnel and suspension systems, with the main interferometer itself through the application of the quantum-teleportation technique. When applying to ETLF, we revealed that it offers a sensitivity advantage around the optical spring resonance which is the primary objective of detuning. This benefits from the long effective length of the arm cavity as the filter cavities and contributes to expand the bandwidth of the whole detector to lower frequencies.

However, we also need to note two drawbacks of the QT squeezing: (1) a factor of 3 (4.8 dB) higher injected squeezing required to reach the same level of detected squeezing at the readout port as achieved by the baseline FDS.
(2) threefold noise contributions from input and output losses, which limit the sensitivity across the entire frequency band (see the figure of noise budget in Appendix E). More generally, if $N$ states participates in the teleportation, the eventual squeezing level is degraded and noise contributions from input and output loss are increased both by a factor of $N$ in power.

In addition, it is essential to highlight the technical flexibility of our scheme. The configuration illustrated in Fig.~\ref{fig:configuration_Bell_measurement} allows for seamless transitions between detuned and tuned configurations. This transition can be achieved by turning off the pumping laser beam for Victor and adjusting the pumping frequency of the other beam, along with the macroscopic lengths of the arm and signal extraction cavity,effectively reverting back to the EPR squeezing in~\cite{Ma_2017}. 
Furthermore, QT squeezing provides the capability to address variations in the SEC detuning by optimizing the parameters of pumping frequencies and macroscopic length tuning. This means we do not need to replace the input mirror of the filter cavity itself to adjust the filter-cavity bandwidth, which might otherwise be required due to practical factors, such as changes of the main laser power or ice formation on the cryogenic mirrors.

\begin{acknowledgments}
Authors are grateful to S.~Hild for stimulating discussions and insight into potential benefits our scheme might bring for Einstein Telescope, and the ANU CGA squeezer group for fruitful discussion about experimental realization. Research by Y. N. is supported by JSPS Grant-in-Aid for JSPS Fellows Grant Number 23KJ0787. T. Z. acknowledges the support of the Institute for Gravitational Wave Astronomy at the University of Birmingham.  S.~D. is grateful for support to the Faculty of Science and Engineering of Maastricht University and to the European Research Council (within the frames of the ERC-2020-AdG agreement No. 101019978).
\end{acknowledgments}

\appendix
\section{Derivation of Wiener filters}\label{m.sec1}
In the ideal lossless case, response of a differential mode cavity of the detuned interferometer for the carrier field  can be written as below according to the scaling-law theorem~\cite{PhysRevD.67.062002}:
\begin{align}
    \begin{pmatrix}
        \hat{B}_1 \\
        \hat{B}_2
    \end{pmatrix}
    &= \frac{1}{\tilde{M}} \begin{pmatrix}
        C_{11} & C_{12} \\
        C_{21} & C_{22}
    \end{pmatrix}
    \begin{pmatrix}
        \hat{b}_1 \\
        \hat{b}_2
    \end{pmatrix},
\end{align}
where
\begin{align*}
    \tilde{M} = \{(\gamma-i\Omega)^2+\delta^2\}\Omega^2-\delta\Theta,\\
    C_{11} = C_{22} = \Omega^2(\Omega^2-\delta^2+\gamma^2)+\delta\Theta,\\
    \quad C_{12} = 2\delta\gamma\Omega^2-2\gamma\Theta, C_{21} = -2\delta\gamma\Omega^2.
\end{align*}
Here $\delta$ and $\gamma$ are the effective detuning and half-bandwidth. $\Theta=\frac{8\omega_0 P_c}{McL}$ is the normalized optical power with the arm circulating power $P_c$, reduced mass $M$ and arm length $L$. From these relations, $\Gamma$, $\theta_b$ and $\beta_b$ in Eq.~(4) in the main text can be derived as:
\begin{align}
    \Gamma = \frac{\sqrt{C_{21}^2+C_{22}^2}}{|\tilde{M}|}&,\quad \theta_b =- \arctan\left({\frac{C_{22}}{C_{21}}}\right), \notag\\
    \beta_b&=\arg{\tilde{M}^*}.
\end{align}
Combining the measurement data with filter gains $(g_1\ g_2)$ such that $\hat{B}_2^g=\hat{B}_2-g_1\hat{\alpha}_1-g_2\hat{\alpha}_2$, we have the noise spectrum:
\begin{align}
    S_{\hat{B}_2^g\hat{B}_2^g} &= S_{\hat{B}_2\hat{B}_2} + |g_1|^2 S_{\hat{\alpha}_1\hat{\alpha}_1} + |g_2|^2 S_{\hat{\alpha}_2\hat{\alpha}_2} \notag \\
    &-g_1^*S_{\hat{B}_2\hat{\alpha}_1}-g_1S_{\hat{\alpha}_1\hat{B}_2} -g_2^*S_{\hat{B}_2\hat{\alpha}_2}-g_2S_{\hat{\alpha}_2\hat{B}_2} \notag \\
    & + g_1g_2^*S_{\hat{\alpha}_1\hat{\alpha}_2} + g_1^*g_2S_{\hat{\alpha}_2\hat{\alpha}_1}, \label{eq.Spectrum}
\end{align}
where
\begin{align*}
    S_{\hat{B}_2\hat{B}_2} &= \Gamma^2\cosh2r,\\
    S_{\hat{\alpha}_1\hat{\alpha}_1} &= \frac{e^{-2r}\cos^2\theta_v+e^{2r}\sin^2\theta_v+\cosh2r}{2},\\   S_{\hat{\alpha}_2\hat{\alpha}_2} &= \frac{e^{-2r}\sin^2\theta_v+e^{2r}\cos^2\theta_v+\cosh2r}{2}, \\
    S_{\hat{\alpha}_1\hat{\alpha}_2} &= \frac{(e^{-2r}-e^{2r})\sin\theta_v\cos\theta_v}{2}, \\
    S_{\hat{B}_2\hat{\alpha}_1} &= S^*_{\hat{\alpha}_1\hat{B}_2} = -\frac{\Gamma e^{i(\beta_b-\beta_a)}\cos\theta_v\sinh2r}{\sqrt{2}}, \\
    S_{\hat{B}_2\hat{\alpha}_2} &= S^*_{\hat{\alpha}_2\hat{B}_2} = -\frac{\Gamma e^{i(\beta_b-\beta_a)}\sin\theta_v\sinh2r}{\sqrt{2}}.
\end{align*}
$S_{\hat{B}_2^g\hat{B}_2^g}$ takes its minimum when $g_1$ and $g_2$ are Wiener filters determined as follows:
\begin{align*}
    g_1 &=  \frac{S_{\hat{B}_2\hat{\alpha}_1}S_{\hat{\alpha}_2\hat{\alpha}_2}-S_{\hat{\alpha}_2\hat{\alpha}_1}S_{\hat{B}_2\hat{\alpha}_2}}{S_{\hat{\alpha}_1\hat{\alpha}_1}S_{\hat{\alpha}_2\hat{\alpha}_2}-|S_{\hat{\alpha}_1\hat{\alpha}_2}|^2} \notag\\
    &= -\frac{\sqrt{2}\Gamma e^{i(\beta_b-\beta_a)}\sinh{2r}\cos\theta_v}{\cosh{2r}+e^{-2r}}, \notag \\
    g_2 &= \frac{S_{\hat{B}_2\hat{\alpha}_2}S_{\hat{\alpha}_1\hat{\alpha}_1}-S_{\hat{\alpha}_1\hat{\alpha}_2}S_{\hat{B}_2\hat{\alpha}_1}}{S_{\hat{\alpha}_1\hat{\alpha}_1}S_{\hat{\alpha}_2\hat{\alpha}_2}-|S_{\hat{\alpha}_1\hat{\alpha}_2}|^2} \notag\\
    &= -\frac{\sqrt{2}\Gamma e^{i(\beta_b-\beta_a)}\sinh{2r}\sin\theta_v}{\cosh{2r}+e^{-2r}}.
\end{align*}
Note that we used the relation $\theta_b+\theta_a=-\theta_v$.
Substituting those into Eq.~(\ref{eq.Spectrum}) leads to the noise spectrum density shown in Eq.~(9) in the main text.

\section{Generalization to N phase rotations}
\begin{figure}
    \centering
    \includegraphics[scale=0.32]{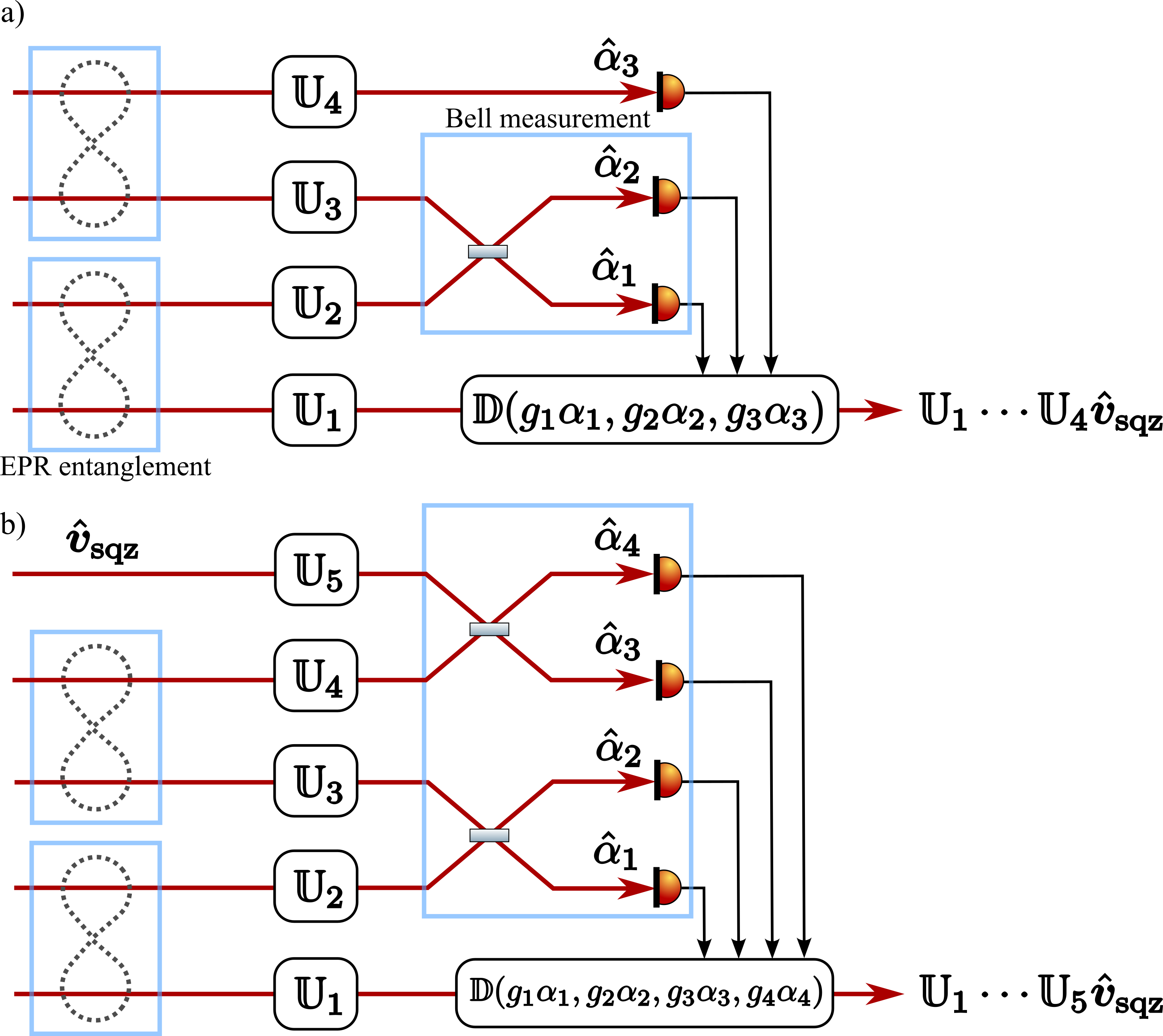}
    \caption{Schematics of the generalized quantum teleportation squeezing with (a) three equivalent-filter cavities and (b) four filter cavities. Here, $\mathbb{U}_1$ represents the ponderomotive squeezing and other operations represent phase rotations, and $\pmb{\hat{v}}_\mathrm{sqz}$ is a squeezed vacuum.}
    \label{fig:Fig_CV_N_teleportation}
\end{figure}

Fig.~\ref{fig:Fig_CV_N_teleportation} illustrates quantum teleportation squeezing to generate arbitrary number of phase rotations. 
There are two possible approaches: connecting EPR squeezing and QT squeezing to the QT squeezing protocol as Fig.~\ref{fig:Fig_CV_N_teleportation}(a) and (b). The first protocol prepares three phase rotations, while the second prepares four. Inductively, one can achieve an arbitrary number of pre-filtering phases by preparing multiple entanglement states and iterating Bell measurements.
When one prepares $N$ phase rotations, Wiener filters, $\mathbf{g} = \begin{pmatrix} g_1 & g_2 & \cdots & g_N \end{pmatrix}^\top$, can be expressed as follows:
\begin{align}
    \mathbf{g} = \mathbf{S}_{\alpha\alpha}^{-1} \mathbf{S}_{B\alpha},
\end{align}
where $\mathbf{S}_{\alpha}$ and $\mathbf{S}_{B\alpha}$ are the cross-spectral density matrix and vector, respectively, defined as:
\begin{align}
\mathbf{S}_{\alpha\alpha}=\begin{pmatrix}
    S_{\alpha_1 \alpha_1} & S_{\alpha_1 \alpha_2} & \cdots & S_{\alpha_1 \alpha_N} \\
S_{\alpha_2 \alpha_1} & S_{\alpha_2 \alpha_2} & \cdots & S_{\alpha_2 \alpha_N} \\
\vdots & \vdots & \ddots & \vdots \\
S_{\alpha_N \alpha_1} & S_{\alpha_N \alpha_2} & \cdots & S_{\alpha_N \alpha_N}
\end{pmatrix},
\end{align}
and
\begin{align}   
    \mathbf{S}_{B\alpha} = \begin{pmatrix}
S_{B\alpha_1} \\
S_{B\alpha_2} \\
\vdots \\
S_{B\alpha_N}
\end{pmatrix}.
\end{align}

\section{Mathematical description of Bell measurement}\label{m.sec2}
We describe the Bell measurement procedure, with specific reference to section II.C in~\cite{PhysRevD.67.122005}. Bell measurement utilizes a coherent laser as the Local Oscillator (LO) to measure the quadratures $\hat{V}_1-\hat{A}_1$ and $\hat{V}_2+\hat{A}_2$ (see Fig. 2 in the main text). The frequency of the LO is precisely tuned to match the central frequency of Victor and Alice. In our specific experimental context, it is necessary to control the LO frequency to $\omega_0+(\Delta_a+\Delta_v)/2$, a frequency regime in the radio frequencies (RF) domain, typically in the megahertz (MHz) range. 

Bell measurement involves several key steps. Firstly, a half beam splitter (HBS) is employed to combine the two idlers with the LO, expressed as:
\begin{align}
    E_{r,t} (t) = \frac{S(t) \pm L(t)}{\sqrt{2}}. \label{eq:E}
\end{align}
Here $E_{r,t}$ are the reflection and transmission of the HBS. $S(t)$ and $L(t)$ are the output from the interferomter and the LO field, respectively, expressed in the sideband picture as:
\begin{align}
    S(t) &= \int_{-\Lambda}^{\Lambda}\frac{\mathrm{d}\Omega}{2\pi}\{\hat{A}_{\omega_0+\Delta_a+\Omega}e^{-i(\omega_0+\Delta_a+\Omega)t}\notag\\
    &\quad +\hat{V}_{\omega_0+\Delta_v+\Omega}e^{-i(\omega_0+\Delta_v+\Omega)t} + \mathrm{h.c.}\} \\ 
    L(t) &= D e^{i\{\omega_0+(\Delta_a+\Delta_v)/2\}t} + \mathrm{h.c.},
\end{align}
where $\Lambda\lesssim(\Delta_a+\Delta_v)/2$ is the demodulation bandwidth, $D$ is the complex amplitude, and h.c. represents Hamiltonian conjugate. 

Secondly, two fields are detected by the photodetectors, combining two outputs to reject classical and quantum fluctuations in the LO field. After the combination, the photocurrent is proportional to the square of the field Eq.~(\ref{eq:E}):
\begin{align}
    i(t) &\propto E_r^2 - E_t^2 \propto S(t)L(t) \notag \\
    &= D \int_{-\Lambda}^{\Lambda}\frac{\mathrm{d}\Omega}{2\pi}\{A_{\omega_0+\Delta_a+\Omega}e^{i\{(\Delta_v-\Delta_a)-\Omega\}t} \notag\\
    &\quad +V_{\omega_0+\Delta_v+\Omega}e^{-i\{(\Delta_v-\Delta_a)+\Omega\}t}\} + \mathrm{h.c.} \notag \\
    &\quad + [\mathrm{irrelevant\ terms\ at\ high\ frequencies}]
\end{align}

Finally, by mixing $\cos\{(\Delta_v+\Delta_a)t/2+\xi_\mathrm{d}\}$ and applying a low-pass filter with the cut-off of $\Lambda$, one obtains:
\begin{align}
    O(\xi_\mathrm{d};t) = D \int_{-\Lambda}^{\Lambda}\frac{\mathrm{d}\Omega}{2\pi}\{A_{\omega_0+\Delta_a+\Omega} e^{-i\xi_\mathrm{d}} e^{-i\Omega t} + \mathrm{h.c.}\} \notag \\
    + D \int_{-\Lambda}^{\Lambda}\frac{\mathrm{d}\Omega}{2\pi}\{V_{\omega_0+\Delta_v+\Omega} e^{i\xi_\mathrm{d}} e^{-i\Omega t} + \mathrm{h.c.}\}. \label{eq:O}
\end{align}
In the quadrature picture, the quadrature operator $A_\zeta$ is defined as:
\begin{align}
    A_\zeta = A_1 \sin \zeta + A_2 \cos \zeta,
\end{align}
where
\begin{align}
    A_1 = \frac{A_{\omega_0+\Omega}+A^\dagger_{\omega_0-\Omega}}{\sqrt{2}},\ A_1 = \frac{A_{\omega_0+\Omega}-A^\dagger_{\omega_0-\Omega}}{i\sqrt{2}}.
\end{align}
Using those relations, Eq.~(\ref{eq:O}) leads to:
\begin{align}
    O(\xi_\mathrm{d};\Omega) = |D| \int_0^\Lambda \frac{\mathrm{d}\Omega}{2\pi} e^{-i\Omega t} \{ A_{\zeta_A}(\Omega)+V_{\zeta_V}(\Omega) \},
\end{align}
where $\zeta_A=-\xi_\mathrm{d}+\frac{\pi}{2}+\arg D$ and $\zeta_V=\xi_\mathrm{d}+\frac{\pi}{2}+\arg D$ (see also Eqs. (9)-(12) in~\cite{PhysRevD.67.122005}). In the frequency domain, one obtains:
\begin{align}
    O(\xi_\mathrm{d};\Omega) = |D| \frac{A_{\zeta_A}(\Omega)+V_{\zeta_V}(\Omega)}{\sqrt{2}}
\end{align}
The LO angle $\arg D$ is considered as a free parameter, determined experimentally, while the demodulation angle $\xi_\mathrm{d}$ can be adjusted after detection. By setting $\arg D=\pi/2$, the outputs become $(V_1-A_1)/\sqrt{2}$ and $(V_2+A_2)/\sqrt{2}$ with $\xi_\mathrm{d}$ of $-\pi/2$ and $\pi$, denoted as I and Q phase in Fig. 2 in the main text, respectively.

\section{Interferometer response for idlers}\label{s.sec3}
In this section, we examine the response of the central interferometer as filter cavities for two idlers. We show the parameter optimization process, crucial for using the interferometer as quantum filter cavities to our specific requirements, then investigate the effect of the arm cavity and SEC losses onto those filter parameters, comparing the conventional filter cavity scheme.
\subsection{Parameter searching}\label{s.sec3.subsec1}
\begin{figure}
    \centering
    \includegraphics[scale=0.45]{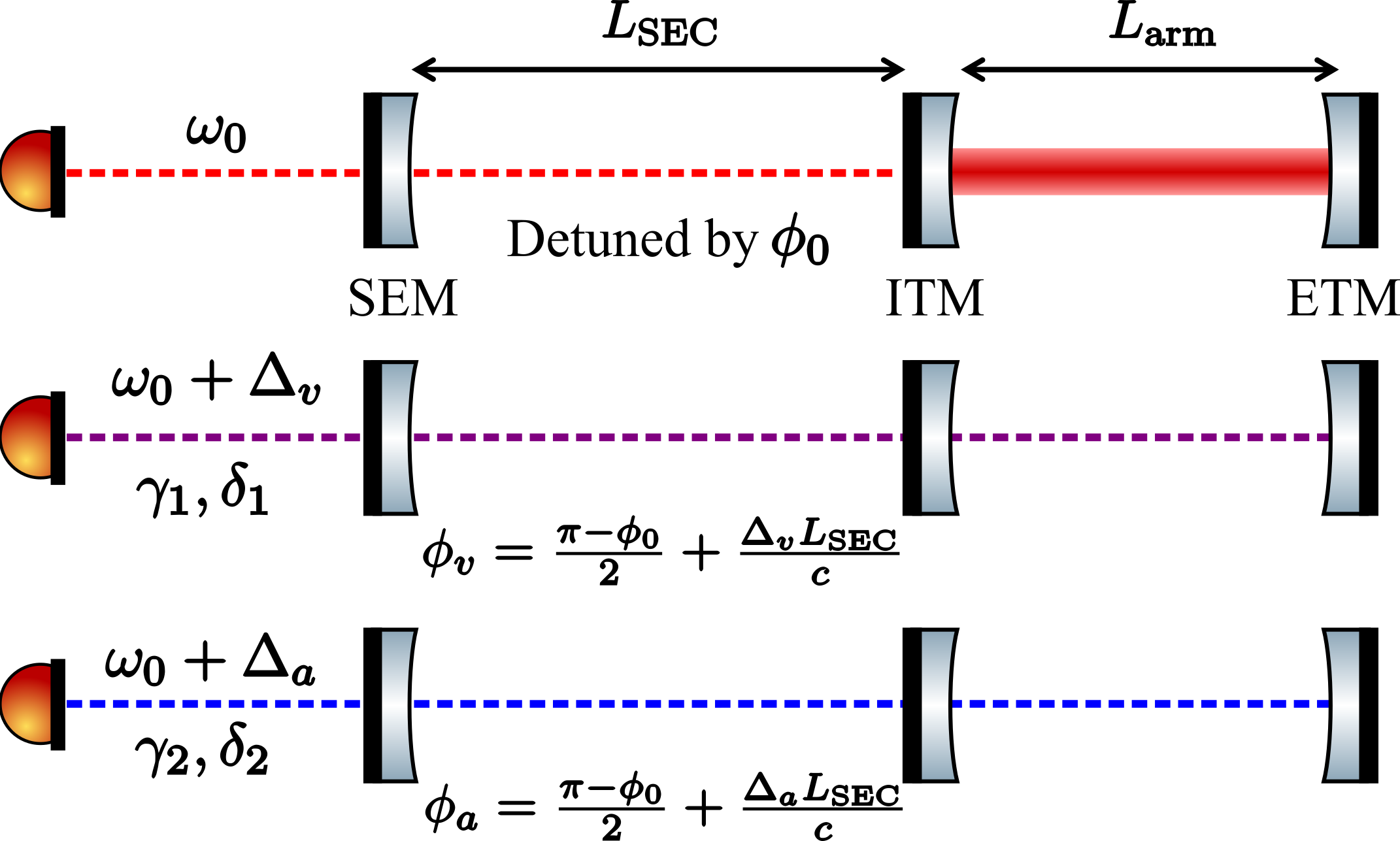}
    \caption{Resonance conditions for differential mode cavities: Shown from top to bottom are the resonance conditions for Bob, Alice and Victor.}
    \label{Fig_differential_mode_cavity}
\end{figure}
The coupled cavity formed by the SEC, ITM and ETM functions as a passive optical cavity for two idler beams. Due to the phase shifts acquired in the SEC, the bandwidths for two idlers can be tuned to meet the requirement of filter cavity bandwidths. To begin, let's examine the bandwidth of the arm cavity, which can be expressed as:
\begin{align}
    \gamma_\mathrm{1arm} = \frac{cT_\mathrm{ITM}}{4L_\mathrm{arm}}
\end{align}
According to the scaling law theorem \cite{PhysRevD.67.062002, Danilishin_2012}, the effective bandwidth can be expressed as follows:
\begin{align}
    \gamma &= \gamma_\mathrm{1arm}\mathrm{Re}\left[\frac{1-\sqrt{R_\mathrm{SEM}}e^{2i\phi_\mathrm{SEC}}}{1+\sqrt{R_\mathrm{SEM}}e^{2i\phi_\mathrm{SEC}}}\right]\notag \\
    &= \frac{\gamma_\mathrm{1arm}T_\mathrm{SEM}}{1+2\sqrt{R_\mathrm{SEM}}\cos2\phi_\mathrm{SEC}+R_\mathrm{SEM}}, \label{eq.gamma}
\end{align}
leading the requirement of round-trip phase $\phi_\mathrm{SEC}$:
\begin{align}
    \phi_\mathrm{SEC} = \frac{1}{2}\left[\arccos\left(\frac{T_\mathrm{SEM}\frac{\gamma_\mathrm{1arm}}{\gamma}-1-R_\mathrm{SEM}}{2\sqrt{R_\mathrm{SEM}}}\right)\right] + n\pi. \label{eq: 1.44}
\end{align}
Here, $n$ is an integer that determines the number of spectral ranges of the SEC. The phase $\phi_\mathrm{SEC}$ for Victor and Alice, denoting as $\phi_{v,a}$, can be described as:
\begin{align}
    \phi_{v,a} &= \frac{(\omega_0+\Delta_{v,a})L_\mathrm{SEC}}{c} = \frac{\pi-\phi_0}{2} + \frac{\Delta_{v,a}L_\mathrm{SEC}}{c}, \label{eq. phi}
\end{align}
where $\phi_0$ represents the detuning phase of the SEC for the main carrier. $\phi_{v,a}$ should be tuned to realize the required filter cavity bandwidth $\gamma_{1,2}$; substituting Eq.~(\ref{eq. phi}) into Eq.~(\ref{eq: 1.44}), $\Delta_{v,a}$ need to  satisfy the following relation:
\begin{align}
    \Delta_{v,a} &= \frac{c}{2L_\mathrm{SEC}}\Big[\arccos\left(\frac{T_\mathrm{SEM}\frac{\gamma_\mathrm{1arm}}{\gamma_{1,2}}-1-R_\mathrm{SEM}}{2\sqrt{R_\mathrm{SEM}}}\right) \notag\\
    &\quad + (2n_{v,a}-1)\pi +\phi_0\Big],
\end{align}
having $n_{1,2}$ as free parameters.
On the other hand, $\phi_{v,a}$ should also satisfy the following relation to realize the effective detunings $\delta_{1,2}$ as:
\begin{align}
    \delta_{1,2} &= \mathrm{Mod}_{\omega_\mathrm{FSR}^\mathrm{arm}}(\Delta_{v,a})-\gamma_\mathrm{1arm}\mathrm{Im}\left[\frac{1-\sqrt{R_\mathrm{SEM}}e^{2i\phi_{v,a}}}{1+\sqrt{R_\mathrm{SEM}}e^{2i\phi_{v,a}}}\right],
\end{align}
where $\omega_\mathrm{FSR}^\mathrm{arm}$ is the FSR of the arm cavity.

\begin{figure*}
    \centering
    \includegraphics[scale=0.35]{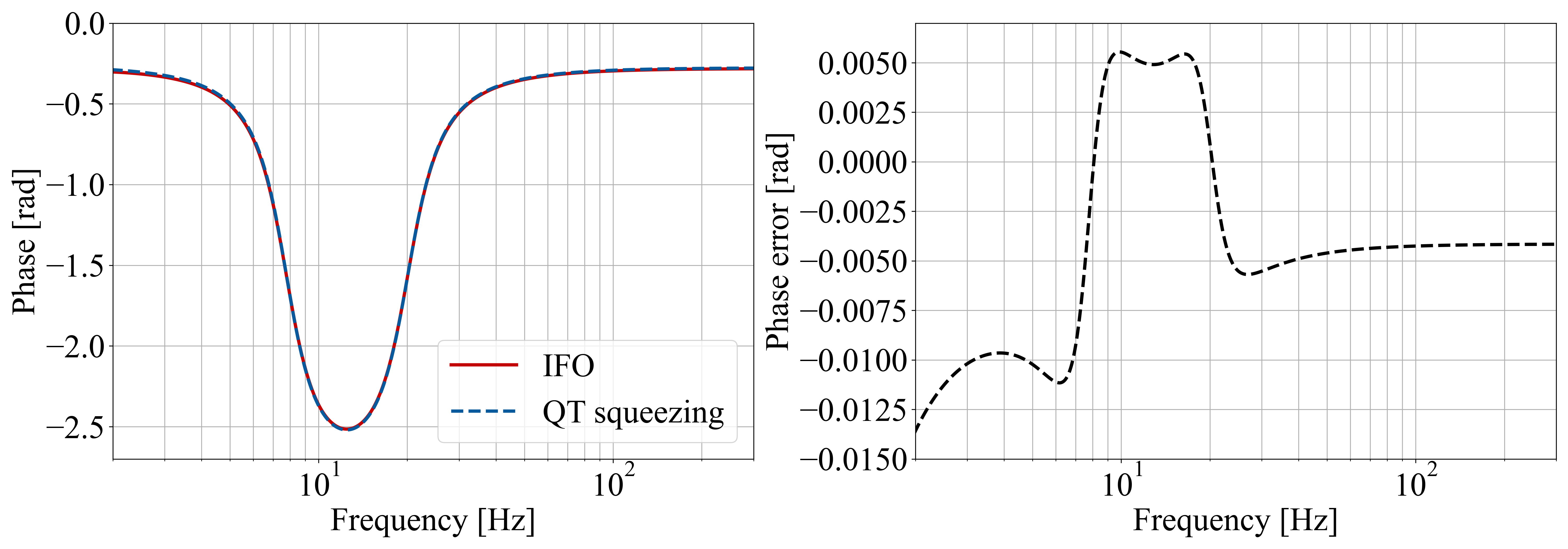}
    \caption{Left panel: Quadrature rotations. The solid red curve is caused by the interferometer (IFO) and the dashed blue curve is rotation given to two idlers. Right panel: Angle error between the two rotations.}
    \label{phase_rotation}
\end{figure*}

We have four tunable parameters: macroscopic arm and SEC length, $\delta L_\mathrm{arm} = q\lambda/2$ and $\delta L_\mathrm{SEC}=p\lambda/2$ where $q$ and $p$ are integer values, and $n_{v,a}$ meaning the number of the SEC spectral ranges contained in $\Delta_{v,a}$.
In Table~\ref{s.Tab1}, we provide the optimal length and frequency tuning parameters. The adjustments required for both the arm and SEC are approximately 4.6 and -2.1 cm, respectively. Detuning frequencies of Victor and Alice are approximately \YN{$\Delta_a \sim 319$ MHz and $\Delta_v \sim 962$ MHz and the corresponding OPA cavity length is $\sim 94$ cm.} Fig~\ref{phase_rotation} illustrates a comparison of phase rotations, highlighting the ponderomotive squeezing and the approximate rotations achieved through the combination of the two idlers.

\begin{table*}[h]
\centering
\begin{tabular}{ p{1cm}||p{6.cm}|p{3.5cm}}
 \hline
 \multicolumn{3}{c}{Parameter List} \\
 \hline
    $\lambda$ & Carrier wavelength & 1550 nm \\ \hline
  $T_\mathrm{arm}$ & ITM power transmittance & 7000 ppm \\ \hline
  $T_\mathrm{SEM}$ & SEM power transmittance & 20 \% \\ \hline
  $m$ & Mirror mass & 211 kg \\ \hline
  $I_0$ & Power at BS & 63 W  \\ \hline
  $\phi_0$ & Detuning of the SEC & 0.75 rad \\ \hline
  $L_\mathrm{arm}^{(0)}$ & Arm initial length & 10 km \footnotemark[1]   \\ \hline
  $L_\mathrm{SEC}^{(0)}$ & SEC initial length & 100 m  \\ \hline
  $\gamma_\mathrm{1arm}$ & Arm cavity bandwidth & 8.35 Hz\\ \hline
  $\gamma_1/\delta_1$ & bandwidth/detuning of the first FC & 4.27/19.54 Hz \\ \hline
  $\gamma_2/\delta_2$ & bandwidth/detuning of the second FC & 1.64/-7.62 Hz \\ \hline
  $\delta L_\mathrm{arm}$ & Arm length tuning & $30000\lambda$  \\ \hline
  $\delta L_\mathrm{SEC}$ & SEC length tuning & $14030\lambda$  \\ \hline
  $\Delta_a$ & Detuning of Alice & $769\ \mathrm{kHz} + 213\ \mathrm{FSR}_\mathrm{SEC}$\footnotemark[2] \\ \hline
  $\Delta_v$ & Detuning of Victor & $656\ \mathrm{kHz} +642\ \mathrm{FSR}_\mathrm{SEC}$ \\ \hline
\end{tabular}
\footnotetext[1]{To be more precise, $L_\mathrm{arm}^{(0)}=6451612903\lambda$ and $L_\mathrm{SEC}^{(0)}=64516129\lambda$, where $\lambda$ is the wavelength of the main laser.}
\footnotetext[2]{$n_a=213$ and $n_v=642$.}
\caption{\label{s.Tab1} Parameters for ETLF~\cite{Hild_2011}}
\end{table*}

\subsection{Effect of arm cavity and SEC loss}\label{s.sec3.subsec2}
The arm cavity loss contains the power loss per each mirror and the transmissivity of the end mirror, which has values of 20 ppm and 5 ppm in the current design of ET. The arm loss is amplified by the arm cavity approximately by a factor of finesse, which expands the bandwidth of the coupled cavity as a filter cavity for idlers. On the other hand, the SEC loss contribution to the cavity bandwidth is not significant even though it has a large value as 1000 ppm. 

According to the scaling-law theorem, the effective bandwidth of the lossy SEC-arm coupled cavity, $\gamma_\mathrm{loss}^\mathrm{eff}$, can be expressed as follows:
\begin{align}
    \gamma_\mathrm{loss}^\mathrm{eff} 
    &= \gamma_\mathrm{loss}+\gamma_2 \notag\\
    &= \frac{(T_\mathrm{SEC}+A_\mathrm{SEC})\gamma_\mathrm{1arm}}{1+2\sqrt{R_\mathrm{SEC}}\cos 2\phi_\mathrm{SEC} + R_\mathrm{SEC}} + \frac{cA_\mathrm{arm}}{4L_\mathrm{arm}} \notag \\
    &= \frac{T_S\gamma_\mathrm{1arm}}{1+2\sqrt{R_\mathrm{SEC}}\cos 2\phi_\mathrm{SEC} + R_\mathrm{SEC}} + \frac{cA^\mathrm{eff}}{4L_\mathrm{arm}},
\end{align}
where
\begin{align}
    A^\mathrm{eff} = \frac{T_\mathrm{arm}A_\mathrm{SEC}}{1+2\sqrt{R_\mathrm{SEC}}\cos 2\phi_\mathrm{SEC} + R_\mathrm{SEC}}+A_\mathrm{arm}. \label{s.eq.A_eff}
\end{align}
Here, $\gamma_\mathrm{loss}$ represents the coupled-cavity bandwidth when arm cavities have no loss, and $\gamma_2$ is the contribution of the arm cavity loss to expansion of the bandwidth and $\phi_\mathrm{SEC}$ is the one-way phase inside the SEC for idlers. Eq.~(\ref{s.eq.A_eff}) is the expansion of the bandwidth due to the losses. Using the current ETLF parameter, the numerator of the first term $T_\mathrm{arm}A_\mathrm{SEC}$ is calculated as $7~\mathrm{ppm}$. Considering the amplification gains by the denominator for two idlers, approximately $1.0$ and $2.5$ respectively, the effective filter cavity losses are $A_v^\mathrm{eff} \sim 52~\mathrm{ppm}$ and $A_a^\mathrm{eff} \sim 63~\mathrm{ppm}$.
Since the length of the filter cavity is that of the arm cavity $L_\mathrm{arm}=10~\mathrm{km}$, the QT squeezing has less noise contribution than conventional filter cavities in terms of the loss per unit length (This is explained also in the references \cite{PhysRevD.81.122002, PhysRevD.90.062006}). Given the round-trip loss in the filter cavity is $A_\mathrm{FC}=20~\mathrm{ppm}$, and the length $L_\mathrm{FC}=1~\mathrm{km}$, discussion above leads:
\begin{align*}
    \frac{A_{v,a}^\mathrm{eff}}{L_\mathrm{arm}}< \frac{A_\mathrm{FC}}{L_\mathrm{FC}},
\end{align*}
showing that the optical loss contribution to the expansion of the filter-cavity bandwidth is smaller in the QT squeezing than the conventional squeezing.

\section{Noise budget}
\begin{figure*}
    \centering
    \includegraphics[scale=0.35]{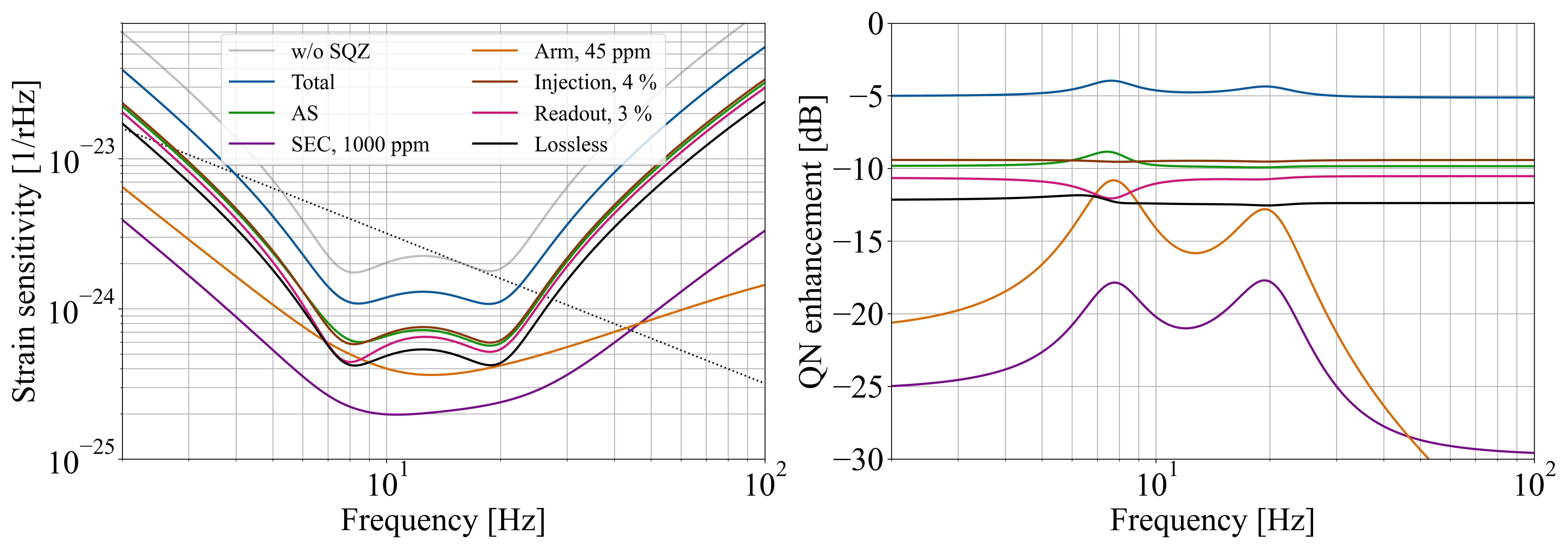}
    \caption{Left panel: Noise contributions of each loss source to the strain sensitivity. Phase noises such as squeezer noise rms, local oscillator rms, and SEC length rms are all encompassed within the AS noise. The black curve depicts the ideal case without imperfections. Right panel: Quantum noise enhancement factor in dB compared to the noise spectrum without squeezing (shown in gray in the left panel).}
    \label{fig:ss_QT_reduced}
\end{figure*}

Fig.~\ref{fig:ss_QT_reduced} displays the quantum noise budget for the QT squeezing with squeezing level of -15 dB, alongside the corresponding ETLF results for conventional squeezing. Vacuum fields are introduced into the beam path by each loss source, and their total contribution is obtained by integrating the fields from the same loss point. Four optical losses have been accounted for, including (1)  4\% injection loss that considers losses in the OPA cavity and Faraday isolator; (2) A loss of 45ppm assumed for the round-trip of light in the arm, as discussed in Section~\ref{s.sec3.subsec2}. (3) A loss of 1000ppm is assumed for the signal extraction mirror, the central beam splitter, and the imperfections of the Michelson interferometer, included in the overall SEC loss. (4) A 3\% loss in readout due to the combined losses in the Faraday isolator, output mode cleaner, and inefficiency of the photo-detector. Losses from any source result in uncorrelated vacuum noise in the beam path that reduces the level of squeezing and correlation between EPR-entangled photons. 

We evaluated three types of phase noise via root-mean-square (RMS) values in our analysis: (1) 10 milliradians of squeezer phase noise, which indicates the relative phase noise between the squeezer and the primary laser. (2) 10 mrad of local oscillator RMS phase uncertainty, which indicates the relative phase fluctuations between the local oscillator and the primary laser. (3) One picometer of SEC RMS length variations, which refers to the fluctuations in the optical length of the signal extraction cavity. These imperfections do not introduce any uncorrelated vacuum noise, rather contaminate the output with the contribution from the quadrature of light orthogonal to the one measured, which should not appear in the detection. 

\begin{figure}
    \centering
    \includegraphics[scale=0.27]{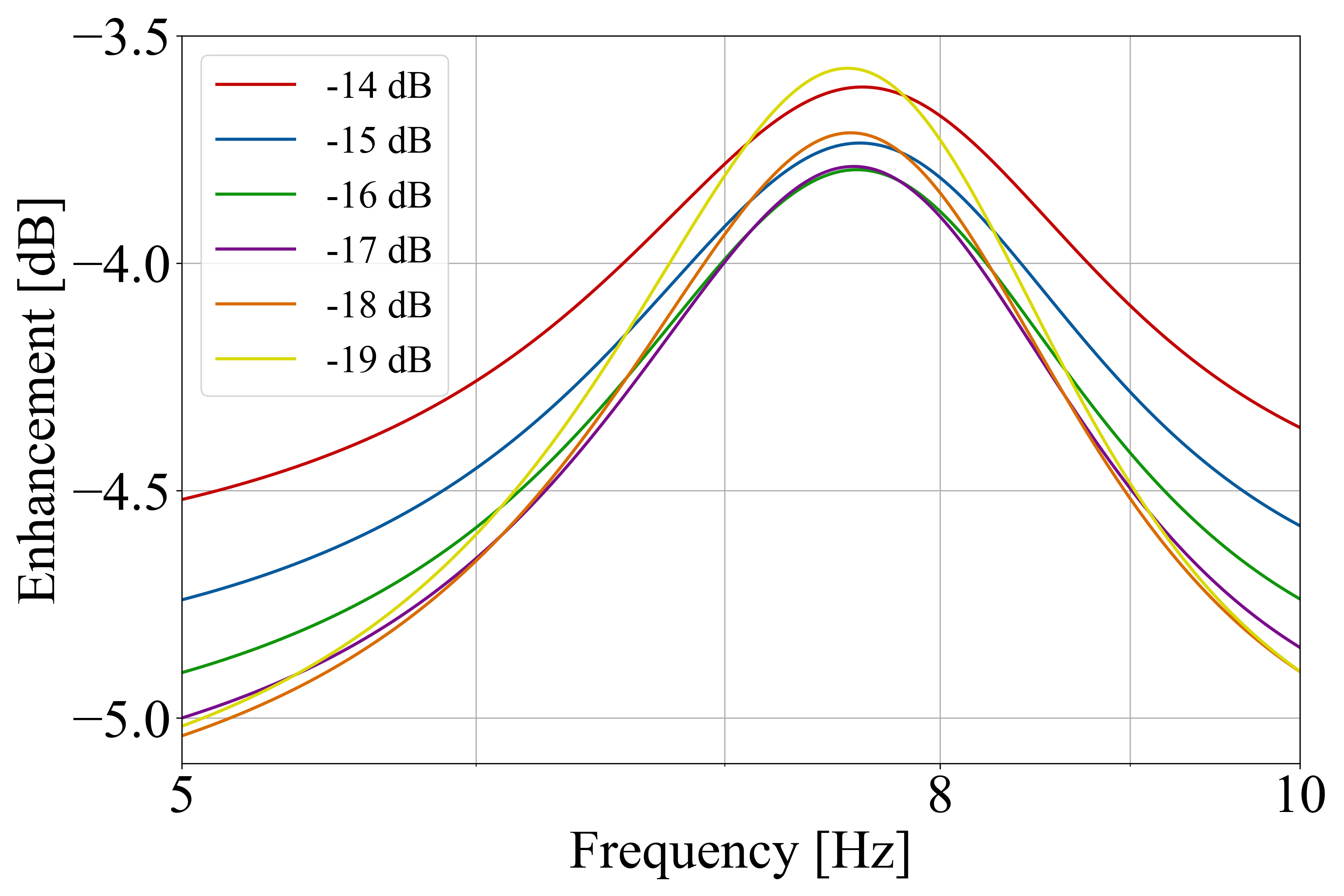}
    \caption{Quantum noise enhancement with various squeezing levels.}
    \label{fig:QN_enhancement_various_SQZ_levels}
\end{figure}

As shown in Fig.~\ref{fig:ss_QT_reduced}, all noises induced by phase fluctuations are included in the anti-symmetric (AS) port vacuum. It will be a subject of further research to break down the phase noise into parameters termed dephasing, as exemplified in \cite{PhysRevD.104.062006}, to elucidate the specific impact of each phase noise. We also plot the noise enhancement curves with various squeezing levels in Fig.~\ref{fig:QN_enhancement_various_SQZ_levels}. It shows that the phase noise does not exceed the merit of squeezing when squeezing level is below -17 dB, while above that the sensitivity deteriorates at the optical-spring resonance.

\bibliography{references}

\begin{thebibliography}{53}%
\makeatletter
\providecommand \@ifxundefined [1]{%
 \@ifx{#1\undefined}
}%
\providecommand \@ifnum [1]{%
 \ifnum #1\expandafter \@firstoftwo
 \else \expandafter \@secondoftwo
 \fi
}%
\providecommand \@ifx [1]{%
 \ifx #1\expandafter \@firstoftwo
 \else \expandafter \@secondoftwo
 \fi
}%
\providecommand \natexlab [1]{#1}%
\providecommand \enquote  [1]{``#1''}%
\providecommand \bibnamefont  [1]{#1}%
\providecommand \bibfnamefont [1]{#1}%
\providecommand \citenamefont [1]{#1}%
\providecommand \href@noop [0]{\@secondoftwo}%
\providecommand \href [0]{\begingroup \@sanitize@url \@href}%
\providecommand \@href[1]{\@@startlink{#1}\@@href}%
\providecommand \@@href[1]{\endgroup#1\@@endlink}%
\providecommand \@sanitize@url [0]{\catcode `\\12\catcode `\$12\catcode `\&12\catcode `\#12\catcode `\^12\catcode `\_12\catcode `\%12\relax}%
\providecommand \@@startlink[1]{}%
\providecommand \@@endlink[0]{}%
\providecommand \url  [0]{\begingroup\@sanitize@url \@url }%
\providecommand \@url [1]{\endgroup\@href {#1}{\urlprefix }}%
\providecommand \urlprefix  [0]{URL }%
\providecommand \Eprint [0]{\href }%
\providecommand \doibase [0]{https://doi.org/}%
\providecommand \selectlanguage [0]{\@gobble}%
\providecommand \bibinfo  [0]{\@secondoftwo}%
\providecommand \bibfield  [0]{\@secondoftwo}%
\providecommand \translation [1]{[#1]}%
\providecommand \BibitemOpen [0]{}%
\providecommand \bibitemStop [0]{}%
\providecommand \bibitemNoStop [0]{.\EOS\space}%
\providecommand \EOS [0]{\spacefactor3000\relax}%
\providecommand \BibitemShut  [1]{\csname bibitem#1\endcsname}%
\let\auto@bib@innerbib\@empty
\bibitem [{\citenamefont {{LIGO Scientific Collaboration}}\ \emph {et~al.}(2015)\citenamefont {{LIGO Scientific Collaboration}} \emph {et~al.}}]{Aasi_2015}%
  \BibitemOpen
  \bibfield  {author} {\bibinfo {author} {\bibnamefont {{LIGO Scientific Collaboration}}} \emph {et~al.},\ }\bibfield  {title} {\bibinfo {title} {Advanced ligo},\ }\href {https://doi.org/10.1088/0264-9381/32/7/074001} {\bibfield  {journal} {\bibinfo  {journal} {Classical and Quantum Gravity}\ }\textbf {\bibinfo {volume} {32}},\ \bibinfo {pages} {074001} (\bibinfo {year} {2015})}\BibitemShut {NoStop}%
\bibitem [{\citenamefont {{LIGO Scientific Collaboration}}\ \emph {et~al.}(2016)\citenamefont {{LIGO Scientific Collaboration}}, \citenamefont {{Virgo Collaboration}} \emph {et~al.}}]{PhysRevLett.116.061102}%
  \BibitemOpen
  \bibfield  {author} {\bibinfo {author} {\bibnamefont {{LIGO Scientific Collaboration}}}, \bibinfo {author} {\bibnamefont {{Virgo Collaboration}}}, \emph {et~al.} (\bibinfo {collaboration} {LIGO Scientific Collaboration and Virgo Collaboration}),\ }\bibfield  {title} {\bibinfo {title} {Observation of gravitational waves from a binary black hole merger},\ }\href {https://doi.org/10.1103/PhysRevLett.116.061102} {\bibfield  {journal} {\bibinfo  {journal} {Phys. Rev. Lett.}\ }\textbf {\bibinfo {volume} {116}},\ \bibinfo {pages} {061102} (\bibinfo {year} {2016})}\BibitemShut {NoStop}%
\bibitem [{\citenamefont {Acernese}\ \emph {et~al.}(2014)\citenamefont {Acernese} \emph {et~al.}}]{Acernese_2015}%
  \BibitemOpen
  \bibfield  {author} {\bibinfo {author} {\bibfnamefont {F.}~\bibnamefont {Acernese}} \emph {et~al.},\ }\bibfield  {title} {\bibinfo {title} {Advanced virgo: a second-generation interferometric gravitational wave detector},\ }\href {https://doi.org/10.1088/0264-9381/32/2/024001} {\bibfield  {journal} {\bibinfo  {journal} {Classical and Quantum Gravity}\ }\textbf {\bibinfo {volume} {32}},\ \bibinfo {pages} {024001} (\bibinfo {year} {2014})}\BibitemShut {NoStop}%
\bibitem [{\citenamefont {{KAGRA Collaboration and others}}(2019)}]{KAGRA_2019}%
  \BibitemOpen
  \bibfield  {author} {\bibinfo {author} {\bibnamefont {{KAGRA Collaboration and others}}},\ }\bibfield  {title} {\bibinfo {title} {{KAGRA}: 2.5 generation interferometric gravitational wave detector},\ }\href {https://doi.org/10.1038/s41550-018-0658-y} {\bibfield  {journal} {\bibinfo  {journal} {Nature Astronomy}\ }\textbf {\bibinfo {volume} {3}},\ \bibinfo {pages} {35} (\bibinfo {year} {2019})}\BibitemShut {NoStop}%
\bibitem [{\citenamefont {{LIGO Scientific Collaboration}}\ \emph {et~al.}(2019)\citenamefont {{LIGO Scientific Collaboration}}, \citenamefont {{Virgo Collaboration}} \emph {et~al.}}]{PhysRevX.9.031040}%
  \BibitemOpen
  \bibfield  {author} {\bibinfo {author} {\bibnamefont {{LIGO Scientific Collaboration}}}, \bibinfo {author} {\bibnamefont {{Virgo Collaboration}}}, \emph {et~al.} (\bibinfo {collaboration} {LIGO Scientific Collaboration, Virgo Collaboration}),\ }\bibfield  {title} {\bibinfo {title} {{GWTC-1}: A gravitational-wave transient catalog of compact binary mergers observed by ligo and virgo during the first and second observing runs},\ }\href {https://doi.org/10.1103/PhysRevX.9.031040} {\bibfield  {journal} {\bibinfo  {journal} {Phys. Rev. X}\ }\textbf {\bibinfo {volume} {9}},\ \bibinfo {pages} {031040} (\bibinfo {year} {2019})}\BibitemShut {NoStop}%
\bibitem [{\citenamefont {{LIGO Scientific Collaboration}}\ \emph {et~al.}(2021{\natexlab{a}})\citenamefont {{LIGO Scientific Collaboration}}, \citenamefont {{Virgo Collaboration}} \emph {et~al.}}]{PhysRevX.11.021053}%
  \BibitemOpen
  \bibfield  {author} {\bibinfo {author} {\bibnamefont {{LIGO Scientific Collaboration}}}, \bibinfo {author} {\bibnamefont {{Virgo Collaboration}}}, \emph {et~al.} (\bibinfo {collaboration} {LIGO Scientific Collaboration and Virgo Collaboration}),\ }\bibfield  {title} {\bibinfo {title} {Gwtc-2: Compact binary coalescences observed by ligo and virgo during the first half of the third observing run},\ }\href {https://doi.org/10.1103/PhysRevX.11.021053} {\bibfield  {journal} {\bibinfo  {journal} {Phys. Rev. X}\ }\textbf {\bibinfo {volume} {11}},\ \bibinfo {pages} {021053} (\bibinfo {year} {2021}{\natexlab{a}})}\BibitemShut {NoStop}%
\bibitem [{\citenamefont {{LIGO Scientific Collaboration}}\ \emph {et~al.}(2021{\natexlab{b}})\citenamefont {{LIGO Scientific Collaboration}}, \citenamefont {{Virgo Collaboration}} \emph {et~al.}}]{theligoscientificcollaboration2021gwtc3}%
  \BibitemOpen
  \bibfield  {author} {\bibinfo {author} {\bibnamefont {{LIGO Scientific Collaboration}}}, \bibinfo {author} {\bibnamefont {{Virgo Collaboration}}}, \emph {et~al.},\ }\href@noop {} {\bibinfo {title} {Gwtc-3: Compact binary coalescences observed by ligo and virgo during the second part of the third observing run}} (\bibinfo {year} {2021}{\natexlab{b}}),\ \Eprint {https://arxiv.org/abs/2111.03606} {arXiv:2111.03606 [gr-qc]} \BibitemShut {NoStop}%
\bibitem [{\citenamefont {Venumadhav}\ \emph {et~al.}(2020)\citenamefont {Venumadhav}, \citenamefont {Zackay}, \citenamefont {Roulet}, \citenamefont {Dai},\ and\ \citenamefont {Zaldarriaga}}]{PhysRevD.101.083030}%
  \BibitemOpen
  \bibfield  {author} {\bibinfo {author} {\bibfnamefont {T.}~\bibnamefont {Venumadhav}}, \bibinfo {author} {\bibfnamefont {B.}~\bibnamefont {Zackay}}, \bibinfo {author} {\bibfnamefont {J.}~\bibnamefont {Roulet}}, \bibinfo {author} {\bibfnamefont {L.}~\bibnamefont {Dai}},\ and\ \bibinfo {author} {\bibfnamefont {M.}~\bibnamefont {Zaldarriaga}},\ }\bibfield  {title} {\bibinfo {title} {New binary black hole mergers in the second observing run of advanced ligo and advanced virgo},\ }\href {https://doi.org/10.1103/PhysRevD.101.083030} {\bibfield  {journal} {\bibinfo  {journal} {Phys. Rev. D}\ }\textbf {\bibinfo {volume} {101}},\ \bibinfo {pages} {083030} (\bibinfo {year} {2020})}\BibitemShut {NoStop}%
\bibitem [{\citenamefont {Abbott}\ \emph {et~al.}(2017)\citenamefont {Abbott} \emph {et~al.}}]{Abbott_2017}%
  \BibitemOpen
  \bibfield  {author} {\bibinfo {author} {\bibfnamefont {B.~P.}\ \bibnamefont {Abbott}} \emph {et~al.},\ }\bibfield  {title} {\bibinfo {title} {Exploring the sensitivity of next generation gravitational wave detectors},\ }\href {https://doi.org/10.1088/1361-6382/aa51f4} {\bibfield  {journal} {\bibinfo  {journal} {Classical and Quantum Gravity}\ }\textbf {\bibinfo {volume} {34}},\ \bibinfo {pages} {044001} (\bibinfo {year} {2017})}\BibitemShut {NoStop}%
\bibitem [{\citenamefont {Hild}\ \emph {et~al.}(2011)\citenamefont {Hild} \emph {et~al.}}]{Hild_2011}%
  \BibitemOpen
  \bibfield  {author} {\bibinfo {author} {\bibfnamefont {S.}~\bibnamefont {Hild}} \emph {et~al.},\ }\bibfield  {title} {\bibinfo {title} {Sensitivity studies for third-generation gravitational wave observatories},\ }\href {https://doi.org/10.1088/0264-9381/28/9/094013} {\bibfield  {journal} {\bibinfo  {journal} {Classical and Quantum Gravity}\ }\textbf {\bibinfo {volume} {28}},\ \bibinfo {pages} {094013} (\bibinfo {year} {2011})}\BibitemShut {NoStop}%
\bibitem [{\citenamefont {Pieroni}\ \emph {et~al.}(2022)\citenamefont {Pieroni}, \citenamefont {Ricciardone},\ and\ \citenamefont {Barausse}}]{Pieroni_2022}%
  \BibitemOpen
  \bibfield  {author} {\bibinfo {author} {\bibfnamefont {M.}~\bibnamefont {Pieroni}}, \bibinfo {author} {\bibfnamefont {A.}~\bibnamefont {Ricciardone}},\ and\ \bibinfo {author} {\bibfnamefont {E.}~\bibnamefont {Barausse}},\ }\bibfield  {title} {\bibinfo {title} {Detectability and parameter estimation of stellar origin black hole binaries with next generation gravitational wave detectors},\ }\bibfield  {journal} {\bibinfo  {journal} {Scientific Reports}\ }\textbf {\bibinfo {volume} {12}},\ \href {https://doi.org/10.1038/s41598-022-19540-7} {10.1038/s41598-022-19540-7} (\bibinfo {year} {2022})\BibitemShut {NoStop}%
\bibitem [{\citenamefont {Maggiore}\ \emph {et~al.}(2020)\citenamefont {Maggiore} \emph {et~al.}}]{Maggiore_2020}%
  \BibitemOpen
  \bibfield  {author} {\bibinfo {author} {\bibfnamefont {M.}~\bibnamefont {Maggiore}} \emph {et~al.},\ }\bibfield  {title} {\bibinfo {title} {Science case for the einstein telescope},\ }\href {https://doi.org/10.1088/1475-7516/2020/03/050} {\bibfield  {journal} {\bibinfo  {journal} {Journal of Cosmology and Astroparticle Physics}\ }\textbf {\bibinfo {volume} {2020}}\bibinfo  {number} { (03)},\ \bibinfo {pages} {050}}\BibitemShut {NoStop}%
\bibitem [{\citenamefont {Sathyaprakash}\ \emph {et~al.}(2012)\citenamefont {Sathyaprakash} \emph {et~al.}}]{Sathyaprakash_2012}%
  \BibitemOpen
\bibfield  {number} {  }\bibfield  {author} {\bibinfo {author} {\bibfnamefont {B.}~\bibnamefont {Sathyaprakash}} \emph {et~al.},\ }\bibfield  {title} {\bibinfo {title} {Scientific objectives of einstein telescope},\ }\href {https://doi.org/10.1088/0264-9381/29/12/124013} {\bibfield  {journal} {\bibinfo  {journal} {Classical and Quantum Gravity}\ }\textbf {\bibinfo {volume} {29}},\ \bibinfo {pages} {124013} (\bibinfo {year} {2012})}\BibitemShut {NoStop}%
\bibitem [{\citenamefont {Shankaranarayanan}\ and\ \citenamefont {Johnson}(2022)}]{Shankaranarayanan_2022}%
  \BibitemOpen
  \bibfield  {author} {\bibinfo {author} {\bibfnamefont {S.}~\bibnamefont {Shankaranarayanan}}\ and\ \bibinfo {author} {\bibfnamefont {J.~P.}\ \bibnamefont {Johnson}},\ }\bibfield  {title} {\bibinfo {title} {Modified theories of gravity: Why, how and what?},\ }\bibfield  {journal} {\bibinfo  {journal} {General Relativity and Gravitation}\ }\textbf {\bibinfo {volume} {54}},\ \href {https://doi.org/10.1007/s10714-022-02927-2} {10.1007/s10714-022-02927-2} (\bibinfo {year} {2022})\BibitemShut {NoStop}%
\bibitem [{\citenamefont {Sathyaprakash}\ \emph {et~al.}(2019)\citenamefont {Sathyaprakash}, \citenamefont {Belgacem}, \citenamefont {Bertacca}, \citenamefont {Caprini}, \citenamefont {Cusin}, \citenamefont {Dirian}, \citenamefont {Fan}, \citenamefont {Figueroa}, \citenamefont {Foffa}, \citenamefont {Hall}, \citenamefont {Harms}, \citenamefont {Maggiore}, \citenamefont {Mandic}, \citenamefont {Matas}, \citenamefont {Regimbau}, \citenamefont {Sakellariadou}, \citenamefont {Tamanini},\ and\ \citenamefont {Thrane}}]{sathyaprakash2019cosmology}%
  \BibitemOpen
  \bibfield  {author} {\bibinfo {author} {\bibfnamefont {B.~S.}\ \bibnamefont {Sathyaprakash}}, \bibinfo {author} {\bibfnamefont {E.}~\bibnamefont {Belgacem}}, \bibinfo {author} {\bibfnamefont {D.}~\bibnamefont {Bertacca}}, \bibinfo {author} {\bibfnamefont {C.}~\bibnamefont {Caprini}}, \bibinfo {author} {\bibfnamefont {G.}~\bibnamefont {Cusin}}, \bibinfo {author} {\bibfnamefont {Y.}~\bibnamefont {Dirian}}, \bibinfo {author} {\bibfnamefont {X.}~\bibnamefont {Fan}}, \bibinfo {author} {\bibfnamefont {D.}~\bibnamefont {Figueroa}}, \bibinfo {author} {\bibfnamefont {S.}~\bibnamefont {Foffa}}, \bibinfo {author} {\bibfnamefont {E.}~\bibnamefont {Hall}}, \bibinfo {author} {\bibfnamefont {J.}~\bibnamefont {Harms}}, \bibinfo {author} {\bibfnamefont {M.}~\bibnamefont {Maggiore}}, \bibinfo {author} {\bibfnamefont {V.}~\bibnamefont {Mandic}}, \bibinfo {author} {\bibfnamefont {A.}~\bibnamefont {Matas}}, \bibinfo {author} {\bibfnamefont {T.}~\bibnamefont {Regimbau}}, \bibinfo {author} {\bibfnamefont {M.}~\bibnamefont
  {Sakellariadou}}, \bibinfo {author} {\bibfnamefont {N.}~\bibnamefont {Tamanini}},\ and\ \bibinfo {author} {\bibfnamefont {E.}~\bibnamefont {Thrane}},\ }\href@noop {} {\bibinfo {title} {Cosmology and the early universe}} (\bibinfo {year} {2019}),\ \Eprint {https://arxiv.org/abs/1903.09260} {arXiv:1903.09260 [astro-ph.HE]} \BibitemShut {NoStop}%
\bibitem [{\citenamefont {{Braginski{\v{i}}}}(1968)}]{1968JETP.26.831}%
  \BibitemOpen
  \bibfield  {author} {\bibinfo {author} {\bibfnamefont {V.~B.}\ \bibnamefont {{Braginski{\v{i}}}}},\ }\bibfield  {title} {\bibinfo {title} {{Classical and Quantum Restrictions on the Detection of Weak Disturbances of a Macroscopic Oscillator}},\ }\href@noop {} {\bibfield  {journal} {\bibinfo  {journal} {Soviet Journal of Experimental and Theoretical Physics}\ }\textbf {\bibinfo {volume} {26}},\ \bibinfo {pages} {831} (\bibinfo {year} {1968})}\BibitemShut {NoStop}%
\bibitem [{\citenamefont {Caves}(1980)}]{PhysRevLett.45.75}%
  \BibitemOpen
  \bibfield  {author} {\bibinfo {author} {\bibfnamefont {C.~M.}\ \bibnamefont {Caves}},\ }\bibfield  {title} {\bibinfo {title} {Quantum-mechanical radiation-pressure fluctuations in an interferometer},\ }\href {https://doi.org/10.1103/PhysRevLett.45.75} {\bibfield  {journal} {\bibinfo  {journal} {Phys. Rev. Lett.}\ }\textbf {\bibinfo {volume} {45}},\ \bibinfo {pages} {75} (\bibinfo {year} {1980})}\BibitemShut {NoStop}%
\bibitem [{\citenamefont {Thorne}\ \emph {et~al.}(1978)\citenamefont {Thorne}, \citenamefont {Drever}, \citenamefont {Caves}, \citenamefont {Zimmermann},\ and\ \citenamefont {Sandberg}}]{PhysRevLett.40.667}%
  \BibitemOpen
  \bibfield  {author} {\bibinfo {author} {\bibfnamefont {K.~S.}\ \bibnamefont {Thorne}}, \bibinfo {author} {\bibfnamefont {R.~W.~P.}\ \bibnamefont {Drever}}, \bibinfo {author} {\bibfnamefont {C.~M.}\ \bibnamefont {Caves}}, \bibinfo {author} {\bibfnamefont {M.}~\bibnamefont {Zimmermann}},\ and\ \bibinfo {author} {\bibfnamefont {V.~D.}\ \bibnamefont {Sandberg}},\ }\bibfield  {title} {\bibinfo {title} {Quantum nondemolition measurements of harmonic oscillators},\ }\href {https://doi.org/10.1103/PhysRevLett.40.667} {\bibfield  {journal} {\bibinfo  {journal} {Phys. Rev. Lett.}\ }\textbf {\bibinfo {volume} {40}},\ \bibinfo {pages} {667} (\bibinfo {year} {1978})}\BibitemShut {NoStop}%
\bibitem [{\citenamefont {Braginsky}\ and\ \citenamefont {Khalili}(1996)}]{RevModPhys.68.1}%
  \BibitemOpen
  \bibfield  {author} {\bibinfo {author} {\bibfnamefont {V.~B.}\ \bibnamefont {Braginsky}}\ and\ \bibinfo {author} {\bibfnamefont {F.~Y.}\ \bibnamefont {Khalili}},\ }\bibfield  {title} {\bibinfo {title} {Quantum nondemolition measurements: the route from toys to tools},\ }\href {https://doi.org/10.1103/RevModPhys.68.1} {\bibfield  {journal} {\bibinfo  {journal} {Rev. Mod. Phys.}\ }\textbf {\bibinfo {volume} {68}},\ \bibinfo {pages} {1} (\bibinfo {year} {1996})}\BibitemShut {NoStop}%
\bibitem [{\citenamefont {Kimble}\ \emph {et~al.}(2001)\citenamefont {Kimble}, \citenamefont {Levin}, \citenamefont {Matsko}, \citenamefont {Thorne},\ and\ \citenamefont {Vyatchanin}}]{PhysRevD.65.022002}%
  \BibitemOpen
  \bibfield  {author} {\bibinfo {author} {\bibfnamefont {H.~J.}\ \bibnamefont {Kimble}}, \bibinfo {author} {\bibfnamefont {Y.}~\bibnamefont {Levin}}, \bibinfo {author} {\bibfnamefont {A.~B.}\ \bibnamefont {Matsko}}, \bibinfo {author} {\bibfnamefont {K.~S.}\ \bibnamefont {Thorne}},\ and\ \bibinfo {author} {\bibfnamefont {S.~P.}\ \bibnamefont {Vyatchanin}},\ }\bibfield  {title} {\bibinfo {title} {Conversion of conventional gravitational-wave interferometers into quantum nondemolition interferometers by modifying their input and/or output optics},\ }\href {https://doi.org/10.1103/PhysRevD.65.022002} {\bibfield  {journal} {\bibinfo  {journal} {Phys. Rev. D}\ }\textbf {\bibinfo {volume} {65}},\ \bibinfo {pages} {022002} (\bibinfo {year} {2001})}\BibitemShut {NoStop}%
\bibitem [{\citenamefont {Caves}(1981)}]{PhysRevD.23.1693}%
  \BibitemOpen
  \bibfield  {author} {\bibinfo {author} {\bibfnamefont {C.~M.}\ \bibnamefont {Caves}},\ }\bibfield  {title} {\bibinfo {title} {Quantum-mechanical noise in an interferometer},\ }\href {https://doi.org/10.1103/PhysRevD.23.1693} {\bibfield  {journal} {\bibinfo  {journal} {Phys. Rev. D}\ }\textbf {\bibinfo {volume} {23}},\ \bibinfo {pages} {1693} (\bibinfo {year} {1981})}\BibitemShut {NoStop}%
\bibitem [{\citenamefont {Mizuno}\ \emph {et~al.}(1993)\citenamefont {Mizuno}, \citenamefont {Strain}, \citenamefont {Nelson}, \citenamefont {Chen}, \citenamefont {Schilling}, \citenamefont {Rüdiger}, \citenamefont {Winkler},\ and\ \citenamefont {Danzmann}}]{MIZUNO1993273}%
  \BibitemOpen
  \bibfield  {author} {\bibinfo {author} {\bibfnamefont {J.}~\bibnamefont {Mizuno}}, \bibinfo {author} {\bibfnamefont {K.}~\bibnamefont {Strain}}, \bibinfo {author} {\bibfnamefont {P.}~\bibnamefont {Nelson}}, \bibinfo {author} {\bibfnamefont {J.}~\bibnamefont {Chen}}, \bibinfo {author} {\bibfnamefont {R.}~\bibnamefont {Schilling}}, \bibinfo {author} {\bibfnamefont {A.}~\bibnamefont {Rüdiger}}, \bibinfo {author} {\bibfnamefont {W.}~\bibnamefont {Winkler}},\ and\ \bibinfo {author} {\bibfnamefont {K.}~\bibnamefont {Danzmann}},\ }\bibfield  {title} {\bibinfo {title} {Resonant sideband extraction: a new configuration for interferometric gravitational wave detectors},\ }\href {https://doi.org/https://doi.org/10.1016/0375-9601(93)90620-F} {\bibfield  {journal} {\bibinfo  {journal} {Physics Letters A}\ }\textbf {\bibinfo {volume} {175}},\ \bibinfo {pages} {273} (\bibinfo {year} {1993})}\BibitemShut {NoStop}%
\bibitem [{\citenamefont {Purdue}\ and\ \citenamefont {Chen}(2002)}]{PhysRevD.66.122004}%
  \BibitemOpen
  \bibfield  {author} {\bibinfo {author} {\bibfnamefont {P.}~\bibnamefont {Purdue}}\ and\ \bibinfo {author} {\bibfnamefont {Y.}~\bibnamefont {Chen}},\ }\bibfield  {title} {\bibinfo {title} {Practical speed meter designs for quantum nondemolition gravitational-wave interferometers},\ }\href {https://doi.org/10.1103/PhysRevD.66.122004} {\bibfield  {journal} {\bibinfo  {journal} {Phys. Rev. D}\ }\textbf {\bibinfo {volume} {66}},\ \bibinfo {pages} {122004} (\bibinfo {year} {2002})}\BibitemShut {NoStop}%
\bibitem [{\citenamefont {Khalili}(2010)}]{PhysRevD.81.122002}%
  \BibitemOpen
  \bibfield  {author} {\bibinfo {author} {\bibfnamefont {F.~Y.}\ \bibnamefont {Khalili}},\ }\bibfield  {title} {\bibinfo {title} {Optimal configurations of filter cavity in future gravitational-wave detectors},\ }\href {https://doi.org/10.1103/PhysRevD.81.122002} {\bibfield  {journal} {\bibinfo  {journal} {Phys. Rev. D}\ }\textbf {\bibinfo {volume} {81}},\ \bibinfo {pages} {122002} (\bibinfo {year} {2010})}\BibitemShut {NoStop}%
\bibitem [{\citenamefont {Harms}\ \emph {et~al.}(2003)\citenamefont {Harms}, \citenamefont {Chen}, \citenamefont {Chelkowski}, \citenamefont {Franzen}, \citenamefont {Vahlbruch}, \citenamefont {Danzmann},\ and\ \citenamefont {Schnabel}}]{PhysRevD.68.042001}%
  \BibitemOpen
  \bibfield  {author} {\bibinfo {author} {\bibfnamefont {J.}~\bibnamefont {Harms}}, \bibinfo {author} {\bibfnamefont {Y.}~\bibnamefont {Chen}}, \bibinfo {author} {\bibfnamefont {S.}~\bibnamefont {Chelkowski}}, \bibinfo {author} {\bibfnamefont {A.}~\bibnamefont {Franzen}}, \bibinfo {author} {\bibfnamefont {H.}~\bibnamefont {Vahlbruch}}, \bibinfo {author} {\bibfnamefont {K.}~\bibnamefont {Danzmann}},\ and\ \bibinfo {author} {\bibfnamefont {R.}~\bibnamefont {Schnabel}},\ }\bibfield  {title} {\bibinfo {title} {Squeezed-input, optical-spring, signal-recycled gravitational-wave detectors},\ }\href {https://doi.org/10.1103/PhysRevD.68.042001} {\bibfield  {journal} {\bibinfo  {journal} {Phys. Rev. D}\ }\textbf {\bibinfo {volume} {68}},\ \bibinfo {pages} {042001} (\bibinfo {year} {2003})}\BibitemShut {NoStop}%
\bibitem [{\citenamefont {Miao}\ \emph {et~al.}(2018)\citenamefont {Miao}, \citenamefont {Yang},\ and\ \citenamefont {Martynov}}]{PhysRevD.98.044044}%
  \BibitemOpen
  \bibfield  {author} {\bibinfo {author} {\bibfnamefont {H.}~\bibnamefont {Miao}}, \bibinfo {author} {\bibfnamefont {H.}~\bibnamefont {Yang}},\ and\ \bibinfo {author} {\bibfnamefont {D.}~\bibnamefont {Martynov}},\ }\bibfield  {title} {\bibinfo {title} {Towards the design of gravitational-wave detectors for probing neutron-star physics},\ }\href {https://doi.org/10.1103/PhysRevD.98.044044} {\bibfield  {journal} {\bibinfo  {journal} {Phys. Rev. D}\ }\textbf {\bibinfo {volume} {98}},\ \bibinfo {pages} {044044} (\bibinfo {year} {2018})}\BibitemShut {NoStop}%
\bibitem [{\citenamefont {Ganapathy}\ \emph {et~al.}(2023)\citenamefont {Ganapathy}, \citenamefont {Jia}, \citenamefont {Nakano}, \citenamefont {Xu}, \citenamefont {Aritomi}, \citenamefont {Cullen}, \citenamefont {Kijbunchoo}, \citenamefont {Dwyer}, \citenamefont {Mullavey}, \citenamefont {McCuller}, \citenamefont {Abbott}, \citenamefont {Abouelfettouh}, \citenamefont {Adhikari}, \citenamefont {Ananyeva}, \citenamefont {Appert}, \citenamefont {Arai}, \citenamefont {Aston}, \citenamefont {Ball}, \citenamefont {Ballmer}, \citenamefont {Barker}, \citenamefont {Barsotti}, \citenamefont {Berger}, \citenamefont {Betzwieser}, \citenamefont {Bhattacharjee}, \citenamefont {Billingsley}, \citenamefont {Biscans}, \citenamefont {Bode}, \citenamefont {Bonilla}, \citenamefont {Bossilkov}, \citenamefont {Branch}, \citenamefont {Brooks}, \citenamefont {Brown}, \citenamefont {Bryant}, \citenamefont {Cahillane}, \citenamefont {Cao}, \citenamefont {Capote}, \citenamefont {Clara}, \citenamefont {Collins}, \citenamefont
  {Compton}, \citenamefont {Cottingham}, \citenamefont {Coyne}, \citenamefont {Crouch}, \citenamefont {Csizmazia}, \citenamefont {Dartez}, \citenamefont {Demos}, \citenamefont {Dohmen}, \citenamefont {Driggers}, \citenamefont {Effler}, \citenamefont {Ejlli}, \citenamefont {Etzel}, \citenamefont {Evans}, \citenamefont {Feicht}, \citenamefont {Frey}, \citenamefont {Frischhertz}, \citenamefont {Fritschel}, \citenamefont {Frolov}, \citenamefont {Fulda}, \citenamefont {Fyffe}, \citenamefont {Gateley}, \citenamefont {Giaime}, \citenamefont {Giardina}, \citenamefont {Glanzer}, \citenamefont {Goetz}, \citenamefont {Goetz}, \citenamefont {Goodwin-Jones}, \citenamefont {Gras}, \citenamefont {Gray}, \citenamefont {Griffith}, \citenamefont {Grote}, \citenamefont {Guidry}, \citenamefont {Hall}, \citenamefont {Hanks}, \citenamefont {Hanson}, \citenamefont {Heintze}, \citenamefont {Helmling-Cornell}, \citenamefont {Holland}, \citenamefont {Hoyland}, \citenamefont {Huang}, \citenamefont {Inoue}, \citenamefont {James},
  \citenamefont {Jennings}, \citenamefont {Karat}, \citenamefont {Karki}, \citenamefont {Kasprzack}, \citenamefont {Kawabe}, \citenamefont {King}, \citenamefont {Kissel}, \citenamefont {Komori}, \citenamefont {Kontos}, \citenamefont {Kumar}, \citenamefont {Kuns}, \citenamefont {Landry}, \citenamefont {Lantz}, \citenamefont {Laxen}, \citenamefont {Lee}, \citenamefont {Lesovsky}, \citenamefont {Llamas}, \citenamefont {Lormand}, \citenamefont {Loughlin}, \citenamefont {Macas}, \citenamefont {MacInnis}, \citenamefont {Makarem}, \citenamefont {Mannix}, \citenamefont {Mansell}, \citenamefont {Martin}, \citenamefont {Mason}, \citenamefont {Matichard}, \citenamefont {Mavalvala}, \citenamefont {Maxwell}, \citenamefont {McCarrol}, \citenamefont {McCarthy}, \citenamefont {McClelland}, \citenamefont {McCormick}, \citenamefont {McRae}, \citenamefont {Mera}, \citenamefont {Merilh}, \citenamefont {Meylahn}, \citenamefont {Mittleman}, \citenamefont {Moraru}, \citenamefont {Moreno}, \citenamefont {Nelson}, \citenamefont
  {Neunzert}, \citenamefont {Notte}, \citenamefont {Oberling}, \citenamefont {O'Hanlon}, \citenamefont {Osthelder}, \citenamefont {Ottaway}, \citenamefont {Overmier}, \citenamefont {Parker}, \citenamefont {Pele}, \citenamefont {Pham}, \citenamefont {Pirello}, \citenamefont {Quetschke}, \citenamefont {Ramirez}, \citenamefont {Reyes}, \citenamefont {Richardson}, \citenamefont {Robinson}, \citenamefont {Rollins}, \citenamefont {Romel}, \citenamefont {Romie}, \citenamefont {Ross}, \citenamefont {Ryan}, \citenamefont {Sadecki}, \citenamefont {Sanchez}, \citenamefont {Sanchez}, \citenamefont {Sanchez}, \citenamefont {Savage}, \citenamefont {Schaetzl}, \citenamefont {Schiworski}, \citenamefont {Schnabel}, \citenamefont {Schofield}, \citenamefont {Schwartz}, \citenamefont {Sellers}, \citenamefont {Shaffer}, \citenamefont {Short}, \citenamefont {Sigg}, \citenamefont {Slagmolen}, \citenamefont {Soike}, \citenamefont {Soni}, \citenamefont {Srivastava}, \citenamefont {Sun}, \citenamefont {Tanner}, \citenamefont {Thomas},
  \citenamefont {Thomas}, \citenamefont {Thorne}, \citenamefont {Torrie}, \citenamefont {Traylor}, \citenamefont {Ubhi}, \citenamefont {Vajente}, \citenamefont {Vanosky}, \citenamefont {Vecchio}, \citenamefont {Veitch}, \citenamefont {Vibhute}, \citenamefont {von Reis}, \citenamefont {Warner}, \citenamefont {Weaver}, \citenamefont {Weiss}, \citenamefont {Whittle}, \citenamefont {Willke}, \citenamefont {Wipf}, \citenamefont {Yamamoto}, \citenamefont {Zhang},\ and\ \citenamefont {Zucker}}]{PhysRevX.13.041021}%
  \BibitemOpen
  \bibfield  {author} {\bibinfo {author} {\bibfnamefont {D.}~\bibnamefont {Ganapathy}}, \bibinfo {author} {\bibfnamefont {W.}~\bibnamefont {Jia}}, \bibinfo {author} {\bibfnamefont {M.}~\bibnamefont {Nakano}}, \bibinfo {author} {\bibfnamefont {V.}~\bibnamefont {Xu}}, \bibinfo {author} {\bibfnamefont {N.}~\bibnamefont {Aritomi}}, \bibinfo {author} {\bibfnamefont {T.}~\bibnamefont {Cullen}}, \bibinfo {author} {\bibfnamefont {N.}~\bibnamefont {Kijbunchoo}}, \bibinfo {author} {\bibfnamefont {S.~E.}\ \bibnamefont {Dwyer}}, \bibinfo {author} {\bibfnamefont {A.}~\bibnamefont {Mullavey}}, \bibinfo {author} {\bibfnamefont {L.}~\bibnamefont {McCuller}}, \bibinfo {author} {\bibfnamefont {R.}~\bibnamefont {Abbott}}, \bibinfo {author} {\bibfnamefont {I.}~\bibnamefont {Abouelfettouh}}, \bibinfo {author} {\bibfnamefont {R.~X.}\ \bibnamefont {Adhikari}}, \bibinfo {author} {\bibfnamefont {A.}~\bibnamefont {Ananyeva}}, \bibinfo {author} {\bibfnamefont {S.}~\bibnamefont {Appert}}, \bibinfo {author} {\bibfnamefont
  {K.}~\bibnamefont {Arai}}, \bibinfo {author} {\bibfnamefont {S.~M.}\ \bibnamefont {Aston}}, \bibinfo {author} {\bibfnamefont {M.}~\bibnamefont {Ball}}, \bibinfo {author} {\bibfnamefont {S.~W.}\ \bibnamefont {Ballmer}}, \bibinfo {author} {\bibfnamefont {D.}~\bibnamefont {Barker}}, \bibinfo {author} {\bibfnamefont {L.}~\bibnamefont {Barsotti}}, \bibinfo {author} {\bibfnamefont {B.~K.}\ \bibnamefont {Berger}}, \bibinfo {author} {\bibfnamefont {J.}~\bibnamefont {Betzwieser}}, \bibinfo {author} {\bibfnamefont {D.}~\bibnamefont {Bhattacharjee}}, \bibinfo {author} {\bibfnamefont {G.}~\bibnamefont {Billingsley}}, \bibinfo {author} {\bibfnamefont {S.}~\bibnamefont {Biscans}}, \bibinfo {author} {\bibfnamefont {N.}~\bibnamefont {Bode}}, \bibinfo {author} {\bibfnamefont {E.}~\bibnamefont {Bonilla}}, \bibinfo {author} {\bibfnamefont {V.}~\bibnamefont {Bossilkov}}, \bibinfo {author} {\bibfnamefont {A.}~\bibnamefont {Branch}}, \bibinfo {author} {\bibfnamefont {A.~F.}\ \bibnamefont {Brooks}}, \bibinfo {author}
  {\bibfnamefont {D.~D.}\ \bibnamefont {Brown}}, \bibinfo {author} {\bibfnamefont {J.}~\bibnamefont {Bryant}}, \bibinfo {author} {\bibfnamefont {C.}~\bibnamefont {Cahillane}}, \bibinfo {author} {\bibfnamefont {H.}~\bibnamefont {Cao}}, \bibinfo {author} {\bibfnamefont {E.}~\bibnamefont {Capote}}, \bibinfo {author} {\bibfnamefont {F.}~\bibnamefont {Clara}}, \bibinfo {author} {\bibfnamefont {J.}~\bibnamefont {Collins}}, \bibinfo {author} {\bibfnamefont {C.~M.}\ \bibnamefont {Compton}}, \bibinfo {author} {\bibfnamefont {R.}~\bibnamefont {Cottingham}}, \bibinfo {author} {\bibfnamefont {D.~C.}\ \bibnamefont {Coyne}}, \bibinfo {author} {\bibfnamefont {R.}~\bibnamefont {Crouch}}, \bibinfo {author} {\bibfnamefont {J.}~\bibnamefont {Csizmazia}}, \bibinfo {author} {\bibfnamefont {L.~P.}\ \bibnamefont {Dartez}}, \bibinfo {author} {\bibfnamefont {N.}~\bibnamefont {Demos}}, \bibinfo {author} {\bibfnamefont {E.}~\bibnamefont {Dohmen}}, \bibinfo {author} {\bibfnamefont {J.~C.}\ \bibnamefont {Driggers}}, \bibinfo {author}
  {\bibfnamefont {A.}~\bibnamefont {Effler}}, \bibinfo {author} {\bibfnamefont {A.}~\bibnamefont {Ejlli}}, \bibinfo {author} {\bibfnamefont {T.}~\bibnamefont {Etzel}}, \bibinfo {author} {\bibfnamefont {M.}~\bibnamefont {Evans}}, \bibinfo {author} {\bibfnamefont {J.}~\bibnamefont {Feicht}}, \bibinfo {author} {\bibfnamefont {R.}~\bibnamefont {Frey}}, \bibinfo {author} {\bibfnamefont {W.}~\bibnamefont {Frischhertz}}, \bibinfo {author} {\bibfnamefont {P.}~\bibnamefont {Fritschel}}, \bibinfo {author} {\bibfnamefont {V.~V.}\ \bibnamefont {Frolov}}, \bibinfo {author} {\bibfnamefont {P.}~\bibnamefont {Fulda}}, \bibinfo {author} {\bibfnamefont {M.}~\bibnamefont {Fyffe}}, \bibinfo {author} {\bibfnamefont {B.}~\bibnamefont {Gateley}}, \bibinfo {author} {\bibfnamefont {J.~A.}\ \bibnamefont {Giaime}}, \bibinfo {author} {\bibfnamefont {K.~D.}\ \bibnamefont {Giardina}}, \bibinfo {author} {\bibfnamefont {J.}~\bibnamefont {Glanzer}}, \bibinfo {author} {\bibfnamefont {E.}~\bibnamefont {Goetz}}, \bibinfo {author} {\bibfnamefont
  {R.}~\bibnamefont {Goetz}}, \bibinfo {author} {\bibfnamefont {A.~W.}\ \bibnamefont {Goodwin-Jones}}, \bibinfo {author} {\bibfnamefont {S.}~\bibnamefont {Gras}}, \bibinfo {author} {\bibfnamefont {C.}~\bibnamefont {Gray}}, \bibinfo {author} {\bibfnamefont {D.}~\bibnamefont {Griffith}}, \bibinfo {author} {\bibfnamefont {H.}~\bibnamefont {Grote}}, \bibinfo {author} {\bibfnamefont {T.}~\bibnamefont {Guidry}}, \bibinfo {author} {\bibfnamefont {E.~D.}\ \bibnamefont {Hall}}, \bibinfo {author} {\bibfnamefont {J.}~\bibnamefont {Hanks}}, \bibinfo {author} {\bibfnamefont {J.}~\bibnamefont {Hanson}}, \bibinfo {author} {\bibfnamefont {M.~C.}\ \bibnamefont {Heintze}}, \bibinfo {author} {\bibfnamefont {A.~F.}\ \bibnamefont {Helmling-Cornell}}, \bibinfo {author} {\bibfnamefont {N.~A.}\ \bibnamefont {Holland}}, \bibinfo {author} {\bibfnamefont {D.}~\bibnamefont {Hoyland}}, \bibinfo {author} {\bibfnamefont {H.~Y.}\ \bibnamefont {Huang}}, \bibinfo {author} {\bibfnamefont {Y.}~\bibnamefont {Inoue}}, \bibinfo {author}
  {\bibfnamefont {A.~L.}\ \bibnamefont {James}}, \bibinfo {author} {\bibfnamefont {A.}~\bibnamefont {Jennings}}, \bibinfo {author} {\bibfnamefont {S.}~\bibnamefont {Karat}}, \bibinfo {author} {\bibfnamefont {S.}~\bibnamefont {Karki}}, \bibinfo {author} {\bibfnamefont {M.}~\bibnamefont {Kasprzack}}, \bibinfo {author} {\bibfnamefont {K.}~\bibnamefont {Kawabe}}, \bibinfo {author} {\bibfnamefont {P.~J.}\ \bibnamefont {King}}, \bibinfo {author} {\bibfnamefont {J.~S.}\ \bibnamefont {Kissel}}, \bibinfo {author} {\bibfnamefont {K.}~\bibnamefont {Komori}}, \bibinfo {author} {\bibfnamefont {A.}~\bibnamefont {Kontos}}, \bibinfo {author} {\bibfnamefont {R.}~\bibnamefont {Kumar}}, \bibinfo {author} {\bibfnamefont {K.}~\bibnamefont {Kuns}}, \bibinfo {author} {\bibfnamefont {M.}~\bibnamefont {Landry}}, \bibinfo {author} {\bibfnamefont {B.}~\bibnamefont {Lantz}}, \bibinfo {author} {\bibfnamefont {M.}~\bibnamefont {Laxen}}, \bibinfo {author} {\bibfnamefont {K.}~\bibnamefont {Lee}}, \bibinfo {author} {\bibfnamefont
  {M.}~\bibnamefont {Lesovsky}}, \bibinfo {author} {\bibfnamefont {F.}~\bibnamefont {Llamas}}, \bibinfo {author} {\bibfnamefont {M.}~\bibnamefont {Lormand}}, \bibinfo {author} {\bibfnamefont {H.~A.}\ \bibnamefont {Loughlin}}, \bibinfo {author} {\bibfnamefont {R.}~\bibnamefont {Macas}}, \bibinfo {author} {\bibfnamefont {M.}~\bibnamefont {MacInnis}}, \bibinfo {author} {\bibfnamefont {C.~N.}\ \bibnamefont {Makarem}}, \bibinfo {author} {\bibfnamefont {B.}~\bibnamefont {Mannix}}, \bibinfo {author} {\bibfnamefont {G.~L.}\ \bibnamefont {Mansell}}, \bibinfo {author} {\bibfnamefont {R.~M.}\ \bibnamefont {Martin}}, \bibinfo {author} {\bibfnamefont {K.}~\bibnamefont {Mason}}, \bibinfo {author} {\bibfnamefont {F.}~\bibnamefont {Matichard}}, \bibinfo {author} {\bibfnamefont {N.}~\bibnamefont {Mavalvala}}, \bibinfo {author} {\bibfnamefont {N.}~\bibnamefont {Maxwell}}, \bibinfo {author} {\bibfnamefont {G.}~\bibnamefont {McCarrol}}, \bibinfo {author} {\bibfnamefont {R.}~\bibnamefont {McCarthy}}, \bibinfo {author}
  {\bibfnamefont {D.~E.}\ \bibnamefont {McClelland}}, \bibinfo {author} {\bibfnamefont {S.}~\bibnamefont {McCormick}}, \bibinfo {author} {\bibfnamefont {T.}~\bibnamefont {McRae}}, \bibinfo {author} {\bibfnamefont {F.}~\bibnamefont {Mera}}, \bibinfo {author} {\bibfnamefont {E.~L.}\ \bibnamefont {Merilh}}, \bibinfo {author} {\bibfnamefont {F.}~\bibnamefont {Meylahn}}, \bibinfo {author} {\bibfnamefont {R.}~\bibnamefont {Mittleman}}, \bibinfo {author} {\bibfnamefont {D.}~\bibnamefont {Moraru}}, \bibinfo {author} {\bibfnamefont {G.}~\bibnamefont {Moreno}}, \bibinfo {author} {\bibfnamefont {T.~J.~N.}\ \bibnamefont {Nelson}}, \bibinfo {author} {\bibfnamefont {A.}~\bibnamefont {Neunzert}}, \bibinfo {author} {\bibfnamefont {J.}~\bibnamefont {Notte}}, \bibinfo {author} {\bibfnamefont {J.}~\bibnamefont {Oberling}}, \bibinfo {author} {\bibfnamefont {T.}~\bibnamefont {O'Hanlon}}, \bibinfo {author} {\bibfnamefont {C.}~\bibnamefont {Osthelder}}, \bibinfo {author} {\bibfnamefont {D.~J.}\ \bibnamefont {Ottaway}}, \bibinfo
  {author} {\bibfnamefont {H.}~\bibnamefont {Overmier}}, \bibinfo {author} {\bibfnamefont {W.}~\bibnamefont {Parker}}, \bibinfo {author} {\bibfnamefont {A.}~\bibnamefont {Pele}}, \bibinfo {author} {\bibfnamefont {H.}~\bibnamefont {Pham}}, \bibinfo {author} {\bibfnamefont {M.}~\bibnamefont {Pirello}}, \bibinfo {author} {\bibfnamefont {V.}~\bibnamefont {Quetschke}}, \bibinfo {author} {\bibfnamefont {K.~E.}\ \bibnamefont {Ramirez}}, \bibinfo {author} {\bibfnamefont {J.}~\bibnamefont {Reyes}}, \bibinfo {author} {\bibfnamefont {J.~W.}\ \bibnamefont {Richardson}}, \bibinfo {author} {\bibfnamefont {M.}~\bibnamefont {Robinson}}, \bibinfo {author} {\bibfnamefont {J.~G.}\ \bibnamefont {Rollins}}, \bibinfo {author} {\bibfnamefont {C.~L.}\ \bibnamefont {Romel}}, \bibinfo {author} {\bibfnamefont {J.~H.}\ \bibnamefont {Romie}}, \bibinfo {author} {\bibfnamefont {M.~P.}\ \bibnamefont {Ross}}, \bibinfo {author} {\bibfnamefont {K.}~\bibnamefont {Ryan}}, \bibinfo {author} {\bibfnamefont {T.}~\bibnamefont {Sadecki}}, \bibinfo
  {author} {\bibfnamefont {A.}~\bibnamefont {Sanchez}}, \bibinfo {author} {\bibfnamefont {E.~J.}\ \bibnamefont {Sanchez}}, \bibinfo {author} {\bibfnamefont {L.~E.}\ \bibnamefont {Sanchez}}, \bibinfo {author} {\bibfnamefont {R.~L.}\ \bibnamefont {Savage}}, \bibinfo {author} {\bibfnamefont {D.}~\bibnamefont {Schaetzl}}, \bibinfo {author} {\bibfnamefont {M.~G.}\ \bibnamefont {Schiworski}}, \bibinfo {author} {\bibfnamefont {R.}~\bibnamefont {Schnabel}}, \bibinfo {author} {\bibfnamefont {R.~M.~S.}\ \bibnamefont {Schofield}}, \bibinfo {author} {\bibfnamefont {E.}~\bibnamefont {Schwartz}}, \bibinfo {author} {\bibfnamefont {D.}~\bibnamefont {Sellers}}, \bibinfo {author} {\bibfnamefont {T.}~\bibnamefont {Shaffer}}, \bibinfo {author} {\bibfnamefont {R.~W.}\ \bibnamefont {Short}}, \bibinfo {author} {\bibfnamefont {D.}~\bibnamefont {Sigg}}, \bibinfo {author} {\bibfnamefont {B.~J.~J.}\ \bibnamefont {Slagmolen}}, \bibinfo {author} {\bibfnamefont {C.}~\bibnamefont {Soike}}, \bibinfo {author} {\bibfnamefont {S.}~\bibnamefont
  {Soni}}, \bibinfo {author} {\bibfnamefont {V.}~\bibnamefont {Srivastava}}, \bibinfo {author} {\bibfnamefont {L.}~\bibnamefont {Sun}}, \bibinfo {author} {\bibfnamefont {D.~B.}\ \bibnamefont {Tanner}}, \bibinfo {author} {\bibfnamefont {M.}~\bibnamefont {Thomas}}, \bibinfo {author} {\bibfnamefont {P.}~\bibnamefont {Thomas}}, \bibinfo {author} {\bibfnamefont {K.~A.}\ \bibnamefont {Thorne}}, \bibinfo {author} {\bibfnamefont {C.~I.}\ \bibnamefont {Torrie}}, \bibinfo {author} {\bibfnamefont {G.}~\bibnamefont {Traylor}}, \bibinfo {author} {\bibfnamefont {A.~S.}\ \bibnamefont {Ubhi}}, \bibinfo {author} {\bibfnamefont {G.}~\bibnamefont {Vajente}}, \bibinfo {author} {\bibfnamefont {J.}~\bibnamefont {Vanosky}}, \bibinfo {author} {\bibfnamefont {A.}~\bibnamefont {Vecchio}}, \bibinfo {author} {\bibfnamefont {P.~J.}\ \bibnamefont {Veitch}}, \bibinfo {author} {\bibfnamefont {A.~M.}\ \bibnamefont {Vibhute}}, \bibinfo {author} {\bibfnamefont {E.~R.~G.}\ \bibnamefont {von Reis}}, \bibinfo {author} {\bibfnamefont
  {J.}~\bibnamefont {Warner}}, \bibinfo {author} {\bibfnamefont {B.}~\bibnamefont {Weaver}}, \bibinfo {author} {\bibfnamefont {R.}~\bibnamefont {Weiss}}, \bibinfo {author} {\bibfnamefont {C.}~\bibnamefont {Whittle}}, \bibinfo {author} {\bibfnamefont {B.}~\bibnamefont {Willke}}, \bibinfo {author} {\bibfnamefont {C.~C.}\ \bibnamefont {Wipf}}, \bibinfo {author} {\bibfnamefont {H.}~\bibnamefont {Yamamoto}}, \bibinfo {author} {\bibfnamefont {L.}~\bibnamefont {Zhang}},\ and\ \bibinfo {author} {\bibfnamefont {M.~E.}\ \bibnamefont {Zucker}} (\bibinfo {collaboration} {LIGO O4 Detector Collaboration}),\ }\bibfield  {title} {\bibinfo {title} {Broadband quantum enhancement of the ligo detectors with frequency-dependent squeezing},\ }\href {https://doi.org/10.1103/PhysRevX.13.041021} {\bibfield  {journal} {\bibinfo  {journal} {Phys. Rev. X}\ }\textbf {\bibinfo {volume} {13}},\ \bibinfo {pages} {041021} (\bibinfo {year} {2023})}\BibitemShut {NoStop}%
\bibitem [{\citenamefont {Acernese}\ \emph {et~al.}(2023)\citenamefont {Acernese}, \citenamefont {Agathos}, \citenamefont {Ain}, \citenamefont {Albanesi}, \citenamefont {All\'en\'e}, \citenamefont {Allocca}, \citenamefont {Amato}, \citenamefont {Amra}, \citenamefont {Andia}, \citenamefont {Andrade}, \citenamefont {Andres}, \citenamefont {Andr\'es-Carcasona}, \citenamefont {Andri\ifmmode~\acute{c}\else \'{c}\fi{}}, \citenamefont {Ansoldi}, \citenamefont {Antier}, \citenamefont {Apostolatos}, \citenamefont {Appavuravther}, \citenamefont {Ar\`ene}, \citenamefont {Arnaud}, \citenamefont {Assiduo}, \citenamefont {Melo}, \citenamefont {Astone}, \citenamefont {Aubin}, \citenamefont {Babak}, \citenamefont {Badaracco}, \citenamefont {Bagnasco}, \citenamefont {Baird}, \citenamefont {Baka}, \citenamefont {Ballardin}, \citenamefont {Baltus}, \citenamefont {Banerjee}, \citenamefont {Barneo}, \citenamefont {Barone}, \citenamefont {Barsuglia}, \citenamefont {Barta}, \citenamefont {Basti}, \citenamefont {Bawaj},
  \citenamefont {Bazzan}, \citenamefont {Beirnaert}, \citenamefont {Bejger}, \citenamefont {Benedetto}, \citenamefont {Berbel}, \citenamefont {Bernuzzi}, \citenamefont {Bersanetti}, \citenamefont {Bertolini}, \citenamefont {Bhardwaj}, \citenamefont {Bianchi}, \citenamefont {Bilicki}, \citenamefont {Bini}, \citenamefont {Bischi}, \citenamefont {Bitossi}, \citenamefont {Bizouard}, \citenamefont {Bobba}, \citenamefont {Bo\"er}, \citenamefont {Bogaert}, \citenamefont {Boileau}, \citenamefont {Boldrini}, \citenamefont {Bonavena}, \citenamefont {Bondarescu}, \citenamefont {Bondu}, \citenamefont {Bonnand}, \citenamefont {Boschi}, \citenamefont {Boudart}, \citenamefont {Bouffanais}, \citenamefont {Bozzi}, \citenamefont {Bradaschia}, \citenamefont {Braglia}, \citenamefont {Branchesi}, \citenamefont {Breschi}, \citenamefont {Briant}, \citenamefont {Brillet}, \citenamefont {Brooks}, \citenamefont {Bruno}, \citenamefont {Bucci}, \citenamefont {Bulashenko}, \citenamefont {Bulik}, \citenamefont {Bulten}, \citenamefont
  {Buscicchio}, \citenamefont {Buskulic}, \citenamefont {Buy}, \citenamefont {Cabras}, \citenamefont {Cabrita}, \citenamefont {Cagnoli}, \citenamefont {Calloni}, \citenamefont {Canepa}, \citenamefont {Santoro}, \citenamefont {Cannavacciuolo}, \citenamefont {Capocasa}, \citenamefont {Carapella}, \citenamefont {Carbognani}, \citenamefont {Carpinelli}, \citenamefont {Carullo}, \citenamefont {Diaz}, \citenamefont {Casentini}, \citenamefont {Caudill}, \citenamefont {Cavalieri}, \citenamefont {Cella}, \citenamefont {Cerd\'a-Dur\'an}, \citenamefont {Cesarini}, \citenamefont {Chaibi}, \citenamefont {Chanial}, \citenamefont {Chassande-Mottin}, \citenamefont {Chaty}, \citenamefont {Chessa}, \citenamefont {Chiadini}, \citenamefont {Chiarini}, \citenamefont {Chierici}, \citenamefont {Chincarini}, \citenamefont {Chiofalo}, \citenamefont {Chiummo}, \citenamefont {Christensen}, \citenamefont {Chua}, \citenamefont {Ciani}, \citenamefont {Ciecielag}, \citenamefont {Cie\ifmmode~\acute{s}\else \'{s}\fi{}lar}, \citenamefont
  {Cifaldi}, \citenamefont {Ciolfi}, \citenamefont {Clesse}, \citenamefont {Cleva}, \citenamefont {Coccia}, \citenamefont {Codazzo}, \citenamefont {Cohadon}, \citenamefont {Colombo}, \citenamefont {Colpi}, \citenamefont {Conti}, \citenamefont {Cordero-Carri\'on}, \citenamefont {Corezzi}, \citenamefont {Cortese}, \citenamefont {Coulon}, \citenamefont {Coupechoux}, \citenamefont {Croquette}, \citenamefont {Cudell}, \citenamefont {Cuoco}, \citenamefont {Cury\l{}o}, \citenamefont {Dabadie}, \citenamefont {Canton}, \citenamefont {Dall'Osso}, \citenamefont {D\'alya}, \citenamefont {D'Angelo}, \citenamefont {Dangoisse}, \citenamefont {Danilishin}, \citenamefont {D'Antonio}, \citenamefont {Dattilo}, \citenamefont {Davier}, \citenamefont {Degallaix}, \citenamefont {De~Laurentis}, \citenamefont {Del\'eglise}, \citenamefont {De~Lillo}, \citenamefont {Dell'Aquila}, \citenamefont {Del~Pozzo}, \citenamefont {De~Matteis}, \citenamefont {Depasse}, \citenamefont {De~Pietri}, \citenamefont {De~Rosa}, \citenamefont {De~Rossi},
  \citenamefont {De~Simone}, \citenamefont {Di~Fiore}, \citenamefont {Di~Giorgio}, \citenamefont {Di~Giovanni}, \citenamefont {Di~Giovanni}, \citenamefont {Di~Girolamo}, \citenamefont {Diksha}, \citenamefont {Di~Lieto}, \citenamefont {Di~Michele}, \citenamefont {Ding}, \citenamefont {Di~Pace}, \citenamefont {Di~Palma}, \citenamefont {Di~Renzo}, \citenamefont {D'Onofrio}, \citenamefont {Dooney}, \citenamefont {Dorosh}, \citenamefont {Drago}, \citenamefont {Ducoin}, \citenamefont {Dupletsa}, \citenamefont {Durante}, \citenamefont {D'Urso}, \citenamefont {Duverne}, \citenamefont {Eisenmann}, \citenamefont {Errico}, \citenamefont {Estevez}, \citenamefont {Fabrizi}, \citenamefont {Faedi}, \citenamefont {Fafone}, \citenamefont {Favaro}, \citenamefont {Fays}, \citenamefont {Fenyvesi}, \citenamefont {Ferrante}, \citenamefont {Fidecaro}, \citenamefont {Figura}, \citenamefont {Fiori}, \citenamefont {Fiori}, \citenamefont {Fittipaldi}, \citenamefont {Fiumara}, \citenamefont {Flaminio}, \citenamefont {Font},
  \citenamefont {Frasca}, \citenamefont {Frasconi}, \citenamefont {Freise}, \citenamefont {Freitas}, \citenamefont {Fronz\'e}, \citenamefont {Gadre}, \citenamefont {Gamba}, \citenamefont {Garaventa}, \citenamefont {Garcia-Bellido}, \citenamefont {Gargiulo}, \citenamefont {Garufi}, \citenamefont {Gasbarra}, \citenamefont {Gemme}, \citenamefont {Gennai}, \citenamefont {Ghosh}, \citenamefont {Giacoppo}, \citenamefont {Giri}, \citenamefont {Gissi}, \citenamefont {Gkaitatzis}, \citenamefont {Glotin}, \citenamefont {Goncharov}, \citenamefont {Gosselin}, \citenamefont {Gouaty}, \citenamefont {Grado}, \citenamefont {Granata}, \citenamefont {Granata}, \citenamefont {Greco}, \citenamefont {Grignani}, \citenamefont {Grimaldi}, \citenamefont {Guerra}, \citenamefont {Guetta}, \citenamefont {Guidi}, \citenamefont {Gulminelli}, \citenamefont {Guo}, \citenamefont {Gupta}, \citenamefont {Gutierrez}, \citenamefont {Haegel}, \citenamefont {Halim}, \citenamefont {Hannuksela}, \citenamefont {Harder}, \citenamefont {Haris},
  \citenamefont {Harmark}, \citenamefont {Harms}, \citenamefont {Haskell}, \citenamefont {Heidmann}, \citenamefont {Heitmann}, \citenamefont {Hello}, \citenamefont {Hemming}, \citenamefont {Hennes}, \citenamefont {Hennig}, \citenamefont {Hennig}, \citenamefont {Hild}, \citenamefont {Hofman}, \citenamefont {Holland}, \citenamefont {Hui}, \citenamefont {Iandolo}, \citenamefont {Idzkowski}, \citenamefont {Iess}, \citenamefont {Iorio}, \citenamefont {Iosif}, \citenamefont {Jacqmin}, \citenamefont {Jacquet}, \citenamefont {Janquart}, \citenamefont {Janssens}, \citenamefont {Jaraba}, \citenamefont {Jaranowski}, \citenamefont {Jasal}, \citenamefont {Juste}, \citenamefont {Kalaghatgi}, \citenamefont {Karathanasis}, \citenamefont {Katsanevas}, \citenamefont {K\'ef\'elian}, \citenamefont {Koekoek}, \citenamefont {Koley}, \citenamefont {Kolstein}, \citenamefont {Kranzhoff}, \citenamefont {Kr\'olak}, \citenamefont {Kuijer}, \citenamefont {Kuroyanagi}, \citenamefont {Lagabbe}, \citenamefont {Laghi}, \citenamefont
  {Lalleman}, \citenamefont {Lamberts}, \citenamefont {La~Rana}, \citenamefont {La~Rosa}, \citenamefont {Lartaux-Vollard}, \citenamefont {Lazzaro}, \citenamefont {Leaci}, \citenamefont {Lema\^{\i}tre}, \citenamefont {Lenti}, \citenamefont {Leonova}, \citenamefont {Lequime}, \citenamefont {Leroy}, \citenamefont {Letendre}, \citenamefont {Lethuillier}, \citenamefont {Leyde}, \citenamefont {Linde}, \citenamefont {London}, \citenamefont {Longo}, \citenamefont {Portilla}, \citenamefont {Lorenzini}, \citenamefont {Loriette}, \citenamefont {Losurdo}, \citenamefont {Lumaca}, \citenamefont {Macquet}, \citenamefont {Magazz\`u}, \citenamefont {Maggiore}, \citenamefont {Magnozzi}, \citenamefont {Majorana}, \citenamefont {Man}, \citenamefont {Mangano}, \citenamefont {Mantovani}, \citenamefont {Mapelli}, \citenamefont {Marchesoni}, \citenamefont {Pina}, \citenamefont {Marion}, \citenamefont {Marquina}, \citenamefont {Marsat}, \citenamefont {Martelli}, \citenamefont {Martinez}, \citenamefont {Martinez}, \citenamefont
  {Masserot}, \citenamefont {Mastrodicasa}, \citenamefont {Mastrogiovanni}, \citenamefont {Meijer}, \citenamefont {Menendez-Vazquez}, \citenamefont {Mereni}, \citenamefont {Merzougui}, \citenamefont {Miani}, \citenamefont {Michel}, \citenamefont {Miller}, \citenamefont {Miller}, \citenamefont {Milotti}, \citenamefont {Minenkov}, \citenamefont {Mir}, \citenamefont {Miravet-Ten\'es}, \citenamefont {Mitchell}, \citenamefont {Mondal}, \citenamefont {Montani}, \citenamefont {Morawski}, \citenamefont {Morras}, \citenamefont {Moscatello}, \citenamefont {Mours}, \citenamefont {Mow-Lowry}, \citenamefont {Msihid}, \citenamefont {Muciaccia}, \citenamefont {Mukherjee}, \citenamefont {Nagar}, \citenamefont {Napolano}, \citenamefont {Nardecchia}, \citenamefont {Narola}, \citenamefont {Naticchioni}, \citenamefont {Neilson}, \citenamefont {Nesseris}, \citenamefont {Nguyen}, \citenamefont {Nieradka}, \citenamefont {Nissanke}, \citenamefont {Nitoglia}, \citenamefont {Nocera}, \citenamefont {Novak}, \citenamefont {no~Siles},
  \citenamefont {Oertel}, \citenamefont {Oganesyan}, \citenamefont {Oliveri}, \citenamefont {Orselli}, \citenamefont {Palomba}, \citenamefont {Pang}, \citenamefont {Pannarale}, \citenamefont {Paoletti}, \citenamefont {Paoli}, \citenamefont {Paolone}, \citenamefont {Pappas}, \citenamefont {Parisi}, \citenamefont {Pascucci}, \citenamefont {Pasqualetti}, \citenamefont {Passaquieti}, \citenamefont {Passuello}, \citenamefont {Patricelli}, \citenamefont {Pedurand}, \citenamefont {Pegna}, \citenamefont {Pegoraro}, \citenamefont {Perego}, \citenamefont {Pereira}, \citenamefont {P\'erigois}, \citenamefont {Perreca}, \citenamefont {Perri\`es}, \citenamefont {Perry}, \citenamefont {Pesios}, \citenamefont {Petrillo}, \citenamefont {Phukon}, \citenamefont {Piccinni}, \citenamefont {Pichot}, \citenamefont {Piendibene}, \citenamefont {Piergiovanni}, \citenamefont {Pierini}, \citenamefont {Pierra}, \citenamefont {Pierro}, \citenamefont {Pillant}, \citenamefont {Pillas}, \citenamefont {Pilo}, \citenamefont {Pinard},
  \citenamefont {Pinto}, \citenamefont {Pinto}, \citenamefont {Pinto}, \citenamefont {Piotrzkowski}, \citenamefont {Placidi}, \citenamefont {Placidi}, \citenamefont {Plastino}, \citenamefont {Poggiani}, \citenamefont {Polini}, \citenamefont {Porcelli}, \citenamefont {Portell}, \citenamefont {Porter}, \citenamefont {Poulton}, \citenamefont {Pracchia}, \citenamefont {Pradier}, \citenamefont {Principe}, \citenamefont {Prodi}, \citenamefont {Prosposito}, \citenamefont {Puecher}, \citenamefont {Punturo}, \citenamefont {Puosi}, \citenamefont {Puppo}, \citenamefont {Raaijmakers}, \citenamefont {Radulesco}, \citenamefont {Rapagnani}, \citenamefont {Razzano}, \citenamefont {Regimbau}, \citenamefont {Rei}, \citenamefont {Rettegno}, \citenamefont {Revenu}, \citenamefont {Reza}, \citenamefont {Rezaei}, \citenamefont {Ricci}, \citenamefont {Rinaldi}, \citenamefont {Robinet}, \citenamefont {Rocchi}, \citenamefont {Rolland}, \citenamefont {Romanelli}, \citenamefont {Romano}, \citenamefont {Romero}, \citenamefont {Ronchini},
  \citenamefont {Rosa}, \citenamefont {Rosi\ifmmode~\acute{n}\else \'{n}\fi{}ska}, \citenamefont {Roy}, \citenamefont {Rozza}, \citenamefont {Ruggi}, \citenamefont {Morales}, \citenamefont {Saffarieh}, \citenamefont {Salafia}, \citenamefont {Salconi}, \citenamefont {Salemi}, \citenamefont {Sall\'e}, \citenamefont {Samajdar}, \citenamefont {Sanchis-Gual}, \citenamefont {Sanuy}, \citenamefont {Sasli}, \citenamefont {Sassi}, \citenamefont {Sassolas}, \citenamefont {Sayah}, \citenamefont {Schmidt}, \citenamefont {Seglar-Arroyo}, \citenamefont {Sentenac}, \citenamefont {Sequino}, \citenamefont {Servignat}, \citenamefont {Setyawati}, \citenamefont {Shcheblanov}, \citenamefont {Sieniawska}, \citenamefont {Silenzi}, \citenamefont {Singh}, \citenamefont {Singha}, \citenamefont {Sipala}, \citenamefont {Soldateschi}, \citenamefont {Sordini}, \citenamefont {Sorrentino}, \citenamefont {Sorrentino}, \citenamefont {Soulard}, \citenamefont {Spagnuolo}, \citenamefont {Spera}, \citenamefont {Spinicelli}, \citenamefont
  {Stachie}, \citenamefont {Steer}, \citenamefont {Steinlechner}, \citenamefont {Steinlechner}, \citenamefont {Stergioulas}, \citenamefont {Stratta}, \citenamefont {Suchenek}, \citenamefont {Sur}, \citenamefont {Suresh}, \citenamefont {Swinkels}, \citenamefont {Syx}, \citenamefont {Szewczyk}, \citenamefont {Tacca}, \citenamefont {Tamanini}, \citenamefont {Tanasijczuk}, \citenamefont {Mart\'{\i}n}, \citenamefont {Taranto}, \citenamefont {Tonelli}, \citenamefont {Torres-Forn\'e}, \citenamefont {e~Melo}, \citenamefont {Tournefier}, \citenamefont {Trapananti}, \citenamefont {Travasso}, \citenamefont {Trenado}, \citenamefont {Tringali}, \citenamefont {Troiano}, \citenamefont {Trovato}, \citenamefont {Trozzo}, \citenamefont {Tsang}, \citenamefont {Turbang}, \citenamefont {Turconi}, \citenamefont {Turski}, \citenamefont {Ubach}, \citenamefont {Utina}, \citenamefont {Valentini}, \citenamefont {Vallero}, \citenamefont {van Bakel}, \citenamefont {van Beuzekom}, \citenamefont {van Dael}, \citenamefont {van~den Brand},
  \citenamefont {Van Den~Broeck}, \citenamefont {van~der Sluys}, \citenamefont {Van~de Walle}, \citenamefont {van Dongen}, \citenamefont {van Haevermaet}, \citenamefont {van Heijningen}, \citenamefont {van Ranst}, \citenamefont {van Remortel}, \citenamefont {Vardaro}, \citenamefont {Vas\'uth}, \citenamefont {Vedovato}, \citenamefont {Verdier}, \citenamefont {Verkindt}, \citenamefont {Verma}, \citenamefont {Vetrano}, \citenamefont {Vicer\'e}, \citenamefont {Vinet}, \citenamefont {Viret}, \citenamefont {Virtuoso}, \citenamefont {Vocca}, \citenamefont {Walet}, \citenamefont {Was}, \citenamefont {Yadav}, \citenamefont {Zadro\ifmmode~\dot{z}\else \.{z}\fi{}ny}, \citenamefont {Zelenova}, \citenamefont {Zendri}, \citenamefont {Zhao}, \citenamefont {Zerrad}, \citenamefont {Vahlbruch}, \citenamefont {Mehmet}, \citenamefont {L\"uck},\ and\ \citenamefont {Danzmann}}]{PhysRevLett.131.041403}%
  \BibitemOpen
  \bibfield  {author} {\bibinfo {author} {\bibfnamefont {F.}~\bibnamefont {Acernese}}, \bibinfo {author} {\bibfnamefont {M.}~\bibnamefont {Agathos}}, \bibinfo {author} {\bibfnamefont {A.}~\bibnamefont {Ain}}, \bibinfo {author} {\bibfnamefont {S.}~\bibnamefont {Albanesi}}, \bibinfo {author} {\bibfnamefont {C.}~\bibnamefont {All\'en\'e}}, \bibinfo {author} {\bibfnamefont {A.}~\bibnamefont {Allocca}}, \bibinfo {author} {\bibfnamefont {A.}~\bibnamefont {Amato}}, \bibinfo {author} {\bibfnamefont {C.}~\bibnamefont {Amra}}, \bibinfo {author} {\bibfnamefont {M.}~\bibnamefont {Andia}}, \bibinfo {author} {\bibfnamefont {T.}~\bibnamefont {Andrade}}, \bibinfo {author} {\bibfnamefont {N.}~\bibnamefont {Andres}}, \bibinfo {author} {\bibfnamefont {M.}~\bibnamefont {Andr\'es-Carcasona}}, \bibinfo {author} {\bibfnamefont {T.}~\bibnamefont {Andri\ifmmode~\acute{c}\else \'{c}\fi{}}}, \bibinfo {author} {\bibfnamefont {S.}~\bibnamefont {Ansoldi}}, \bibinfo {author} {\bibfnamefont {S.}~\bibnamefont {Antier}}, \bibinfo {author}
  {\bibfnamefont {T.}~\bibnamefont {Apostolatos}}, \bibinfo {author} {\bibfnamefont {E.~Z.}\ \bibnamefont {Appavuravther}}, \bibinfo {author} {\bibfnamefont {M.}~\bibnamefont {Ar\`ene}}, \bibinfo {author} {\bibfnamefont {N.}~\bibnamefont {Arnaud}}, \bibinfo {author} {\bibfnamefont {M.}~\bibnamefont {Assiduo}}, \bibinfo {author} {\bibfnamefont {S.~A. d.~S.}\ \bibnamefont {Melo}}, \bibinfo {author} {\bibfnamefont {P.}~\bibnamefont {Astone}}, \bibinfo {author} {\bibfnamefont {F.}~\bibnamefont {Aubin}}, \bibinfo {author} {\bibfnamefont {S.}~\bibnamefont {Babak}}, \bibinfo {author} {\bibfnamefont {F.}~\bibnamefont {Badaracco}}, \bibinfo {author} {\bibfnamefont {S.}~\bibnamefont {Bagnasco}}, \bibinfo {author} {\bibfnamefont {J.}~\bibnamefont {Baird}}, \bibinfo {author} {\bibfnamefont {T.}~\bibnamefont {Baka}}, \bibinfo {author} {\bibfnamefont {G.}~\bibnamefont {Ballardin}}, \bibinfo {author} {\bibfnamefont {G.}~\bibnamefont {Baltus}}, \bibinfo {author} {\bibfnamefont {B.}~\bibnamefont {Banerjee}}, \bibinfo {author}
  {\bibfnamefont {P.}~\bibnamefont {Barneo}}, \bibinfo {author} {\bibfnamefont {F.}~\bibnamefont {Barone}}, \bibinfo {author} {\bibfnamefont {M.}~\bibnamefont {Barsuglia}}, \bibinfo {author} {\bibfnamefont {D.}~\bibnamefont {Barta}}, \bibinfo {author} {\bibfnamefont {A.}~\bibnamefont {Basti}}, \bibinfo {author} {\bibfnamefont {M.}~\bibnamefont {Bawaj}}, \bibinfo {author} {\bibfnamefont {M.}~\bibnamefont {Bazzan}}, \bibinfo {author} {\bibfnamefont {F.}~\bibnamefont {Beirnaert}}, \bibinfo {author} {\bibfnamefont {M.}~\bibnamefont {Bejger}}, \bibinfo {author} {\bibfnamefont {V.}~\bibnamefont {Benedetto}}, \bibinfo {author} {\bibfnamefont {M.}~\bibnamefont {Berbel}}, \bibinfo {author} {\bibfnamefont {S.}~\bibnamefont {Bernuzzi}}, \bibinfo {author} {\bibfnamefont {D.}~\bibnamefont {Bersanetti}}, \bibinfo {author} {\bibfnamefont {A.}~\bibnamefont {Bertolini}}, \bibinfo {author} {\bibfnamefont {U.}~\bibnamefont {Bhardwaj}}, \bibinfo {author} {\bibfnamefont {A.}~\bibnamefont {Bianchi}}, \bibinfo {author}
  {\bibfnamefont {M.}~\bibnamefont {Bilicki}}, \bibinfo {author} {\bibfnamefont {S.}~\bibnamefont {Bini}}, \bibinfo {author} {\bibfnamefont {M.}~\bibnamefont {Bischi}}, \bibinfo {author} {\bibfnamefont {M.}~\bibnamefont {Bitossi}}, \bibinfo {author} {\bibfnamefont {M.-A.}\ \bibnamefont {Bizouard}}, \bibinfo {author} {\bibfnamefont {F.}~\bibnamefont {Bobba}}, \bibinfo {author} {\bibfnamefont {M.}~\bibnamefont {Bo\"er}}, \bibinfo {author} {\bibfnamefont {G.}~\bibnamefont {Bogaert}}, \bibinfo {author} {\bibfnamefont {G.}~\bibnamefont {Boileau}}, \bibinfo {author} {\bibfnamefont {M.}~\bibnamefont {Boldrini}}, \bibinfo {author} {\bibfnamefont {L.~D.}\ \bibnamefont {Bonavena}}, \bibinfo {author} {\bibfnamefont {R.}~\bibnamefont {Bondarescu}}, \bibinfo {author} {\bibfnamefont {F.}~\bibnamefont {Bondu}}, \bibinfo {author} {\bibfnamefont {R.}~\bibnamefont {Bonnand}}, \bibinfo {author} {\bibfnamefont {V.}~\bibnamefont {Boschi}}, \bibinfo {author} {\bibfnamefont {V.}~\bibnamefont {Boudart}}, \bibinfo {author}
  {\bibfnamefont {Y.}~\bibnamefont {Bouffanais}}, \bibinfo {author} {\bibfnamefont {A.}~\bibnamefont {Bozzi}}, \bibinfo {author} {\bibfnamefont {C.}~\bibnamefont {Bradaschia}}, \bibinfo {author} {\bibfnamefont {M.}~\bibnamefont {Braglia}}, \bibinfo {author} {\bibfnamefont {M.}~\bibnamefont {Branchesi}}, \bibinfo {author} {\bibfnamefont {M.}~\bibnamefont {Breschi}}, \bibinfo {author} {\bibfnamefont {T.}~\bibnamefont {Briant}}, \bibinfo {author} {\bibfnamefont {A.}~\bibnamefont {Brillet}}, \bibinfo {author} {\bibfnamefont {J.}~\bibnamefont {Brooks}}, \bibinfo {author} {\bibfnamefont {G.}~\bibnamefont {Bruno}}, \bibinfo {author} {\bibfnamefont {F.}~\bibnamefont {Bucci}}, \bibinfo {author} {\bibfnamefont {O.}~\bibnamefont {Bulashenko}}, \bibinfo {author} {\bibfnamefont {T.}~\bibnamefont {Bulik}}, \bibinfo {author} {\bibfnamefont {H.~J.}\ \bibnamefont {Bulten}}, \bibinfo {author} {\bibfnamefont {R.}~\bibnamefont {Buscicchio}}, \bibinfo {author} {\bibfnamefont {D.}~\bibnamefont {Buskulic}}, \bibinfo {author}
  {\bibfnamefont {C.}~\bibnamefont {Buy}}, \bibinfo {author} {\bibfnamefont {G.}~\bibnamefont {Cabras}}, \bibinfo {author} {\bibfnamefont {R.}~\bibnamefont {Cabrita}}, \bibinfo {author} {\bibfnamefont {G.}~\bibnamefont {Cagnoli}}, \bibinfo {author} {\bibfnamefont {E.}~\bibnamefont {Calloni}}, \bibinfo {author} {\bibfnamefont {M.}~\bibnamefont {Canepa}}, \bibinfo {author} {\bibfnamefont {G.~C.}\ \bibnamefont {Santoro}}, \bibinfo {author} {\bibfnamefont {M.}~\bibnamefont {Cannavacciuolo}}, \bibinfo {author} {\bibfnamefont {E.}~\bibnamefont {Capocasa}}, \bibinfo {author} {\bibfnamefont {G.}~\bibnamefont {Carapella}}, \bibinfo {author} {\bibfnamefont {F.}~\bibnamefont {Carbognani}}, \bibinfo {author} {\bibfnamefont {M.}~\bibnamefont {Carpinelli}}, \bibinfo {author} {\bibfnamefont {G.}~\bibnamefont {Carullo}}, \bibinfo {author} {\bibfnamefont {J.~C.}\ \bibnamefont {Diaz}}, \bibinfo {author} {\bibfnamefont {C.}~\bibnamefont {Casentini}}, \bibinfo {author} {\bibfnamefont {S.}~\bibnamefont {Caudill}}, \bibinfo
  {author} {\bibfnamefont {R.}~\bibnamefont {Cavalieri}}, \bibinfo {author} {\bibfnamefont {G.}~\bibnamefont {Cella}}, \bibinfo {author} {\bibfnamefont {P.}~\bibnamefont {Cerd\'a-Dur\'an}}, \bibinfo {author} {\bibfnamefont {E.}~\bibnamefont {Cesarini}}, \bibinfo {author} {\bibfnamefont {W.}~\bibnamefont {Chaibi}}, \bibinfo {author} {\bibfnamefont {P.}~\bibnamefont {Chanial}}, \bibinfo {author} {\bibfnamefont {E.}~\bibnamefont {Chassande-Mottin}}, \bibinfo {author} {\bibfnamefont {S.}~\bibnamefont {Chaty}}, \bibinfo {author} {\bibfnamefont {P.}~\bibnamefont {Chessa}}, \bibinfo {author} {\bibfnamefont {F.}~\bibnamefont {Chiadini}}, \bibinfo {author} {\bibfnamefont {G.}~\bibnamefont {Chiarini}}, \bibinfo {author} {\bibfnamefont {R.}~\bibnamefont {Chierici}}, \bibinfo {author} {\bibfnamefont {A.}~\bibnamefont {Chincarini}}, \bibinfo {author} {\bibfnamefont {M.~L.}\ \bibnamefont {Chiofalo}}, \bibinfo {author} {\bibfnamefont {A.}~\bibnamefont {Chiummo}}, \bibinfo {author} {\bibfnamefont {N.}~\bibnamefont
  {Christensen}}, \bibinfo {author} {\bibfnamefont {S.}~\bibnamefont {Chua}}, \bibinfo {author} {\bibfnamefont {G.}~\bibnamefont {Ciani}}, \bibinfo {author} {\bibfnamefont {P.}~\bibnamefont {Ciecielag}}, \bibinfo {author} {\bibfnamefont {M.}~\bibnamefont {Cie\ifmmode~\acute{s}\else \'{s}\fi{}lar}}, \bibinfo {author} {\bibfnamefont {M.}~\bibnamefont {Cifaldi}}, \bibinfo {author} {\bibfnamefont {R.}~\bibnamefont {Ciolfi}}, \bibinfo {author} {\bibfnamefont {S.}~\bibnamefont {Clesse}}, \bibinfo {author} {\bibfnamefont {F.}~\bibnamefont {Cleva}}, \bibinfo {author} {\bibfnamefont {E.}~\bibnamefont {Coccia}}, \bibinfo {author} {\bibfnamefont {E.}~\bibnamefont {Codazzo}}, \bibinfo {author} {\bibfnamefont {P.-F.}\ \bibnamefont {Cohadon}}, \bibinfo {author} {\bibfnamefont {A.}~\bibnamefont {Colombo}}, \bibinfo {author} {\bibfnamefont {M.}~\bibnamefont {Colpi}}, \bibinfo {author} {\bibfnamefont {L.}~\bibnamefont {Conti}}, \bibinfo {author} {\bibfnamefont {I.}~\bibnamefont {Cordero-Carri\'on}}, \bibinfo {author}
  {\bibfnamefont {S.}~\bibnamefont {Corezzi}}, \bibinfo {author} {\bibfnamefont {S.}~\bibnamefont {Cortese}}, \bibinfo {author} {\bibfnamefont {J.-P.}\ \bibnamefont {Coulon}}, \bibinfo {author} {\bibfnamefont {J.-F.}\ \bibnamefont {Coupechoux}}, \bibinfo {author} {\bibfnamefont {M.}~\bibnamefont {Croquette}}, \bibinfo {author} {\bibfnamefont {J.~R.}\ \bibnamefont {Cudell}}, \bibinfo {author} {\bibfnamefont {E.}~\bibnamefont {Cuoco}}, \bibinfo {author} {\bibfnamefont {M.}~\bibnamefont {Cury\l{}o}}, \bibinfo {author} {\bibfnamefont {P.}~\bibnamefont {Dabadie}}, \bibinfo {author} {\bibfnamefont {T.~D.}\ \bibnamefont {Canton}}, \bibinfo {author} {\bibfnamefont {S.}~\bibnamefont {Dall'Osso}}, \bibinfo {author} {\bibfnamefont {G.}~\bibnamefont {D\'alya}}, \bibinfo {author} {\bibfnamefont {B.}~\bibnamefont {D'Angelo}}, \bibinfo {author} {\bibfnamefont {G.}~\bibnamefont {Dangoisse}}, \bibinfo {author} {\bibfnamefont {S.}~\bibnamefont {Danilishin}}, \bibinfo {author} {\bibfnamefont {S.}~\bibnamefont {D'Antonio}},
  \bibinfo {author} {\bibfnamefont {V.}~\bibnamefont {Dattilo}}, \bibinfo {author} {\bibfnamefont {M.}~\bibnamefont {Davier}}, \bibinfo {author} {\bibfnamefont {J.}~\bibnamefont {Degallaix}}, \bibinfo {author} {\bibfnamefont {M.}~\bibnamefont {De~Laurentis}}, \bibinfo {author} {\bibfnamefont {S.}~\bibnamefont {Del\'eglise}}, \bibinfo {author} {\bibfnamefont {F.}~\bibnamefont {De~Lillo}}, \bibinfo {author} {\bibfnamefont {D.}~\bibnamefont {Dell'Aquila}}, \bibinfo {author} {\bibfnamefont {W.}~\bibnamefont {Del~Pozzo}}, \bibinfo {author} {\bibfnamefont {F.}~\bibnamefont {De~Matteis}}, \bibinfo {author} {\bibfnamefont {A.}~\bibnamefont {Depasse}}, \bibinfo {author} {\bibfnamefont {R.}~\bibnamefont {De~Pietri}}, \bibinfo {author} {\bibfnamefont {R.}~\bibnamefont {De~Rosa}}, \bibinfo {author} {\bibfnamefont {C.}~\bibnamefont {De~Rossi}}, \bibinfo {author} {\bibfnamefont {R.}~\bibnamefont {De~Simone}}, \bibinfo {author} {\bibfnamefont {L.}~\bibnamefont {Di~Fiore}}, \bibinfo {author} {\bibfnamefont {C.}~\bibnamefont
  {Di~Giorgio}}, \bibinfo {author} {\bibfnamefont {F.}~\bibnamefont {Di~Giovanni}}, \bibinfo {author} {\bibfnamefont {M.}~\bibnamefont {Di~Giovanni}}, \bibinfo {author} {\bibfnamefont {T.}~\bibnamefont {Di~Girolamo}}, \bibinfo {author} {\bibfnamefont {D.}~\bibnamefont {Diksha}}, \bibinfo {author} {\bibfnamefont {A.}~\bibnamefont {Di~Lieto}}, \bibinfo {author} {\bibfnamefont {A.}~\bibnamefont {Di~Michele}}, \bibinfo {author} {\bibfnamefont {J.}~\bibnamefont {Ding}}, \bibinfo {author} {\bibfnamefont {S.}~\bibnamefont {Di~Pace}}, \bibinfo {author} {\bibfnamefont {I.}~\bibnamefont {Di~Palma}}, \bibinfo {author} {\bibfnamefont {F.}~\bibnamefont {Di~Renzo}}, \bibinfo {author} {\bibfnamefont {L.}~\bibnamefont {D'Onofrio}}, \bibinfo {author} {\bibfnamefont {T.}~\bibnamefont {Dooney}}, \bibinfo {author} {\bibfnamefont {O.}~\bibnamefont {Dorosh}}, \bibinfo {author} {\bibfnamefont {M.}~\bibnamefont {Drago}}, \bibinfo {author} {\bibfnamefont {J.-G.}\ \bibnamefont {Ducoin}}, \bibinfo {author} {\bibfnamefont
  {U.}~\bibnamefont {Dupletsa}}, \bibinfo {author} {\bibfnamefont {O.}~\bibnamefont {Durante}}, \bibinfo {author} {\bibfnamefont {D.}~\bibnamefont {D'Urso}}, \bibinfo {author} {\bibfnamefont {P.-A.}\ \bibnamefont {Duverne}}, \bibinfo {author} {\bibfnamefont {M.}~\bibnamefont {Eisenmann}}, \bibinfo {author} {\bibfnamefont {L.}~\bibnamefont {Errico}}, \bibinfo {author} {\bibfnamefont {D.}~\bibnamefont {Estevez}}, \bibinfo {author} {\bibfnamefont {F.}~\bibnamefont {Fabrizi}}, \bibinfo {author} {\bibfnamefont {F.}~\bibnamefont {Faedi}}, \bibinfo {author} {\bibfnamefont {V.}~\bibnamefont {Fafone}}, \bibinfo {author} {\bibfnamefont {G.}~\bibnamefont {Favaro}}, \bibinfo {author} {\bibfnamefont {M.}~\bibnamefont {Fays}}, \bibinfo {author} {\bibfnamefont {E.}~\bibnamefont {Fenyvesi}}, \bibinfo {author} {\bibfnamefont {I.}~\bibnamefont {Ferrante}}, \bibinfo {author} {\bibfnamefont {F.}~\bibnamefont {Fidecaro}}, \bibinfo {author} {\bibfnamefont {P.}~\bibnamefont {Figura}}, \bibinfo {author} {\bibfnamefont
  {A.}~\bibnamefont {Fiori}}, \bibinfo {author} {\bibfnamefont {I.}~\bibnamefont {Fiori}}, \bibinfo {author} {\bibfnamefont {R.}~\bibnamefont {Fittipaldi}}, \bibinfo {author} {\bibfnamefont {V.}~\bibnamefont {Fiumara}}, \bibinfo {author} {\bibfnamefont {R.}~\bibnamefont {Flaminio}}, \bibinfo {author} {\bibfnamefont {J.~A.}\ \bibnamefont {Font}}, \bibinfo {author} {\bibfnamefont {S.}~\bibnamefont {Frasca}}, \bibinfo {author} {\bibfnamefont {F.}~\bibnamefont {Frasconi}}, \bibinfo {author} {\bibfnamefont {A.}~\bibnamefont {Freise}}, \bibinfo {author} {\bibfnamefont {O.}~\bibnamefont {Freitas}}, \bibinfo {author} {\bibfnamefont {G.~G.}\ \bibnamefont {Fronz\'e}}, \bibinfo {author} {\bibfnamefont {B.}~\bibnamefont {Gadre}}, \bibinfo {author} {\bibfnamefont {R.}~\bibnamefont {Gamba}}, \bibinfo {author} {\bibfnamefont {B.}~\bibnamefont {Garaventa}}, \bibinfo {author} {\bibfnamefont {J.}~\bibnamefont {Garcia-Bellido}}, \bibinfo {author} {\bibfnamefont {J.}~\bibnamefont {Gargiulo}}, \bibinfo {author} {\bibfnamefont
  {F.}~\bibnamefont {Garufi}}, \bibinfo {author} {\bibfnamefont {C.}~\bibnamefont {Gasbarra}}, \bibinfo {author} {\bibfnamefont {G.}~\bibnamefont {Gemme}}, \bibinfo {author} {\bibfnamefont {A.}~\bibnamefont {Gennai}}, \bibinfo {author} {\bibfnamefont {A.}~\bibnamefont {Ghosh}}, \bibinfo {author} {\bibfnamefont {L.}~\bibnamefont {Giacoppo}}, \bibinfo {author} {\bibfnamefont {P.}~\bibnamefont {Giri}}, \bibinfo {author} {\bibfnamefont {F.}~\bibnamefont {Gissi}}, \bibinfo {author} {\bibfnamefont {S.}~\bibnamefont {Gkaitatzis}}, \bibinfo {author} {\bibfnamefont {F.}~\bibnamefont {Glotin}}, \bibinfo {author} {\bibfnamefont {B.}~\bibnamefont {Goncharov}}, \bibinfo {author} {\bibfnamefont {M.}~\bibnamefont {Gosselin}}, \bibinfo {author} {\bibfnamefont {R.}~\bibnamefont {Gouaty}}, \bibinfo {author} {\bibfnamefont {A.}~\bibnamefont {Grado}}, \bibinfo {author} {\bibfnamefont {M.}~\bibnamefont {Granata}}, \bibinfo {author} {\bibfnamefont {V.}~\bibnamefont {Granata}}, \bibinfo {author} {\bibfnamefont {G.}~\bibnamefont
  {Greco}}, \bibinfo {author} {\bibfnamefont {G.}~\bibnamefont {Grignani}}, \bibinfo {author} {\bibfnamefont {A.}~\bibnamefont {Grimaldi}}, \bibinfo {author} {\bibfnamefont {D.}~\bibnamefont {Guerra}}, \bibinfo {author} {\bibfnamefont {D.}~\bibnamefont {Guetta}}, \bibinfo {author} {\bibfnamefont {G.~M.}\ \bibnamefont {Guidi}}, \bibinfo {author} {\bibfnamefont {F.}~\bibnamefont {Gulminelli}}, \bibinfo {author} {\bibfnamefont {Y.}~\bibnamefont {Guo}}, \bibinfo {author} {\bibfnamefont {P.}~\bibnamefont {Gupta}}, \bibinfo {author} {\bibfnamefont {N.}~\bibnamefont {Gutierrez}}, \bibinfo {author} {\bibfnamefont {L.}~\bibnamefont {Haegel}}, \bibinfo {author} {\bibfnamefont {O.}~\bibnamefont {Halim}}, \bibinfo {author} {\bibfnamefont {O.}~\bibnamefont {Hannuksela}}, \bibinfo {author} {\bibfnamefont {T.}~\bibnamefont {Harder}}, \bibinfo {author} {\bibfnamefont {K.}~\bibnamefont {Haris}}, \bibinfo {author} {\bibfnamefont {T.}~\bibnamefont {Harmark}}, \bibinfo {author} {\bibfnamefont {J.}~\bibnamefont {Harms}}, \bibinfo
  {author} {\bibfnamefont {B.}~\bibnamefont {Haskell}}, \bibinfo {author} {\bibfnamefont {A.}~\bibnamefont {Heidmann}}, \bibinfo {author} {\bibfnamefont {H.}~\bibnamefont {Heitmann}}, \bibinfo {author} {\bibfnamefont {P.}~\bibnamefont {Hello}}, \bibinfo {author} {\bibfnamefont {G.}~\bibnamefont {Hemming}}, \bibinfo {author} {\bibfnamefont {E.}~\bibnamefont {Hennes}}, \bibinfo {author} {\bibfnamefont {J.-S.}\ \bibnamefont {Hennig}}, \bibinfo {author} {\bibfnamefont {M.}~\bibnamefont {Hennig}}, \bibinfo {author} {\bibfnamefont {S.}~\bibnamefont {Hild}}, \bibinfo {author} {\bibfnamefont {D.}~\bibnamefont {Hofman}}, \bibinfo {author} {\bibfnamefont {N.~A.}\ \bibnamefont {Holland}}, \bibinfo {author} {\bibfnamefont {V.}~\bibnamefont {Hui}}, \bibinfo {author} {\bibfnamefont {G.~A.}\ \bibnamefont {Iandolo}}, \bibinfo {author} {\bibfnamefont {B.}~\bibnamefont {Idzkowski}}, \bibinfo {author} {\bibfnamefont {A.}~\bibnamefont {Iess}}, \bibinfo {author} {\bibfnamefont {G.}~\bibnamefont {Iorio}}, \bibinfo {author}
  {\bibfnamefont {P.}~\bibnamefont {Iosif}}, \bibinfo {author} {\bibfnamefont {T.}~\bibnamefont {Jacqmin}}, \bibinfo {author} {\bibfnamefont {P.-E.}\ \bibnamefont {Jacquet}}, \bibinfo {author} {\bibfnamefont {J.}~\bibnamefont {Janquart}}, \bibinfo {author} {\bibfnamefont {K.}~\bibnamefont {Janssens}}, \bibinfo {author} {\bibfnamefont {S.}~\bibnamefont {Jaraba}}, \bibinfo {author} {\bibfnamefont {P.}~\bibnamefont {Jaranowski}}, \bibinfo {author} {\bibfnamefont {P.}~\bibnamefont {Jasal}}, \bibinfo {author} {\bibfnamefont {V.}~\bibnamefont {Juste}}, \bibinfo {author} {\bibfnamefont {C.}~\bibnamefont {Kalaghatgi}}, \bibinfo {author} {\bibfnamefont {C.}~\bibnamefont {Karathanasis}}, \bibinfo {author} {\bibfnamefont {S.}~\bibnamefont {Katsanevas}}, \bibinfo {author} {\bibfnamefont {F.}~\bibnamefont {K\'ef\'elian}}, \bibinfo {author} {\bibfnamefont {G.}~\bibnamefont {Koekoek}}, \bibinfo {author} {\bibfnamefont {S.}~\bibnamefont {Koley}}, \bibinfo {author} {\bibfnamefont {M.}~\bibnamefont {Kolstein}}, \bibinfo
  {author} {\bibfnamefont {S.~L.}\ \bibnamefont {Kranzhoff}}, \bibinfo {author} {\bibfnamefont {A.}~\bibnamefont {Kr\'olak}}, \bibinfo {author} {\bibfnamefont {P.}~\bibnamefont {Kuijer}}, \bibinfo {author} {\bibfnamefont {S.}~\bibnamefont {Kuroyanagi}}, \bibinfo {author} {\bibfnamefont {P.}~\bibnamefont {Lagabbe}}, \bibinfo {author} {\bibfnamefont {D.}~\bibnamefont {Laghi}}, \bibinfo {author} {\bibfnamefont {M.}~\bibnamefont {Lalleman}}, \bibinfo {author} {\bibfnamefont {A.}~\bibnamefont {Lamberts}}, \bibinfo {author} {\bibfnamefont {A.}~\bibnamefont {La~Rana}}, \bibinfo {author} {\bibfnamefont {I.}~\bibnamefont {La~Rosa}}, \bibinfo {author} {\bibfnamefont {A.}~\bibnamefont {Lartaux-Vollard}}, \bibinfo {author} {\bibfnamefont {C.}~\bibnamefont {Lazzaro}}, \bibinfo {author} {\bibfnamefont {P.}~\bibnamefont {Leaci}}, \bibinfo {author} {\bibfnamefont {A.}~\bibnamefont {Lema\^{\i}tre}}, \bibinfo {author} {\bibfnamefont {M.}~\bibnamefont {Lenti}}, \bibinfo {author} {\bibfnamefont {E.}~\bibnamefont {Leonova}},
  \bibinfo {author} {\bibfnamefont {M.}~\bibnamefont {Lequime}}, \bibinfo {author} {\bibfnamefont {N.}~\bibnamefont {Leroy}}, \bibinfo {author} {\bibfnamefont {N.}~\bibnamefont {Letendre}}, \bibinfo {author} {\bibfnamefont {M.}~\bibnamefont {Lethuillier}}, \bibinfo {author} {\bibfnamefont {K.}~\bibnamefont {Leyde}}, \bibinfo {author} {\bibfnamefont {F.}~\bibnamefont {Linde}}, \bibinfo {author} {\bibfnamefont {L.}~\bibnamefont {London}}, \bibinfo {author} {\bibfnamefont {A.}~\bibnamefont {Longo}}, \bibinfo {author} {\bibfnamefont {M.~L.}\ \bibnamefont {Portilla}}, \bibinfo {author} {\bibfnamefont {M.}~\bibnamefont {Lorenzini}}, \bibinfo {author} {\bibfnamefont {V.}~\bibnamefont {Loriette}}, \bibinfo {author} {\bibfnamefont {G.}~\bibnamefont {Losurdo}}, \bibinfo {author} {\bibfnamefont {D.}~\bibnamefont {Lumaca}}, \bibinfo {author} {\bibfnamefont {A.}~\bibnamefont {Macquet}}, \bibinfo {author} {\bibfnamefont {C.}~\bibnamefont {Magazz\`u}}, \bibinfo {author} {\bibfnamefont {R.}~\bibnamefont {Maggiore}}, \bibinfo
  {author} {\bibfnamefont {M.}~\bibnamefont {Magnozzi}}, \bibinfo {author} {\bibfnamefont {E.}~\bibnamefont {Majorana}}, \bibinfo {author} {\bibfnamefont {N.}~\bibnamefont {Man}}, \bibinfo {author} {\bibfnamefont {V.}~\bibnamefont {Mangano}}, \bibinfo {author} {\bibfnamefont {M.}~\bibnamefont {Mantovani}}, \bibinfo {author} {\bibfnamefont {M.}~\bibnamefont {Mapelli}}, \bibinfo {author} {\bibfnamefont {F.}~\bibnamefont {Marchesoni}}, \bibinfo {author} {\bibfnamefont {D.~M.}\ \bibnamefont {Pina}}, \bibinfo {author} {\bibfnamefont {F.}~\bibnamefont {Marion}}, \bibinfo {author} {\bibfnamefont {A.}~\bibnamefont {Marquina}}, \bibinfo {author} {\bibfnamefont {S.}~\bibnamefont {Marsat}}, \bibinfo {author} {\bibfnamefont {F.}~\bibnamefont {Martelli}}, \bibinfo {author} {\bibfnamefont {M.}~\bibnamefont {Martinez}}, \bibinfo {author} {\bibfnamefont {V.}~\bibnamefont {Martinez}}, \bibinfo {author} {\bibfnamefont {A.}~\bibnamefont {Masserot}}, \bibinfo {author} {\bibfnamefont {M.}~\bibnamefont {Mastrodicasa}}, \bibinfo
  {author} {\bibfnamefont {S.}~\bibnamefont {Mastrogiovanni}}, \bibinfo {author} {\bibfnamefont {Q.}~\bibnamefont {Meijer}}, \bibinfo {author} {\bibfnamefont {A.}~\bibnamefont {Menendez-Vazquez}}, \bibinfo {author} {\bibfnamefont {L.}~\bibnamefont {Mereni}}, \bibinfo {author} {\bibfnamefont {M.}~\bibnamefont {Merzougui}}, \bibinfo {author} {\bibfnamefont {A.}~\bibnamefont {Miani}}, \bibinfo {author} {\bibfnamefont {C.}~\bibnamefont {Michel}}, \bibinfo {author} {\bibfnamefont {A.}~\bibnamefont {Miller}}, \bibinfo {author} {\bibfnamefont {B.}~\bibnamefont {Miller}}, \bibinfo {author} {\bibfnamefont {E.}~\bibnamefont {Milotti}}, \bibinfo {author} {\bibfnamefont {Y.}~\bibnamefont {Minenkov}}, \bibinfo {author} {\bibfnamefont {L.~M.}\ \bibnamefont {Mir}}, \bibinfo {author} {\bibfnamefont {M.}~\bibnamefont {Miravet-Ten\'es}}, \bibinfo {author} {\bibfnamefont {A.~L.}\ \bibnamefont {Mitchell}}, \bibinfo {author} {\bibfnamefont {C.}~\bibnamefont {Mondal}}, \bibinfo {author} {\bibfnamefont {M.}~\bibnamefont {Montani}},
  \bibinfo {author} {\bibfnamefont {F.}~\bibnamefont {Morawski}}, \bibinfo {author} {\bibfnamefont {G.}~\bibnamefont {Morras}}, \bibinfo {author} {\bibfnamefont {A.}~\bibnamefont {Moscatello}}, \bibinfo {author} {\bibfnamefont {B.}~\bibnamefont {Mours}}, \bibinfo {author} {\bibfnamefont {C.~M.}\ \bibnamefont {Mow-Lowry}}, \bibinfo {author} {\bibfnamefont {E.}~\bibnamefont {Msihid}}, \bibinfo {author} {\bibfnamefont {F.}~\bibnamefont {Muciaccia}}, \bibinfo {author} {\bibfnamefont {S.}~\bibnamefont {Mukherjee}}, \bibinfo {author} {\bibfnamefont {A.}~\bibnamefont {Nagar}}, \bibinfo {author} {\bibfnamefont {V.}~\bibnamefont {Napolano}}, \bibinfo {author} {\bibfnamefont {I.}~\bibnamefont {Nardecchia}}, \bibinfo {author} {\bibfnamefont {H.}~\bibnamefont {Narola}}, \bibinfo {author} {\bibfnamefont {L.}~\bibnamefont {Naticchioni}}, \bibinfo {author} {\bibfnamefont {J.}~\bibnamefont {Neilson}}, \bibinfo {author} {\bibfnamefont {S.}~\bibnamefont {Nesseris}}, \bibinfo {author} {\bibfnamefont {C.}~\bibnamefont {Nguyen}},
  \bibinfo {author} {\bibfnamefont {G.}~\bibnamefont {Nieradka}}, \bibinfo {author} {\bibfnamefont {S.}~\bibnamefont {Nissanke}}, \bibinfo {author} {\bibfnamefont {E.}~\bibnamefont {Nitoglia}}, \bibinfo {author} {\bibfnamefont {F.}~\bibnamefont {Nocera}}, \bibinfo {author} {\bibfnamefont {J.}~\bibnamefont {Novak}}, \bibinfo {author} {\bibfnamefont {J.~F.~N.}\ \bibnamefont {no~Siles}}, \bibinfo {author} {\bibfnamefont {M.}~\bibnamefont {Oertel}}, \bibinfo {author} {\bibfnamefont {G.}~\bibnamefont {Oganesyan}}, \bibinfo {author} {\bibfnamefont {R.}~\bibnamefont {Oliveri}}, \bibinfo {author} {\bibfnamefont {M.}~\bibnamefont {Orselli}}, \bibinfo {author} {\bibfnamefont {C.}~\bibnamefont {Palomba}}, \bibinfo {author} {\bibfnamefont {P.~T.~H.}\ \bibnamefont {Pang}}, \bibinfo {author} {\bibfnamefont {F.}~\bibnamefont {Pannarale}}, \bibinfo {author} {\bibfnamefont {F.}~\bibnamefont {Paoletti}}, \bibinfo {author} {\bibfnamefont {A.}~\bibnamefont {Paoli}}, \bibinfo {author} {\bibfnamefont {A.}~\bibnamefont {Paolone}},
  \bibinfo {author} {\bibfnamefont {G.}~\bibnamefont {Pappas}}, \bibinfo {author} {\bibfnamefont {A.}~\bibnamefont {Parisi}}, \bibinfo {author} {\bibfnamefont {D.}~\bibnamefont {Pascucci}}, \bibinfo {author} {\bibfnamefont {A.}~\bibnamefont {Pasqualetti}}, \bibinfo {author} {\bibfnamefont {R.}~\bibnamefont {Passaquieti}}, \bibinfo {author} {\bibfnamefont {D.}~\bibnamefont {Passuello}}, \bibinfo {author} {\bibfnamefont {B.}~\bibnamefont {Patricelli}}, \bibinfo {author} {\bibfnamefont {R.}~\bibnamefont {Pedurand}}, \bibinfo {author} {\bibfnamefont {R.}~\bibnamefont {Pegna}}, \bibinfo {author} {\bibfnamefont {M.}~\bibnamefont {Pegoraro}}, \bibinfo {author} {\bibfnamefont {A.}~\bibnamefont {Perego}}, \bibinfo {author} {\bibfnamefont {A.}~\bibnamefont {Pereira}}, \bibinfo {author} {\bibfnamefont {C.}~\bibnamefont {P\'erigois}}, \bibinfo {author} {\bibfnamefont {A.}~\bibnamefont {Perreca}}, \bibinfo {author} {\bibfnamefont {S.}~\bibnamefont {Perri\`es}}, \bibinfo {author} {\bibfnamefont {J.~W.}\ \bibnamefont
  {Perry}}, \bibinfo {author} {\bibfnamefont {D.}~\bibnamefont {Pesios}}, \bibinfo {author} {\bibfnamefont {C.}~\bibnamefont {Petrillo}}, \bibinfo {author} {\bibfnamefont {K.~S.}\ \bibnamefont {Phukon}}, \bibinfo {author} {\bibfnamefont {O.~J.}\ \bibnamefont {Piccinni}}, \bibinfo {author} {\bibfnamefont {M.}~\bibnamefont {Pichot}}, \bibinfo {author} {\bibfnamefont {M.}~\bibnamefont {Piendibene}}, \bibinfo {author} {\bibfnamefont {F.}~\bibnamefont {Piergiovanni}}, \bibinfo {author} {\bibfnamefont {L.}~\bibnamefont {Pierini}}, \bibinfo {author} {\bibfnamefont {G.}~\bibnamefont {Pierra}}, \bibinfo {author} {\bibfnamefont {V.}~\bibnamefont {Pierro}}, \bibinfo {author} {\bibfnamefont {G.}~\bibnamefont {Pillant}}, \bibinfo {author} {\bibfnamefont {M.}~\bibnamefont {Pillas}}, \bibinfo {author} {\bibfnamefont {F.}~\bibnamefont {Pilo}}, \bibinfo {author} {\bibfnamefont {L.}~\bibnamefont {Pinard}}, \bibinfo {author} {\bibfnamefont {I.~M.}\ \bibnamefont {Pinto}}, \bibinfo {author} {\bibfnamefont {M.}~\bibnamefont
  {Pinto}}, \bibinfo {author} {\bibfnamefont {M.}~\bibnamefont {Pinto}}, \bibinfo {author} {\bibfnamefont {K.}~\bibnamefont {Piotrzkowski}}, \bibinfo {author} {\bibfnamefont {A.}~\bibnamefont {Placidi}}, \bibinfo {author} {\bibfnamefont {E.}~\bibnamefont {Placidi}}, \bibinfo {author} {\bibfnamefont {W.}~\bibnamefont {Plastino}}, \bibinfo {author} {\bibfnamefont {R.}~\bibnamefont {Poggiani}}, \bibinfo {author} {\bibfnamefont {E.}~\bibnamefont {Polini}}, \bibinfo {author} {\bibfnamefont {E.}~\bibnamefont {Porcelli}}, \bibinfo {author} {\bibfnamefont {J.}~\bibnamefont {Portell}}, \bibinfo {author} {\bibfnamefont {E.~K.}\ \bibnamefont {Porter}}, \bibinfo {author} {\bibfnamefont {R.}~\bibnamefont {Poulton}}, \bibinfo {author} {\bibfnamefont {M.}~\bibnamefont {Pracchia}}, \bibinfo {author} {\bibfnamefont {T.}~\bibnamefont {Pradier}}, \bibinfo {author} {\bibfnamefont {M.}~\bibnamefont {Principe}}, \bibinfo {author} {\bibfnamefont {G.~A.}\ \bibnamefont {Prodi}}, \bibinfo {author} {\bibfnamefont {P.}~\bibnamefont
  {Prosposito}}, \bibinfo {author} {\bibfnamefont {A.}~\bibnamefont {Puecher}}, \bibinfo {author} {\bibfnamefont {M.}~\bibnamefont {Punturo}}, \bibinfo {author} {\bibfnamefont {F.}~\bibnamefont {Puosi}}, \bibinfo {author} {\bibfnamefont {P.}~\bibnamefont {Puppo}}, \bibinfo {author} {\bibfnamefont {G.}~\bibnamefont {Raaijmakers}}, \bibinfo {author} {\bibfnamefont {N.}~\bibnamefont {Radulesco}}, \bibinfo {author} {\bibfnamefont {P.}~\bibnamefont {Rapagnani}}, \bibinfo {author} {\bibfnamefont {M.}~\bibnamefont {Razzano}}, \bibinfo {author} {\bibfnamefont {T.}~\bibnamefont {Regimbau}}, \bibinfo {author} {\bibfnamefont {L.}~\bibnamefont {Rei}}, \bibinfo {author} {\bibfnamefont {P.}~\bibnamefont {Rettegno}}, \bibinfo {author} {\bibfnamefont {B.}~\bibnamefont {Revenu}}, \bibinfo {author} {\bibfnamefont {A.}~\bibnamefont {Reza}}, \bibinfo {author} {\bibfnamefont {A.~S.}\ \bibnamefont {Rezaei}}, \bibinfo {author} {\bibfnamefont {F.}~\bibnamefont {Ricci}}, \bibinfo {author} {\bibfnamefont {S.}~\bibnamefont {Rinaldi}},
  \bibinfo {author} {\bibfnamefont {F.}~\bibnamefont {Robinet}}, \bibinfo {author} {\bibfnamefont {A.}~\bibnamefont {Rocchi}}, \bibinfo {author} {\bibfnamefont {L.}~\bibnamefont {Rolland}}, \bibinfo {author} {\bibfnamefont {M.}~\bibnamefont {Romanelli}}, \bibinfo {author} {\bibfnamefont {R.}~\bibnamefont {Romano}}, \bibinfo {author} {\bibfnamefont {A.}~\bibnamefont {Romero}}, \bibinfo {author} {\bibfnamefont {S.}~\bibnamefont {Ronchini}}, \bibinfo {author} {\bibfnamefont {L.}~\bibnamefont {Rosa}}, \bibinfo {author} {\bibfnamefont {D.}~\bibnamefont {Rosi\ifmmode~\acute{n}\else \'{n}\fi{}ska}}, \bibinfo {author} {\bibfnamefont {S.}~\bibnamefont {Roy}}, \bibinfo {author} {\bibfnamefont {D.}~\bibnamefont {Rozza}}, \bibinfo {author} {\bibfnamefont {P.}~\bibnamefont {Ruggi}}, \bibinfo {author} {\bibfnamefont {E.~R.}\ \bibnamefont {Morales}}, \bibinfo {author} {\bibfnamefont {P.}~\bibnamefont {Saffarieh}}, \bibinfo {author} {\bibfnamefont {O.~S.}\ \bibnamefont {Salafia}}, \bibinfo {author} {\bibfnamefont
  {L.}~\bibnamefont {Salconi}}, \bibinfo {author} {\bibfnamefont {F.}~\bibnamefont {Salemi}}, \bibinfo {author} {\bibfnamefont {M.}~\bibnamefont {Sall\'e}}, \bibinfo {author} {\bibfnamefont {A.}~\bibnamefont {Samajdar}}, \bibinfo {author} {\bibfnamefont {N.}~\bibnamefont {Sanchis-Gual}}, \bibinfo {author} {\bibfnamefont {A.}~\bibnamefont {Sanuy}}, \bibinfo {author} {\bibfnamefont {A.}~\bibnamefont {Sasli}}, \bibinfo {author} {\bibfnamefont {P.}~\bibnamefont {Sassi}}, \bibinfo {author} {\bibfnamefont {B.}~\bibnamefont {Sassolas}}, \bibinfo {author} {\bibfnamefont {S.}~\bibnamefont {Sayah}}, \bibinfo {author} {\bibfnamefont {S.}~\bibnamefont {Schmidt}}, \bibinfo {author} {\bibfnamefont {M.}~\bibnamefont {Seglar-Arroyo}}, \bibinfo {author} {\bibfnamefont {D.}~\bibnamefont {Sentenac}}, \bibinfo {author} {\bibfnamefont {V.}~\bibnamefont {Sequino}}, \bibinfo {author} {\bibfnamefont {G.}~\bibnamefont {Servignat}}, \bibinfo {author} {\bibfnamefont {Y.}~\bibnamefont {Setyawati}}, \bibinfo {author} {\bibfnamefont
  {N.~S.}\ \bibnamefont {Shcheblanov}}, \bibinfo {author} {\bibfnamefont {M.}~\bibnamefont {Sieniawska}}, \bibinfo {author} {\bibfnamefont {L.}~\bibnamefont {Silenzi}}, \bibinfo {author} {\bibfnamefont {N.}~\bibnamefont {Singh}}, \bibinfo {author} {\bibfnamefont {A.}~\bibnamefont {Singha}}, \bibinfo {author} {\bibfnamefont {V.}~\bibnamefont {Sipala}}, \bibinfo {author} {\bibfnamefont {J.}~\bibnamefont {Soldateschi}}, \bibinfo {author} {\bibfnamefont {V.}~\bibnamefont {Sordini}}, \bibinfo {author} {\bibfnamefont {F.}~\bibnamefont {Sorrentino}}, \bibinfo {author} {\bibfnamefont {N.}~\bibnamefont {Sorrentino}}, \bibinfo {author} {\bibfnamefont {R.}~\bibnamefont {Soulard}}, \bibinfo {author} {\bibfnamefont {V.}~\bibnamefont {Spagnuolo}}, \bibinfo {author} {\bibfnamefont {M.}~\bibnamefont {Spera}}, \bibinfo {author} {\bibfnamefont {P.}~\bibnamefont {Spinicelli}}, \bibinfo {author} {\bibfnamefont {C.}~\bibnamefont {Stachie}}, \bibinfo {author} {\bibfnamefont {D.~A.}\ \bibnamefont {Steer}}, \bibinfo {author}
  {\bibfnamefont {J.}~\bibnamefont {Steinlechner}}, \bibinfo {author} {\bibfnamefont {S.}~\bibnamefont {Steinlechner}}, \bibinfo {author} {\bibfnamefont {N.}~\bibnamefont {Stergioulas}}, \bibinfo {author} {\bibfnamefont {G.}~\bibnamefont {Stratta}}, \bibinfo {author} {\bibfnamefont {M.}~\bibnamefont {Suchenek}}, \bibinfo {author} {\bibfnamefont {A.}~\bibnamefont {Sur}}, \bibinfo {author} {\bibfnamefont {J.}~\bibnamefont {Suresh}}, \bibinfo {author} {\bibfnamefont {B.~L.}\ \bibnamefont {Swinkels}}, \bibinfo {author} {\bibfnamefont {A.}~\bibnamefont {Syx}}, \bibinfo {author} {\bibfnamefont {P.}~\bibnamefont {Szewczyk}}, \bibinfo {author} {\bibfnamefont {M.}~\bibnamefont {Tacca}}, \bibinfo {author} {\bibfnamefont {N.}~\bibnamefont {Tamanini}}, \bibinfo {author} {\bibfnamefont {A.~J.}\ \bibnamefont {Tanasijczuk}}, \bibinfo {author} {\bibfnamefont {E.~N. T.~S.}\ \bibnamefont {Mart\'{\i}n}}, \bibinfo {author} {\bibfnamefont {C.}~\bibnamefont {Taranto}}, \bibinfo {author} {\bibfnamefont {M.}~\bibnamefont {Tonelli}},
  \bibinfo {author} {\bibfnamefont {A.}~\bibnamefont {Torres-Forn\'e}}, \bibinfo {author} {\bibfnamefont {I.~T.}\ \bibnamefont {e~Melo}}, \bibinfo {author} {\bibfnamefont {E.}~\bibnamefont {Tournefier}}, \bibinfo {author} {\bibfnamefont {A.}~\bibnamefont {Trapananti}}, \bibinfo {author} {\bibfnamefont {F.}~\bibnamefont {Travasso}}, \bibinfo {author} {\bibfnamefont {J.}~\bibnamefont {Trenado}}, \bibinfo {author} {\bibfnamefont {M.~C.}\ \bibnamefont {Tringali}}, \bibinfo {author} {\bibfnamefont {L.}~\bibnamefont {Troiano}}, \bibinfo {author} {\bibfnamefont {A.}~\bibnamefont {Trovato}}, \bibinfo {author} {\bibfnamefont {L.}~\bibnamefont {Trozzo}}, \bibinfo {author} {\bibfnamefont {K.~W.}\ \bibnamefont {Tsang}}, \bibinfo {author} {\bibfnamefont {K.}~\bibnamefont {Turbang}}, \bibinfo {author} {\bibfnamefont {M.}~\bibnamefont {Turconi}}, \bibinfo {author} {\bibfnamefont {C.}~\bibnamefont {Turski}}, \bibinfo {author} {\bibfnamefont {H.}~\bibnamefont {Ubach}}, \bibinfo {author} {\bibfnamefont {A.}~\bibnamefont
  {Utina}}, \bibinfo {author} {\bibfnamefont {M.}~\bibnamefont {Valentini}}, \bibinfo {author} {\bibfnamefont {S.}~\bibnamefont {Vallero}}, \bibinfo {author} {\bibfnamefont {N.}~\bibnamefont {van Bakel}}, \bibinfo {author} {\bibfnamefont {M.}~\bibnamefont {van Beuzekom}}, \bibinfo {author} {\bibfnamefont {M.}~\bibnamefont {van Dael}}, \bibinfo {author} {\bibfnamefont {J.~F.~J.}\ \bibnamefont {van~den Brand}}, \bibinfo {author} {\bibfnamefont {C.}~\bibnamefont {Van Den~Broeck}}, \bibinfo {author} {\bibfnamefont {M.}~\bibnamefont {van~der Sluys}}, \bibinfo {author} {\bibfnamefont {A.}~\bibnamefont {Van~de Walle}}, \bibinfo {author} {\bibfnamefont {J.}~\bibnamefont {van Dongen}}, \bibinfo {author} {\bibfnamefont {H.}~\bibnamefont {van Haevermaet}}, \bibinfo {author} {\bibfnamefont {J.~V.}\ \bibnamefont {van Heijningen}}, \bibinfo {author} {\bibfnamefont {Z.}~\bibnamefont {van Ranst}}, \bibinfo {author} {\bibfnamefont {N.}~\bibnamefont {van Remortel}}, \bibinfo {author} {\bibfnamefont {M.}~\bibnamefont
  {Vardaro}}, \bibinfo {author} {\bibfnamefont {M.}~\bibnamefont {Vas\'uth}}, \bibinfo {author} {\bibfnamefont {G.}~\bibnamefont {Vedovato}}, \bibinfo {author} {\bibfnamefont {P.}~\bibnamefont {Verdier}}, \bibinfo {author} {\bibfnamefont {D.}~\bibnamefont {Verkindt}}, \bibinfo {author} {\bibfnamefont {P.}~\bibnamefont {Verma}}, \bibinfo {author} {\bibfnamefont {F.}~\bibnamefont {Vetrano}}, \bibinfo {author} {\bibfnamefont {A.}~\bibnamefont {Vicer\'e}}, \bibinfo {author} {\bibfnamefont {J.-Y.}\ \bibnamefont {Vinet}}, \bibinfo {author} {\bibfnamefont {S.}~\bibnamefont {Viret}}, \bibinfo {author} {\bibfnamefont {A.}~\bibnamefont {Virtuoso}}, \bibinfo {author} {\bibfnamefont {H.}~\bibnamefont {Vocca}}, \bibinfo {author} {\bibfnamefont {R.~C.}\ \bibnamefont {Walet}}, \bibinfo {author} {\bibfnamefont {M.}~\bibnamefont {Was}}, \bibinfo {author} {\bibfnamefont {N.}~\bibnamefont {Yadav}}, \bibinfo {author} {\bibfnamefont {A.}~\bibnamefont {Zadro\ifmmode~\dot{z}\else \.{z}\fi{}ny}}, \bibinfo {author} {\bibfnamefont
  {T.}~\bibnamefont {Zelenova}}, \bibinfo {author} {\bibfnamefont {J.-P.}\ \bibnamefont {Zendri}}, \bibinfo {author} {\bibfnamefont {Y.}~\bibnamefont {Zhao}}, \bibinfo {author} {\bibfnamefont {M.}~\bibnamefont {Zerrad}}, \bibinfo {author} {\bibfnamefont {H.}~\bibnamefont {Vahlbruch}}, \bibinfo {author} {\bibfnamefont {M.}~\bibnamefont {Mehmet}}, \bibinfo {author} {\bibfnamefont {H.}~\bibnamefont {L\"uck}},\ and\ \bibinfo {author} {\bibfnamefont {K.}~\bibnamefont {Danzmann}} (\bibinfo {collaboration} {Virgo Collaboration}),\ }\bibfield  {title} {\bibinfo {title} {Frequency-dependent squeezed vacuum source for the advanced virgo gravitational-wave detector},\ }\href {https://doi.org/10.1103/PhysRevLett.131.041403} {\bibfield  {journal} {\bibinfo  {journal} {Phys. Rev. Lett.}\ }\textbf {\bibinfo {volume} {131}},\ \bibinfo {pages} {041403} (\bibinfo {year} {2023})}\BibitemShut {NoStop}%
\bibitem [{\citenamefont {Mikhailov}\ \emph {et~al.}(2006)\citenamefont {Mikhailov}, \citenamefont {Goda}, \citenamefont {Corbitt},\ and\ \citenamefont {Mavalvala}}]{PhysRevA.73.053810}%
  \BibitemOpen
  \bibfield  {author} {\bibinfo {author} {\bibfnamefont {E.~E.}\ \bibnamefont {Mikhailov}}, \bibinfo {author} {\bibfnamefont {K.}~\bibnamefont {Goda}}, \bibinfo {author} {\bibfnamefont {T.}~\bibnamefont {Corbitt}},\ and\ \bibinfo {author} {\bibfnamefont {N.}~\bibnamefont {Mavalvala}},\ }\bibfield  {title} {\bibinfo {title} {Frequency-dependent squeeze-amplitude attenuation and squeeze-angle rotation by electromagnetically induced transparency for gravitational-wave interferometers},\ }\href {https://doi.org/10.1103/PhysRevA.73.053810} {\bibfield  {journal} {\bibinfo  {journal} {Phys. Rev. A}\ }\textbf {\bibinfo {volume} {73}},\ \bibinfo {pages} {053810} (\bibinfo {year} {2006})}\BibitemShut {NoStop}%
\bibitem [{\citenamefont {{Khalili}}\ and\ \citenamefont {{Polzik}}(2018)}]{2018_negative_mass_spin_GWD}%
  \BibitemOpen
  \bibfield  {author} {\bibinfo {author} {\bibfnamefont {F.~Y.}\ \bibnamefont {{Khalili}}}\ and\ \bibinfo {author} {\bibfnamefont {E.~S.}\ \bibnamefont {{Polzik}}},\ }\bibfield  {title} {\bibinfo {title} {{Overcoming the Standard Quantum Limit in Gravitational Wave Detectors Using Spin Systems with a Negative Effective Mass}},\ }\href {https://doi.org/10.1103/PhysRevLett.121.031101} {\bibfield  {journal} {\bibinfo  {journal} {Phys. Rev. Lett.}\ }\textbf {\bibinfo {volume} {121}},\ \bibinfo {eid} {031101} (\bibinfo {year} {2018})},\ \Eprint {https://arxiv.org/abs/1710.10405} {arXiv:1710.10405 [quant-ph]} \BibitemShut {NoStop}%
\bibitem [{\citenamefont {{Zeuthen}}\ \emph {et~al.}(2019)\citenamefont {{Zeuthen}}, \citenamefont {{Polzik}},\ and\ \citenamefont {{Khalili}}}]{2019_negative_mass_spin_GWD}%
  \BibitemOpen
  \bibfield  {author} {\bibinfo {author} {\bibfnamefont {E.}~\bibnamefont {{Zeuthen}}}, \bibinfo {author} {\bibfnamefont {E.~S.}\ \bibnamefont {{Polzik}}},\ and\ \bibinfo {author} {\bibfnamefont {F.~Y.}\ \bibnamefont {{Khalili}}},\ }\bibfield  {title} {\bibinfo {title} {{Gravitational wave detection beyond the standard quantum limit using a negative-mass spin system and virtual rigidity}},\ }\href {https://doi.org/10.1103/PhysRevD.100.062004} {\bibfield  {journal} {\bibinfo  {journal} {Phys. Rev. D}\ }\textbf {\bibinfo {volume} {100}},\ \bibinfo {eid} {062004} (\bibinfo {year} {2019})},\ \Eprint {https://arxiv.org/abs/1908.03416} {arXiv:1908.03416 [gr-qc]} \BibitemShut {NoStop}%
\bibitem [{\citenamefont {Ma}\ \emph {et~al.}(2014)\citenamefont {Ma}, \citenamefont {Danilishin}, \citenamefont {Zhao}, \citenamefont {Miao}, \citenamefont {Korth}, \citenamefont {Chen}, \citenamefont {Ward},\ and\ \citenamefont {Blair}}]{PhysRevLett.113.151102}%
  \BibitemOpen
  \bibfield  {author} {\bibinfo {author} {\bibfnamefont {Y.}~\bibnamefont {Ma}}, \bibinfo {author} {\bibfnamefont {S.~L.}\ \bibnamefont {Danilishin}}, \bibinfo {author} {\bibfnamefont {C.}~\bibnamefont {Zhao}}, \bibinfo {author} {\bibfnamefont {H.}~\bibnamefont {Miao}}, \bibinfo {author} {\bibfnamefont {W.~Z.}\ \bibnamefont {Korth}}, \bibinfo {author} {\bibfnamefont {Y.}~\bibnamefont {Chen}}, \bibinfo {author} {\bibfnamefont {R.~L.}\ \bibnamefont {Ward}},\ and\ \bibinfo {author} {\bibfnamefont {D.~G.}\ \bibnamefont {Blair}},\ }\bibfield  {title} {\bibinfo {title} {Narrowing the filter-cavity bandwidth in gravitational-wave detectors via optomechanical interaction},\ }\href {https://doi.org/10.1103/PhysRevLett.113.151102} {\bibfield  {journal} {\bibinfo  {journal} {Phys. Rev. Lett.}\ }\textbf {\bibinfo {volume} {113}},\ \bibinfo {pages} {151102} (\bibinfo {year} {2014})}\BibitemShut {NoStop}%
\bibitem [{\citenamefont {Ma}\ \emph {et~al.}(2017)\citenamefont {Ma}, \citenamefont {Miao}, \citenamefont {Pang}, \citenamefont {Evans}, \citenamefont {Zhao}, \citenamefont {Harms}, \citenamefont {Schnabel},\ and\ \citenamefont {Chen}}]{Ma_2017}%
  \BibitemOpen
  \bibfield  {author} {\bibinfo {author} {\bibfnamefont {Y.}~\bibnamefont {Ma}}, \bibinfo {author} {\bibfnamefont {H.}~\bibnamefont {Miao}}, \bibinfo {author} {\bibfnamefont {B.~H.}\ \bibnamefont {Pang}}, \bibinfo {author} {\bibfnamefont {M.}~\bibnamefont {Evans}}, \bibinfo {author} {\bibfnamefont {C.}~\bibnamefont {Zhao}}, \bibinfo {author} {\bibfnamefont {J.}~\bibnamefont {Harms}}, \bibinfo {author} {\bibfnamefont {R.}~\bibnamefont {Schnabel}},\ and\ \bibinfo {author} {\bibfnamefont {Y.}~\bibnamefont {Chen}},\ }\bibfield  {title} {\bibinfo {title} {Proposal for gravitational-wave detection beyond the standard quantum limit through {EPR}~entanglement},\ }\href {https://doi.org/10.1038/nphys4118} {\bibfield  {journal} {\bibinfo  {journal} {Nature Physics}\ }\textbf {\bibinfo {volume} {13}},\ \bibinfo {pages} {776} (\bibinfo {year} {2017})}\BibitemShut {NoStop}%
\bibitem [{\citenamefont {Brown}\ \emph {et~al.}(2017)\citenamefont {Brown}, \citenamefont {Miao}, \citenamefont {Collins}, \citenamefont {Mow-Lowry}, \citenamefont {T\"oyr\"a},\ and\ \citenamefont {Freise}}]{PhysRevD.96.062003}%
  \BibitemOpen
  \bibfield  {author} {\bibinfo {author} {\bibfnamefont {D.~D.}\ \bibnamefont {Brown}}, \bibinfo {author} {\bibfnamefont {H.}~\bibnamefont {Miao}}, \bibinfo {author} {\bibfnamefont {C.}~\bibnamefont {Collins}}, \bibinfo {author} {\bibfnamefont {C.}~\bibnamefont {Mow-Lowry}}, \bibinfo {author} {\bibfnamefont {D.}~\bibnamefont {T\"oyr\"a}},\ and\ \bibinfo {author} {\bibfnamefont {A.}~\bibnamefont {Freise}},\ }\bibfield  {title} {\bibinfo {title} {Broadband sensitivity enhancement of detuned dual-recycled michelson interferometers with epr entanglement},\ }\href {https://doi.org/10.1103/PhysRevD.96.062003} {\bibfield  {journal} {\bibinfo  {journal} {Phys. Rev. D}\ }\textbf {\bibinfo {volume} {96}},\ \bibinfo {pages} {062003} (\bibinfo {year} {2017})}\BibitemShut {NoStop}%
\bibitem [{\citenamefont {{Yap}}\ \emph {et~al.}(2020)\citenamefont {{Yap}}, \citenamefont {{Altin}}, \citenamefont {{McRae}}, \citenamefont {{Slagmolen}}, \citenamefont {{Ward}},\ and\ \citenamefont {{McClelland}}}]{2020NaPho..14..223Y}%
  \BibitemOpen
  \bibfield  {author} {\bibinfo {author} {\bibfnamefont {M.~J.}\ \bibnamefont {{Yap}}}, \bibinfo {author} {\bibfnamefont {P.}~\bibnamefont {{Altin}}}, \bibinfo {author} {\bibfnamefont {T.~G.}\ \bibnamefont {{McRae}}}, \bibinfo {author} {\bibfnamefont {B.~J.~J.}\ \bibnamefont {{Slagmolen}}}, \bibinfo {author} {\bibfnamefont {R.~L.}\ \bibnamefont {{Ward}}},\ and\ \bibinfo {author} {\bibfnamefont {D.~E.}\ \bibnamefont {{McClelland}}},\ }\bibfield  {title} {\bibinfo {title} {{Generation and control of frequency-dependent squeezing via Einstein-Podolsky-Rosen entanglement}},\ }\href {https://doi.org/10.1038/s41566-019-0582-4} {\bibfield  {journal} {\bibinfo  {journal} {Nature Photonics}\ }\textbf {\bibinfo {volume} {14}},\ \bibinfo {pages} {223} (\bibinfo {year} {2020})}\BibitemShut {NoStop}%
\bibitem [{\citenamefont {Südbeck}\ \emph {et~al.}(2020)\citenamefont {Südbeck}, \citenamefont {Steinlechner}, \citenamefont {Korobko},\ and\ \citenamefont {Schnabel}}]{S_dbeck_2020}%
  \BibitemOpen
  \bibfield  {author} {\bibinfo {author} {\bibfnamefont {J.}~\bibnamefont {Südbeck}}, \bibinfo {author} {\bibfnamefont {S.}~\bibnamefont {Steinlechner}}, \bibinfo {author} {\bibfnamefont {M.}~\bibnamefont {Korobko}},\ and\ \bibinfo {author} {\bibfnamefont {R.}~\bibnamefont {Schnabel}},\ }\bibfield  {title} {\bibinfo {title} {Demonstration of interferometer enhancement through einstein{\textendash}podolsky{\textendash}rosen entanglement},\ }\href {https://doi.org/10.1038/s41566-019-0583-3} {\bibfield  {journal} {\bibinfo  {journal} {Nature Photonics}\ }\textbf {\bibinfo {volume} {14}},\ \bibinfo {pages} {240} (\bibinfo {year} {2020})}\BibitemShut {NoStop}%
\bibitem [{\citenamefont {Gould}\ \emph {et~al.}(2021)\citenamefont {Gould}, \citenamefont {Yap}, \citenamefont {Adya}, \citenamefont {Slagmolen}, \citenamefont {Ward},\ and\ \citenamefont {McClelland}}]{PhysRevResearch.3.043079}%
  \BibitemOpen
  \bibfield  {author} {\bibinfo {author} {\bibfnamefont {D.~W.}\ \bibnamefont {Gould}}, \bibinfo {author} {\bibfnamefont {M.~J.}\ \bibnamefont {Yap}}, \bibinfo {author} {\bibfnamefont {V.~B.}\ \bibnamefont {Adya}}, \bibinfo {author} {\bibfnamefont {B.~J.~J.}\ \bibnamefont {Slagmolen}}, \bibinfo {author} {\bibfnamefont {R.~L.}\ \bibnamefont {Ward}},\ and\ \bibinfo {author} {\bibfnamefont {D.~E.}\ \bibnamefont {McClelland}},\ }\bibfield  {title} {\bibinfo {title} {Optimal quantum noise cancellation with an entangled witness channel},\ }\href {https://doi.org/10.1103/PhysRevResearch.3.043079} {\bibfield  {journal} {\bibinfo  {journal} {Phys. Rev. Res.}\ }\textbf {\bibinfo {volume} {3}},\ \bibinfo {pages} {043079} (\bibinfo {year} {2021})}\BibitemShut {NoStop}%
\bibitem [{\citenamefont {Wiseman}\ \emph {et~al.}(2007)\citenamefont {Wiseman}, \citenamefont {Jones},\ and\ \citenamefont {Doherty}}]{PhysRevLett.98.140402}%
  \BibitemOpen
  \bibfield  {author} {\bibinfo {author} {\bibfnamefont {H.~M.}\ \bibnamefont {Wiseman}}, \bibinfo {author} {\bibfnamefont {S.~J.}\ \bibnamefont {Jones}},\ and\ \bibinfo {author} {\bibfnamefont {A.~C.}\ \bibnamefont {Doherty}},\ }\bibfield  {title} {\bibinfo {title} {Steering, entanglement, nonlocality, and the einstein-podolsky-rosen paradox},\ }\href {https://doi.org/10.1103/PhysRevLett.98.140402} {\bibfield  {journal} {\bibinfo  {journal} {Phys. Rev. Lett.}\ }\textbf {\bibinfo {volume} {98}},\ \bibinfo {pages} {140402} (\bibinfo {year} {2007})}\BibitemShut {NoStop}%
\bibitem [{\citenamefont {Cavalcanti}\ \emph {et~al.}(2009)\citenamefont {Cavalcanti}, \citenamefont {Jones}, \citenamefont {Wiseman},\ and\ \citenamefont {Reid}}]{PhysRevA.80.032112}%
  \BibitemOpen
  \bibfield  {author} {\bibinfo {author} {\bibfnamefont {E.~G.}\ \bibnamefont {Cavalcanti}}, \bibinfo {author} {\bibfnamefont {S.~J.}\ \bibnamefont {Jones}}, \bibinfo {author} {\bibfnamefont {H.~M.}\ \bibnamefont {Wiseman}},\ and\ \bibinfo {author} {\bibfnamefont {M.~D.}\ \bibnamefont {Reid}},\ }\bibfield  {title} {\bibinfo {title} {Experimental criteria for steering and the einstein-podolsky-rosen paradox},\ }\href {https://doi.org/10.1103/PhysRevA.80.032112} {\bibfield  {journal} {\bibinfo  {journal} {Phys. Rev. A}\ }\textbf {\bibinfo {volume} {80}},\ \bibinfo {pages} {032112} (\bibinfo {year} {2009})}\BibitemShut {NoStop}%
\bibitem [{\citenamefont {Furusawa}\ \emph {et~al.}(1998)\citenamefont {Furusawa}, \citenamefont {Sørensen}, \citenamefont {Braunstein}, \citenamefont {Fuchs}, \citenamefont {Kimble},\ and\ \citenamefont {Polzik}}]{doi:10.1126/science.282.5389.706}%
  \BibitemOpen
  \bibfield  {author} {\bibinfo {author} {\bibfnamefont {A.}~\bibnamefont {Furusawa}}, \bibinfo {author} {\bibfnamefont {J.~L.}\ \bibnamefont {Sørensen}}, \bibinfo {author} {\bibfnamefont {S.~L.}\ \bibnamefont {Braunstein}}, \bibinfo {author} {\bibfnamefont {C.~A.}\ \bibnamefont {Fuchs}}, \bibinfo {author} {\bibfnamefont {H.~J.}\ \bibnamefont {Kimble}},\ and\ \bibinfo {author} {\bibfnamefont {E.~S.}\ \bibnamefont {Polzik}},\ }\bibfield  {title} {\bibinfo {title} {Unconditional quantum teleportation},\ }\href {https://doi.org/10.1126/science.282.5389.706} {\bibfield  {journal} {\bibinfo  {journal} {Science}\ }\textbf {\bibinfo {volume} {282}},\ \bibinfo {pages} {706} (\bibinfo {year} {1998})}\BibitemShut {NoStop}%
\bibitem [{\citenamefont {Braunstein}\ and\ \citenamefont {Kimble}(1998)}]{PhysRevLett.80.869}%
  \BibitemOpen
  \bibfield  {author} {\bibinfo {author} {\bibfnamefont {S.~L.}\ \bibnamefont {Braunstein}}\ and\ \bibinfo {author} {\bibfnamefont {H.~J.}\ \bibnamefont {Kimble}},\ }\bibfield  {title} {\bibinfo {title} {Teleportation of continuous quantum variables},\ }\href {https://doi.org/10.1103/PhysRevLett.80.869} {\bibfield  {journal} {\bibinfo  {journal} {Phys. Rev. Lett.}\ }\textbf {\bibinfo {volume} {80}},\ \bibinfo {pages} {869} (\bibinfo {year} {1998})}\BibitemShut {NoStop}%
\bibitem [{\citenamefont {Einstein}\ \emph {et~al.}(1935)\citenamefont {Einstein}, \citenamefont {Podolsky},\ and\ \citenamefont {Rosen}}]{PhysRev.47.777}%
  \BibitemOpen
  \bibfield  {author} {\bibinfo {author} {\bibfnamefont {A.}~\bibnamefont {Einstein}}, \bibinfo {author} {\bibfnamefont {B.}~\bibnamefont {Podolsky}},\ and\ \bibinfo {author} {\bibfnamefont {N.}~\bibnamefont {Rosen}},\ }\bibfield  {title} {\bibinfo {title} {Can quantum-mechanical description of physical reality be considered complete?},\ }\href {https://doi.org/10.1103/PhysRev.47.777} {\bibfield  {journal} {\bibinfo  {journal} {Phys. Rev.}\ }\textbf {\bibinfo {volume} {47}},\ \bibinfo {pages} {777} (\bibinfo {year} {1935})}\BibitemShut {NoStop}%
\bibitem [{\citenamefont {Buonanno}\ and\ \citenamefont {Chen}(2002)}]{PhysRevD.65.042001}%
  \BibitemOpen
  \bibfield  {author} {\bibinfo {author} {\bibfnamefont {A.}~\bibnamefont {Buonanno}}\ and\ \bibinfo {author} {\bibfnamefont {Y.}~\bibnamefont {Chen}},\ }\bibfield  {title} {\bibinfo {title} {Signal recycled laser-interferometer gravitational-wave detectors as optical springs},\ }\href {https://doi.org/10.1103/PhysRevD.65.042001} {\bibfield  {journal} {\bibinfo  {journal} {Phys. Rev. D}\ }\textbf {\bibinfo {volume} {65}},\ \bibinfo {pages} {042001} (\bibinfo {year} {2002})}\BibitemShut {NoStop}%
\bibitem [{\citenamefont {Zhang}\ \emph {et~al.}(2020)\citenamefont {Zhang}, \citenamefont {Martynov}, \citenamefont {Freise},\ and\ \citenamefont {Miao}}]{PhysRevD.101.124052}%
  \BibitemOpen
  \bibfield  {author} {\bibinfo {author} {\bibfnamefont {T.}~\bibnamefont {Zhang}}, \bibinfo {author} {\bibfnamefont {D.}~\bibnamefont {Martynov}}, \bibinfo {author} {\bibfnamefont {A.}~\bibnamefont {Freise}},\ and\ \bibinfo {author} {\bibfnamefont {H.}~\bibnamefont {Miao}},\ }\bibfield  {title} {\bibinfo {title} {Quantum squeezing schemes for heterodyne readout},\ }\href {https://doi.org/10.1103/PhysRevD.101.124052} {\bibfield  {journal} {\bibinfo  {journal} {Phys. Rev. D}\ }\textbf {\bibinfo {volume} {101}},\ \bibinfo {pages} {124052} (\bibinfo {year} {2020})}\BibitemShut {NoStop}%
\bibitem [{\citenamefont {Schumaker}\ and\ \citenamefont {Caves}(1985)}]{PhysRevA.31.3093}%
  \BibitemOpen
  \bibfield  {author} {\bibinfo {author} {\bibfnamefont {B.~L.}\ \bibnamefont {Schumaker}}\ and\ \bibinfo {author} {\bibfnamefont {C.~M.}\ \bibnamefont {Caves}},\ }\bibfield  {title} {\bibinfo {title} {New formalism for two-photon quantum optics. ii. mathematical foundation and compact notation},\ }\href {https://doi.org/10.1103/PhysRevA.31.3093} {\bibfield  {journal} {\bibinfo  {journal} {Phys. Rev. A}\ }\textbf {\bibinfo {volume} {31}},\ \bibinfo {pages} {3093} (\bibinfo {year} {1985})}\BibitemShut {NoStop}%
\bibitem [{\citenamefont {Caves}\ and\ \citenamefont {Schumaker}(1985)}]{PhysRevA.31.3068}%
  \BibitemOpen
  \bibfield  {author} {\bibinfo {author} {\bibfnamefont {C.~M.}\ \bibnamefont {Caves}}\ and\ \bibinfo {author} {\bibfnamefont {B.~L.}\ \bibnamefont {Schumaker}},\ }\bibfield  {title} {\bibinfo {title} {New formalism for two-photon quantum optics. i. quadrature phases and squeezed states},\ }\href {https://doi.org/10.1103/PhysRevA.31.3068} {\bibfield  {journal} {\bibinfo  {journal} {Phys. Rev. A}\ }\textbf {\bibinfo {volume} {31}},\ \bibinfo {pages} {3068} (\bibinfo {year} {1985})}\BibitemShut {NoStop}%
\bibitem [{\citenamefont {Duan}\ \emph {et~al.}(2000)\citenamefont {Duan}, \citenamefont {Giedke}, \citenamefont {Cirac},\ and\ \citenamefont {Zoller}}]{2000_Duan}%
  \BibitemOpen
  \bibfield  {author} {\bibinfo {author} {\bibfnamefont {L.-M.}\ \bibnamefont {Duan}}, \bibinfo {author} {\bibfnamefont {G.}~\bibnamefont {Giedke}}, \bibinfo {author} {\bibfnamefont {J.~I.}\ \bibnamefont {Cirac}},\ and\ \bibinfo {author} {\bibfnamefont {P.}~\bibnamefont {Zoller}},\ }\bibfield  {title} {\bibinfo {title} {Inseparability criterion for continuous variable systems},\ }\href {https://doi.org/10.1103/PhysRevLett.84.2722} {\bibfield  {journal} {\bibinfo  {journal} {Phys. Rev. Lett.}\ }\textbf {\bibinfo {volume} {84}},\ \bibinfo {pages} {2722} (\bibinfo {year} {2000})}\BibitemShut {NoStop}%
\bibitem [{\citenamefont {McCuller}\ \emph {et~al.}(2021)\citenamefont {McCuller} \emph {et~al.}}]{PhysRevD.104.062006}%
  \BibitemOpen
  \bibfield  {author} {\bibinfo {author} {\bibfnamefont {L.}~\bibnamefont {McCuller}} \emph {et~al.},\ }\bibfield  {title} {\bibinfo {title} {{LIGO}'s quantum response to squeezed states},\ }\href {https://doi.org/10.1103/PhysRevD.104.062006} {\bibfield  {journal} {\bibinfo  {journal} {Phys. Rev. D}\ }\textbf {\bibinfo {volume} {104}},\ \bibinfo {pages} {062006} (\bibinfo {year} {2021})}\BibitemShut {NoStop}%
\bibitem [{\citenamefont {Kwee}\ \emph {et~al.}(2014)\citenamefont {Kwee}, \citenamefont {Miller}, \citenamefont {Isogai}, \citenamefont {Barsotti},\ and\ \citenamefont {Evans}}]{PhysRevD.90.062006}%
  \BibitemOpen
  \bibfield  {author} {\bibinfo {author} {\bibfnamefont {P.}~\bibnamefont {Kwee}}, \bibinfo {author} {\bibfnamefont {J.}~\bibnamefont {Miller}}, \bibinfo {author} {\bibfnamefont {T.}~\bibnamefont {Isogai}}, \bibinfo {author} {\bibfnamefont {L.}~\bibnamefont {Barsotti}},\ and\ \bibinfo {author} {\bibfnamefont {M.}~\bibnamefont {Evans}},\ }\bibfield  {title} {\bibinfo {title} {Decoherence and degradation of squeezed states in quantum filter cavities},\ }\href {https://doi.org/10.1103/PhysRevD.90.062006} {\bibfield  {journal} {\bibinfo  {journal} {Phys. Rev. D}\ }\textbf {\bibinfo {volume} {90}},\ \bibinfo {pages} {062006} (\bibinfo {year} {2014})}\BibitemShut {NoStop}%
\bibitem [{\citenamefont {{Branchesi}}\ \emph {et~al.}(2023)\citenamefont {{Branchesi}}, \citenamefont {{Maggiore}}, \citenamefont {{Alonso}}, \citenamefont {{Badger}}, \citenamefont {{Banerjee}}, \citenamefont {{Beirnaert}}, \citenamefont {{Belgacem}}, \citenamefont {{Bhagwat}}, \citenamefont {{Boileau}}, \citenamefont {{Borhanian}}, \citenamefont {{Brown}}, \citenamefont {{Leong Chan}}, \citenamefont {{Cusin}}, \citenamefont {{Danilishin}}, \citenamefont {{Degallaix}}, \citenamefont {{De Luca}}, \citenamefont {{Dhani}}, \citenamefont {{Dietrich}}, \citenamefont {{Dupletsa}}, \citenamefont {{Foffa}}, \citenamefont {{Franciolini}}, \citenamefont {{Freise}}, \citenamefont {{Gemme}}, \citenamefont {{Goncharov}}, \citenamefont {{Ghosh}}, \citenamefont {{Gulminelli}}, \citenamefont {{Gupta}}, \citenamefont {{Kumar Gupta}}, \citenamefont {{Harms}}, \citenamefont {{Hazra}}, \citenamefont {{Hild}}, \citenamefont {{Hinderer}}, \citenamefont {{Siong Heng}}, \citenamefont {{Iacovelli}}, \citenamefont {{Janquart}},
  \citenamefont {{Janssens}}, \citenamefont {{Jenkins}}, \citenamefont {{Kalaghatgi}}, \citenamefont {{Koroveshi}}, \citenamefont {{Li}}, \citenamefont {{Li}}, \citenamefont {{Loffredo}}, \citenamefont {{Maggio}}, \citenamefont {{Mancarella}}, \citenamefont {{Mapelli}}, \citenamefont {{Martinovic}}, \citenamefont {{Maselli}}, \citenamefont {{Meyers}}, \citenamefont {{Miller}}, \citenamefont {{Mondal}}, \citenamefont {{Muttoni}}, \citenamefont {{Narola}}, \citenamefont {{Oertel}}, \citenamefont {{Oganesyan}}, \citenamefont {{Pacilio}}, \citenamefont {{Palomba}}, \citenamefont {{Pani}}, \citenamefont {{Pasqualetti}}, \citenamefont {{Perego}}, \citenamefont {{P{\'e}rigois}}, \citenamefont {{Pieroni}}, \citenamefont {{Piccinni}}, \citenamefont {{Puecher}}, \citenamefont {{Puppo}}, \citenamefont {{Ricciardone}}, \citenamefont {{Riotto}}, \citenamefont {{Ronchini}}, \citenamefont {{Sakellariadou}}, \citenamefont {{Samajdar}}, \citenamefont {{Santoliquido}}, \citenamefont {{Sathyaprakash}}, \citenamefont
  {{Steinlechner}}, \citenamefont {{Steinlechner}}, \citenamefont {{Utina}}, \citenamefont {{Van Den Broeck}},\ and\ \citenamefont {{Zhang}}}]{2023_ET_CoBA_paper}%
  \BibitemOpen
  \bibfield  {author} {\bibinfo {author} {\bibfnamefont {M.}~\bibnamefont {{Branchesi}}}, \bibinfo {author} {\bibfnamefont {M.}~\bibnamefont {{Maggiore}}}, \bibinfo {author} {\bibfnamefont {D.}~\bibnamefont {{Alonso}}}, \bibinfo {author} {\bibfnamefont {C.}~\bibnamefont {{Badger}}}, \bibinfo {author} {\bibfnamefont {B.}~\bibnamefont {{Banerjee}}}, \bibinfo {author} {\bibfnamefont {F.}~\bibnamefont {{Beirnaert}}}, \bibinfo {author} {\bibfnamefont {E.}~\bibnamefont {{Belgacem}}}, \bibinfo {author} {\bibfnamefont {S.}~\bibnamefont {{Bhagwat}}}, \bibinfo {author} {\bibfnamefont {G.}~\bibnamefont {{Boileau}}}, \bibinfo {author} {\bibfnamefont {S.}~\bibnamefont {{Borhanian}}}, \bibinfo {author} {\bibfnamefont {D.~D.}\ \bibnamefont {{Brown}}}, \bibinfo {author} {\bibfnamefont {M.}~\bibnamefont {{Leong Chan}}}, \bibinfo {author} {\bibfnamefont {G.}~\bibnamefont {{Cusin}}}, \bibinfo {author} {\bibfnamefont {S.~L.}\ \bibnamefont {{Danilishin}}}, \bibinfo {author} {\bibfnamefont {J.}~\bibnamefont {{Degallaix}}},
  \bibinfo {author} {\bibfnamefont {V.}~\bibnamefont {{De Luca}}}, \bibinfo {author} {\bibfnamefont {A.}~\bibnamefont {{Dhani}}}, \bibinfo {author} {\bibfnamefont {T.}~\bibnamefont {{Dietrich}}}, \bibinfo {author} {\bibfnamefont {U.}~\bibnamefont {{Dupletsa}}}, \bibinfo {author} {\bibfnamefont {S.}~\bibnamefont {{Foffa}}}, \bibinfo {author} {\bibfnamefont {G.}~\bibnamefont {{Franciolini}}}, \bibinfo {author} {\bibfnamefont {A.}~\bibnamefont {{Freise}}}, \bibinfo {author} {\bibfnamefont {G.}~\bibnamefont {{Gemme}}}, \bibinfo {author} {\bibfnamefont {B.}~\bibnamefont {{Goncharov}}}, \bibinfo {author} {\bibfnamefont {A.}~\bibnamefont {{Ghosh}}}, \bibinfo {author} {\bibfnamefont {F.}~\bibnamefont {{Gulminelli}}}, \bibinfo {author} {\bibfnamefont {I.}~\bibnamefont {{Gupta}}}, \bibinfo {author} {\bibfnamefont {P.}~\bibnamefont {{Kumar Gupta}}}, \bibinfo {author} {\bibfnamefont {J.}~\bibnamefont {{Harms}}}, \bibinfo {author} {\bibfnamefont {N.}~\bibnamefont {{Hazra}}}, \bibinfo {author} {\bibfnamefont
  {S.}~\bibnamefont {{Hild}}}, \bibinfo {author} {\bibfnamefont {T.}~\bibnamefont {{Hinderer}}}, \bibinfo {author} {\bibfnamefont {I.}~\bibnamefont {{Siong Heng}}}, \bibinfo {author} {\bibfnamefont {F.}~\bibnamefont {{Iacovelli}}}, \bibinfo {author} {\bibfnamefont {J.}~\bibnamefont {{Janquart}}}, \bibinfo {author} {\bibfnamefont {K.}~\bibnamefont {{Janssens}}}, \bibinfo {author} {\bibfnamefont {A.~C.}\ \bibnamefont {{Jenkins}}}, \bibinfo {author} {\bibfnamefont {C.}~\bibnamefont {{Kalaghatgi}}}, \bibinfo {author} {\bibfnamefont {X.}~\bibnamefont {{Koroveshi}}}, \bibinfo {author} {\bibfnamefont {T.~G.~F.}\ \bibnamefont {{Li}}}, \bibinfo {author} {\bibfnamefont {Y.}~\bibnamefont {{Li}}}, \bibinfo {author} {\bibfnamefont {E.}~\bibnamefont {{Loffredo}}}, \bibinfo {author} {\bibfnamefont {E.}~\bibnamefont {{Maggio}}}, \bibinfo {author} {\bibfnamefont {M.}~\bibnamefont {{Mancarella}}}, \bibinfo {author} {\bibfnamefont {M.}~\bibnamefont {{Mapelli}}}, \bibinfo {author} {\bibfnamefont {K.}~\bibnamefont
  {{Martinovic}}}, \bibinfo {author} {\bibfnamefont {A.}~\bibnamefont {{Maselli}}}, \bibinfo {author} {\bibfnamefont {P.}~\bibnamefont {{Meyers}}}, \bibinfo {author} {\bibfnamefont {A.~L.}\ \bibnamefont {{Miller}}}, \bibinfo {author} {\bibfnamefont {C.}~\bibnamefont {{Mondal}}}, \bibinfo {author} {\bibfnamefont {N.}~\bibnamefont {{Muttoni}}}, \bibinfo {author} {\bibfnamefont {H.}~\bibnamefont {{Narola}}}, \bibinfo {author} {\bibfnamefont {M.}~\bibnamefont {{Oertel}}}, \bibinfo {author} {\bibfnamefont {G.}~\bibnamefont {{Oganesyan}}}, \bibinfo {author} {\bibfnamefont {C.}~\bibnamefont {{Pacilio}}}, \bibinfo {author} {\bibfnamefont {C.}~\bibnamefont {{Palomba}}}, \bibinfo {author} {\bibfnamefont {P.}~\bibnamefont {{Pani}}}, \bibinfo {author} {\bibfnamefont {A.}~\bibnamefont {{Pasqualetti}}}, \bibinfo {author} {\bibfnamefont {A.}~\bibnamefont {{Perego}}}, \bibinfo {author} {\bibfnamefont {C.}~\bibnamefont {{P{\'e}rigois}}}, \bibinfo {author} {\bibfnamefont {M.}~\bibnamefont {{Pieroni}}}, \bibinfo {author}
  {\bibfnamefont {O.~J.}\ \bibnamefont {{Piccinni}}}, \bibinfo {author} {\bibfnamefont {A.}~\bibnamefont {{Puecher}}}, \bibinfo {author} {\bibfnamefont {P.}~\bibnamefont {{Puppo}}}, \bibinfo {author} {\bibfnamefont {A.}~\bibnamefont {{Ricciardone}}}, \bibinfo {author} {\bibfnamefont {A.}~\bibnamefont {{Riotto}}}, \bibinfo {author} {\bibfnamefont {S.}~\bibnamefont {{Ronchini}}}, \bibinfo {author} {\bibfnamefont {M.}~\bibnamefont {{Sakellariadou}}}, \bibinfo {author} {\bibfnamefont {A.}~\bibnamefont {{Samajdar}}}, \bibinfo {author} {\bibfnamefont {F.}~\bibnamefont {{Santoliquido}}}, \bibinfo {author} {\bibfnamefont {B.~S.}\ \bibnamefont {{Sathyaprakash}}}, \bibinfo {author} {\bibfnamefont {J.}~\bibnamefont {{Steinlechner}}}, \bibinfo {author} {\bibfnamefont {S.}~\bibnamefont {{Steinlechner}}}, \bibinfo {author} {\bibfnamefont {A.}~\bibnamefont {{Utina}}}, \bibinfo {author} {\bibfnamefont {C.}~\bibnamefont {{Van Den Broeck}}},\ and\ \bibinfo {author} {\bibfnamefont {T.}~\bibnamefont {{Zhang}}},\ }\bibfield
  {title} {\bibinfo {title} {{Science with the Einstein Telescope: a comparison of different designs}},\ }\href {https://doi.org/10.1088/1475-7516/2023/07/068} {\bibfield  {journal} {\bibinfo  {journal} {Journal of Cosmology and Astroparticle Physics}\ }\textbf {\bibinfo {volume} {2023}}\bibfield  {number} {\bibinfo  {number} { (7)},\ \bibinfo {eid} {068}},\ }\Eprint {https://arxiv.org/abs/2303.15923} {arXiv:2303.15923 [gr-qc]} \BibitemShut {NoStop}%
\bibitem [{\citenamefont {Buonanno}\ and\ \citenamefont {Chen}(2003)}]{PhysRevD.67.062002}%
  \BibitemOpen
  \bibfield  {author} {\bibinfo {author} {\bibfnamefont {A.}~\bibnamefont {Buonanno}}\ and\ \bibinfo {author} {\bibfnamefont {Y.}~\bibnamefont {Chen}},\ }\bibfield  {title} {\bibinfo {title} {Scaling law in signal recycled laser-interferometer gravitational-wave detectors},\ }\href {https://doi.org/10.1103/PhysRevD.67.062002} {\bibfield  {journal} {\bibinfo  {journal} {Phys. Rev. D}\ }\textbf {\bibinfo {volume} {67}},\ \bibinfo {pages} {062002} (\bibinfo {year} {2003})}\BibitemShut {NoStop}%
\bibitem [{\citenamefont {Buonanno}\ \emph {et~al.}(2003)\citenamefont {Buonanno}, \citenamefont {Chen},\ and\ \citenamefont {Mavalvala}}]{PhysRevD.67.122005}%
  \BibitemOpen
  \bibfield  {author} {\bibinfo {author} {\bibfnamefont {A.}~\bibnamefont {Buonanno}}, \bibinfo {author} {\bibfnamefont {Y.}~\bibnamefont {Chen}},\ and\ \bibinfo {author} {\bibfnamefont {N.}~\bibnamefont {Mavalvala}},\ }\bibfield  {title} {\bibinfo {title} {Quantum noise in laser-interferometer gravitational-wave detectors with a heterodyne readout scheme},\ }\href {https://doi.org/10.1103/PhysRevD.67.122005} {\bibfield  {journal} {\bibinfo  {journal} {Phys. Rev. D}\ }\textbf {\bibinfo {volume} {67}},\ \bibinfo {pages} {122005} (\bibinfo {year} {2003})}\BibitemShut {NoStop}%
\bibitem [{\citenamefont {Danilishin}\ and\ \citenamefont {Khalili}(2012)}]{Danilishin_2012}%
  \BibitemOpen
  \bibfield  {author} {\bibinfo {author} {\bibfnamefont {S.~L.}\ \bibnamefont {Danilishin}}\ and\ \bibinfo {author} {\bibfnamefont {F.~Y.}\ \bibnamefont {Khalili}},\ }\bibfield  {title} {\bibinfo {title} {Quantum measurement theory in gravitational-wave detectors},\ }\bibfield  {journal} {\bibinfo  {journal} {Living Reviews in Relativity}\ }\textbf {\bibinfo {volume} {15}},\ \href {https://doi.org/10.12942/lrr-2012-5} {10.12942/lrr-2012-5} (\bibinfo {year} {2012})\BibitemShut {NoStop}%
\end{thebibliography}%

\end{document}